\documentclass[12pt,preprint]{aastex}
\shorttitle{Seyfert 1.8s and 1.9s}
\shortauthors{Trippe et al.}
\usepackage{natbib}
\usepackage{psfig}
\bibpunct[ ]{(}{)}{;}{a}{}{,}
\begin{document}
\title{A Multi-Wavelength Study of the Nature of Type 1.8/1.9 Seyfert Galaxies}
\author{M. L. Trippe\footnotemark[1], D. M. Crenshaw\footnotemark[2], R. P. Deo\footnotemark[3], M. Dietrich\footnotemark[4], S. B. Kraemer\footnotemark[5], S. E. Rafter\footnotemark[6], and T.J. Turner\footnotemark[7]}
\footnotetext[1]{Department of Astronomy, University of Maryland, College Park, MD 20742}
\footnotetext[2]{Department of Physics and Astronomy, Georgia State University, One Park Place South SE, Ste. 700, Atlanta, GA 30303}
\footnotetext[3]{Drexel University Department of Physics, Disque Hall, South 32nd St., Rm. 813, Philadelphia, PA 19104}
\footnotetext[4]{Department of Astronomy The Ohio State University 4055 McPherson Lab 140 W. 18th Ave. Columbus, OH, 43210}
\footnotetext[5]{Institute for Astrophysics and Computational Sciences, Department of Physics, The Catholic University
of America, Washington, DC 20064; and Code 667, Astrophysics Science Division, NASA Goddard Space Flight Center, Greenbelt, MD 20771}
\footnotetext[6]{Dept. of Physics, Technion, Haifa 32000, Israel}
\footnotetext[7]{Dept. of Physics, University of Maryland Baltimore County, 1000 Hilltop Circle, Baltimore, MD 21250}
\begin{abstract}
We focus on determining the underlying physical cause of a Seyfert galaxy's appearance as type a 1.8 or 1.9. Are these ``intermediate'' Seyfert types typical Seyfert 1 nuclei with reddened broad-line regions? Or are they objects with intrinsically weak continua and broad emission lines? We compare measurements of the optical reddening of the narrow and broad-line regions with each other and with the X-ray column derived from {\it XMM-Newton} 0.5-10 keV spectra to determine the presence and location of dust in the line of sight. We also searched the literature to see if the objects showed evidence for broad-line variability, and determined if the changes were consistent with a change in reddening or a change in the intrinsic ionizing continuum flux.  We find that 10 of 19 objects previously classified as Seyfert 1.8/1.9s received this designation due to their low continuum flux. In four objects the classification was due to BLR reddening, either by the torus or dust structures in the vicinity of the NLR; in the remaining five objects there is not sufficient evidence to favor one scenario over the other. These findings imply that, in general, samples of 1.8/1.9s are not suitable for use in studies of the gas and dust in the central torus. 
\end{abstract}
\section{Introduction}
\indent The unified model for Seyfert galaxies attributes the diversity of their observed characteristics to their orientation with respect to our line of sight. According to this theory, a parsec-scale torus-shaped region of dusty molecular gas surrounds the innermost region of the nucleus where the broad emission lines originate around the continuum source, which is thought to be an accretion disk surrounding a central supermassive black hole. All Seyferts have optical spectra that show narrow permitted and forbidden emission lines from an extended (hundreds of parsecs scale) low-density narrow emission-line region (NLR). AGN classed as ``type 1'', based on the presence of broad components of the permitted lines in their optical spectra, are believed to have an orientation that allows a clear view to the central region. On the other hand, ``type 2'' AGN (whose optical spectra do not show broad permitted line components) are thought to have an orientation such that the material in this torus intercepts our line of sight and obstructs our view of the compact (light-days scale) broad emission-line region (BLR) and central continuum source \citep[see][for a review of AGN classification and the unified model]{tad08}. \\
\indent Some Seyferts, however, have characteristics that fall in between these two main types. \citet{ost81} added two important intermediate types to our classification system to describe these galaxies with composite spectral features. Seyfert 1.8 galaxies show weak broad components of H$\alpha$ and H$\beta$, while Seyfert 1.9 galaxies show just weak broad H$\alpha$. These AGN also show weaker featureless continua than galaxies classified as type 1. Osterbrock initially suggested the observed characteristics of Seyfert 1.8s and 1.9s could be due to the reddening of the continuum and BLR by dust \citep{ost81}. In terms of the unified model, this would indicate a line of sight that grazes the outer edge of the obscuring dusty torus. This idea was supported by \citet{goo95}, who showed variations in the Balmer decrements of a few of these objects to be consistent with changes in the amount of reddening $E(B-V)$, suggesting the motion of clouds of dust (presumably in the torus) in front of the nucleus. \\
\indent Nuclear dust lanes and spirals may also play a role in the reddening of the BLR in intermediate types. \citet{mal98} performed a {\it Hubble Space Telescope} ({\it HST}) imaging survey of nearby Seyferts using WFPC2 with a broadband (F606W) filter. Their study found that the Seyfert 2 nuclei in their sample were more frequently blocked by lanes/patches of host galaxy dust than Seyfert 1 nuclei, and they proposed that these galactic dust structures on scales of hundreds of parsecs could be a viable alternative to the classical parsec-scale torus model. \citet{pog02}, using the technique of ``structure mapping'' to enhance image contrast and draw out fine dust structures in the WFPC2 images, found that almost all of their sample of 43 nearby Seyferts show circumnuclear dust structures. However, they found there to be very few Seyfert 2 nuclei obscured by large-scale dust structures, and did not see any significant differences in the circumnuclear dust morphologies of Seyfert 1s and 2s. In any case, both these studies highlight the prevalence of dust structures on the size scale of the NLR, and emphasize that they are important reddening sources to consider. \\
\indent However, NGC 2992 is an example of an object that doesn't fit this picture of weak broad lines due to reddening. Although it was originally identified by \citet{war80} as a Seyfert 1.9, it has varied in its apparent type from a Seyfert 1 to a Seyfert 2 over the last few decades, and these variations are due to changes in the flux of the ionizing continuum, and the reaction of the BLR to these changes \citep{tri08,gil00}. It therefore seems probable that this AGN was in a low continuum state when observed by Ward et al., causing it to be classified as a 1.9 even though it is unobscured by a central torus. NGC 2992 does have a reddened nucleus, but the reddening originates from dust in the plane of the host galaxy, as evidenced by similar $E(B-V)$ values for the BLR and NLR, a large silicate 9.7 $\mu$m absorption feature in {\it Spitzer} IRS data, and the presence of a large dust lane running across the nucleus in its {\it HST} WFPC2 image \citep{tri08}. This particular case brought our attention to the fact that the weak broad lines in some 1.8/1.9s may be due to low ionizing continuum flux, in addition to reddening from the host galaxy. This idea is also supported by the study of \citet{goo95},  in which some 1.8/1.9s showed variability inconsistent with a change in reddening. Are most Seyfert 1.8/1.9s AGN with reddened BLRs? Or are they instead mostly objects with intrinsically weak continua? \\
\indent In our current study, we attempt to answer these questions by combining new optical spectra with archival data from other wavelength regimes in order to understand the presence and location of obscuration in a sample of Seyfert 1.8/1.9s. In doing so, we also hope to isolate those Seyferts whose broad lines are likely reddened by the torus, objects suitable for use in a study of the dust and gas in the torus, from those 1.8/1.9s in which the optical extinction, if present, is unrelated to the torus. \\ 
\subsection{Diagnostics}
\indent As mentioned in the last section, there are two main physical situations that could cause an AGN's appearance as a 1.8/1.9. In the first case, a type 1 Seyfert appears to be a 1.8/1.9 because reddening by dust extinguishes most of the BLR flux in the optical except the strongest line, H$\alpha$, and perhaps the next strongest, H$\beta$. The extinction may be from a dusty structure on the size scale of the NLR or smaller (i.e., ``internal reddening''), or it may be from larger-scale sources, such as kiloparsec-scale dust lanes or the plane of the host galaxy (i.e., ``external reddening''). In the other case, which we refer to as the ``low flux'' case, the continuum source has at least temporarily diminished, leaving only weak broad lines and causing the Seyfert to be classified as a 1.8/1.9 even if the nucleus is entirely unobscured by dust. While there is not a single definitive method to discern between these two possibilities, there are several observational clues that can help determine which is the most likely. \\
\indent The reddening $E(B-V)$ affecting the NLR and BLR may be determined by comparing the measured line ratios from the different regions with their theoretically predicted intrinsic values. Internal reddening is indicated by a reddening source between the BLR and NLR (i.e., $E(B-V)_{BLR}>E(B-V)_{NLR}$). As mentioned previously, this could be due to either 1) dust between the BLR and NLR, such as the torus, or 2) dust on the same scale as the NLR, such as nuclear dust spirals. On the other hand, if $E(B-V)_{BLR}\approx E(B-V)_{NLR}$, the reddening of the BLR is most likely due to dust in the host galaxy. For Seyfert 1.8s, whose spectra contain both broad H$\alpha$ and H$\beta$, both the BLR and NLR Balmer decrements are measurable, and one may easily determine if the BLR is being affected by internal or external reddening. However, for 1.9s, only a lower limit to the reddening of the BLR can be found from an upper limit to the broad H$\beta$ flux, ruling out this simple determination of the dust location with only an optical spectrum. For these objects an X-ray spectrum can help determine the location of a reddening source. X-ray spectra give information about the total line-of-sight gas column ($N_{H}$) to the very center of AGN, which can then be compared with the column to the NLR, as measured by the NLR reddening with an assumed dust/gas ratio.  For example, if the reddening of the NLR corresponds to a column density similar to the column derived from the the X-ray spectrum, it is unlikely that the central source is being reddened by a torus. Furthermore, for either type 1.8 or 1.9 a study of optical spectral variability can provide useful information; if variations in the strengths of broad H$\alpha$ and H$\beta$ are consistent with the same change in reddening, we can again infer the presence of a source of internal reddening \citep[see][]{goo95}. \\ 
\indent On the other hand, observed broad-line variations inconsistent with variable reddening indicate an intrinsically variable ionizing continuum, and could mean that the 1.8/1.9 is a low-flux state of an object that would normally appear as a Seyfert 1. For example, if the continuum flux an AGN that normally appears as a Seyfert 1 were to drop off by a factor of ten, broad H$\alpha$ would appear with only 1/10th of its normal strength. Thus, in such objects, the weakness of the broad line can be entirely independent of the amount of reddening. \\
\subsection{Overview}
\indent In this study, we consider a sample of 34 Seyferts classified as 1.8 or 1.9 by at least one of several sources: \citet{ost81}, \citet{ost93}, \citet{goo95}, \citet{mai95}, or \citet{qui01}. Basic data and previous classifications of our sample objects are listed in Table~\ref{sample}. The Seyferts in our sample are bright and relatively well-known, allowing us to establish the classification history of each based on previously published spectra in the literature. Section 2 describes the data reduction procedures for the three data sets we used to measure the absorption in these objects: ground-based optical spectra, {\it XMM-Newton} X-ray spectra, and {\it Spitzer} mid-IR IRS spectra. Section 3 describes the optical type classification criteria we used and gives our type classification of each object in the sample. Section 4 discusses the details of how we measured absorption from each set of data. Section 5 gives brief notes on the spectra of the Seyferts in our sample, and our classification of each object into one of the above categories, based on the evidence from our own data and from previously published sources. Section 6 is a summary and discussion of the findings of Section 5. \\
\section{Data}
\indent To utilize the diagnostics mentioned in the introduction and determine where in the AGN most of the absorption occurs, we collected data from three bands: optical, X-ray, and mid-IR. This section discusses the data acquisition and reduction process for each of these three data sets. \\
\subsection{Optical Spectra}
\indent The first step of our study was to obtain new optical spectra of each of the 1.8/1.9s in our sample to check their classification, analyze their Balmer line ratios, and look for evidence of variability in their broad lines. Nine of the AGN in our sample were observed by the Sloan Digital Sky Survey (SDSS); we retrieved these spectra and included them in our study.  Most of the other galaxies were observed between 2007 June and 2009 January at the R-C spectrograph on the Cerro Tololo Inter-American Observatory (CTIO) 1.5-m telescope in Chile, or at the DeVeny spectrograph on the 72$''$ Perkins telescope at Lowell Observatory in Arizona. Table~\ref{obslog} gives the observing log of these optical observations. \\
\indent Spectra from CTIO were observed using two different settings. The first used a grating with a resolution of 4.3 \AA\ (giving a dispersion of $\sim$1.47 \AA$/$pixel) to take blue spectra from approximately 3660-5420 \AA\ to include the H$\beta$ line. The second used a grating with a resolution of 3.1 \AA\ $($$\sim$1.10 \AA$/$pixel dispersion) and a Schott GG 495 filter to take red spectra from approximately 5630-6940 \AA, to include the H$\alpha$ line. We used a slit at a 90$^{\circ}$ P.A. centered on the galaxy's nucleus to obtain accurate fluxes on photometric nights. The majority of the spectra were taken using a 2$''$ slit width, but observations were also taken on nights when other slit widths were in use, as noted in Table~\ref{obslog}. To eliminate cosmic ray hits, we took three exposures each time each AGN was observed. The stars LTT 4364 and Feige 110 were also observed with these setting for the purpose of flux calibration. The spectra were then reduced and flux calibrated using standard IRAF reduction packages for long-slit spectroscopy. \\
\indent The same observing procedure was used during the observing runs at Lowell. There, most of the objects were observed with a grating with a resolution of 4.5 \AA\ ($\sim$2.18 \AA$/$pixel dispersion) and a 2$''$ slit aperture each night. A slight modification to the grating tilt was made after the first night of the observing run, and therefore several of the spectra have a wavelength range of 3185-6715 \AA, while the others range from 3255-6785 \AA. Mrk 1320 was additionally observed with a separate red setup, as its relatively high redshift meant H$\alpha$ was shifted off the spectrum taken with the first setup. The red setup used to observe Mrk 1320 used a grating with a resolution of 3.4 \AA\ ($\sim$1.67 \AA$/$pixel dispersion), and a Schott OG 570 filter to take a spectrum from 6300-9000 \AA. The standard stars EG 247 and Feige 34 were used for flux calibration of the objects observed at Lowell. \\
\indent We also include a spectrum of NGC 5273 taken using the GoldCam spectrograph on the 2.1-m telescope at Kitt Peak National Observatory (KPNO). The blue part of the spectrum, from approximately 3300-6000 \AA, was observed with a grating with a resolution of 3.1 \AA\ ($\sim$1.25 \AA$/$pixel dispersion), and the red part of the spectrum, from approximately 5500-7000 \AA, was observed with a grating with a resolution of 3.1 \AA\ ($\sim$1.26 \AA$/$pixel dispersion) and a Schott GG 495 filter. The spectrum was flux calibrated using observations of the standard stars BD 28+4211 (blue) and Feige 98 (red). \\
\indent To correct the AGN emission-line fluxes for contamination by stellar absorption lines in the host galaxy, we removed the host galaxy spectrum from the data by subtracting off a normal galaxy spectrum from \citet{kin96}, scaled to give the optimum fit to the observed object's absorption line features. Although most of our sample reside in spiral galaxies, the galaxies' nuclear spectra are dominated by the bulge, and therefore an elliptical galaxy spectrum with an old stellar population provided a good fit to the continuum of many of the objects. The remaining continuum was fit by a power-law and subtracted off, though this component was usually rather weak. Table~\ref{reddenings} gives the Hubble type of the host galaxy template spectrum used for each object, and the fraction of the continuum at 5100 \AA\ attributable to the host. The continuum-subtracted spectra are displayed in Appendix 1. \\
\subsection{X-ray Spectra}
\indent To investigate the X-ray properties of our sample, we retrieved the available archived X-ray observations taken by the European Photon Imaging Camera (EPIC) pn detector on board {\it XMM-Newton} (hereafter referred to as just {\it XMM}). Observations were available for 22 of our 34 objects. Two of these, Mrk 883 and NGC 2622 were observed by us through the {\it XMM} Guest Observer program (P.I.: Crenshaw). If the object was observed multiple times, we retrieved the spectra taken closest in time to our optical observations, but none of the observations were simultaneous. The observation date and ID number of each spectrum used is given in Table~\ref{xmodels}. \\
\indent The {\it XMM} data were reduced using the standard EPCHAIN processing script included in version 7.1.0 of the XMM-SAS (Science Analysis System) software. The data were filtered to exclude times of high background, and the source spectra were then extracted from circles 32$''$ in radius. Background spectra were extracted from an area free of any background sources on the same chip as the source, also from 32$''$-circles. Response matrices and ancillary response matrices were generated using the XMM-SAS tasks $rmfgen$ and $arfgen$. \\
\subsection{{\it Spitzer} IRS Spectra}
\label{spitzer_obs}
\indent Observations in the mid-IR give further insight into the dust properties of these galaxies, and to this end we retrieved and processed the available low-resolution {\it Spitzer} IRS spectra from the {\it Spitzer} archives (see Deo et al. 2007 for processing details) for the objects in our sample. These spectra are displayed in Appendix 4.
\section{Optical Type Classification}
\begin{figure}
\includegraphics[angle=90,scale=0.75]{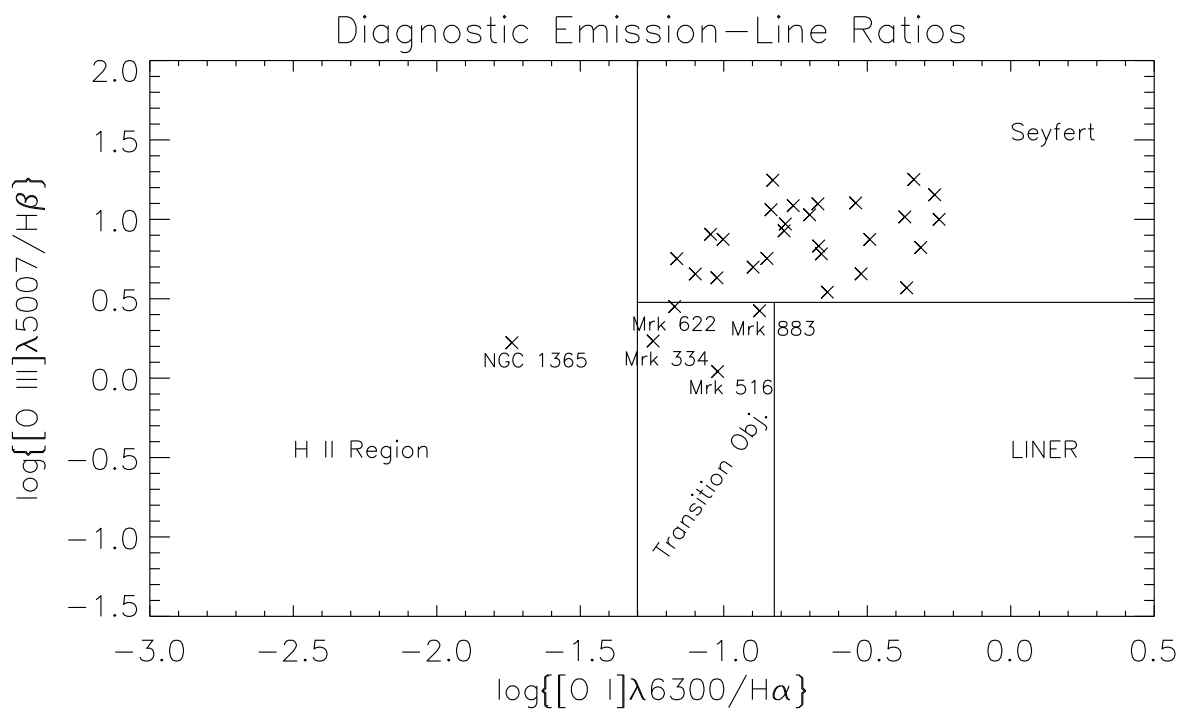}
\caption{A diagnostic diagram in the manner of of \citet{vei87} of the line ratios of the objects in our sample. Most of our objects fall into the Seyfert regime of this diagram, indicating they have high levels of ionization and large partially-ionized zones. Mrk 334, Mrk 516, Mrk 622, and Mrk 883 fall into the ``transition object'' region, indicating their less powerful continua and lower ionization states (though note we still refer to these AGN as ``Seyferts'' in the text), while the position of NGC 1365 indicates its emission includes a strong starburst component.\label{bpt}}
\end{figure}

\indent The optical spectra were first used to determine the classification type of the objects in our sample. Overall type classifications (starburst, LINER, or Seyfert) were made by plotting the position of these objects on a diagnostic diagram in the manner of \citet{vei87}, as shown in Figure~\ref{bpt}. The abscissa on this diagram is the ratio of the flux of the [O~I] $\lambda$6300 line to the flux of narrow H$\alpha$, a measure of the extent of the ``partially-ionized zone'' created by high-energy photons. The ordinate is the ratio of the flux of the [O~III] $\lambda$5007 line to the flux of narrow H$\beta$, a proxy for the level of ionization. Most of our objects do indeed fall into the Seyfert region of this diagram, indicating they have the high levels of ionization and large partially-ionized zones characteristic of Seyfert nuclei. Mrk 334, Mrk 516, Mrk 622, and Mrk 883 fall into the ``transition object'' region, indicating their less powerful continua and thus lower ionization states (however, note that we still refer to these AGN as ``Seyferts'' throughout the rest of this paper). Though it is not plotted in Figure~\ref{bpt} because it does not have measurable H$\beta$, we also classify NGC 2639 as a LINER on the strength of [O~I] $\lambda$3727 relative to [O~III] $\lambda$4959 and $\lambda$5007 \citep[in the manner of][]{hec80}. NGC 1365 has a very strong starburst component, yet shows strong X-ray emission and evidence for weak broad H$\alpha$ emission, and we therefore classify it as a Seyfert plus starburst compound. \\
\indent  The objects were then further classified into numerical sub-types, according to the definitions of these groups by  \citet{mai95} (see Appendix 2). Seyfert 1.8 galaxies are defined as Seyferts whose optical spectra show only weak broad H$\alpha$ and H$\beta$ components, while Seyfert 1.9 galaxies are defined by just weak broad H$\alpha$. However, these definitions (and others) are ambiguous about what constitutes a ``broad'' component, and have led to confusion about which objects should be designated as these types. To keep our classifications self-consistent, we decided on the following definition: ``broad'' emission is line emission that is significantly wider than the [O~III] $\lambda$5007 line, at least in the wings of the profiles, indicating that the emission is not from the inner NLR, but is generated in a distinct high-density region where [O~III] emission is collisionally suppressed (i.e., the BLR). \\
\indent Several factors make the detection faint broad lines a non-trivial process. The first major source of concern is the contamination of the AGN spectrum by starlight. Weak broad emission from the AGN can be hidden by pressure-broadened stellar absorption lines (for example, from a population of A stars) in the host galaxy. Or, the reverse may happen, as the continuum features due to K and M giants may lead to a false detection of broad H$\alpha$ \citep{fil85}. Our subtraction of the host galaxy continuum, described in the last section, is adequate to remove the worst of these problems. \\
\indent Secondly, H$\alpha$ must be very carefully disentangled from the adjacent [N~II] $\lambda\lambda$6548, 6583 doublet lines in order to detect a weak broad component. The [N~II] lines, along with the other narrow lines in the spectrum, are not usually simple Gaussians; they are often asymmetric and frequently have faint extended wings at their bases, and these features must be taken into account to avoid confusing their emission with that of weak broad H$\alpha$. We used the following procedure as our best attempt to de-blend the [N~II]-H$\alpha$ complex. First, we created a template profile from one of the other optical narrow lines, usually [O~III] $\lambda$5007 because it is the strongest and cleanest forbidden line in most of our spectra. If the [O~III] $\lambda$5007 line was obviously an unsuitable template for the lower-ionization [N~II] lines, for example if the lines showed very different shapes, the template was constructed instead from the [S~II] $\lambda$6716 and $\lambda$6731 lines, using the blue half of the $\lambda$6716 line and the red half of the $\lambda$6731 line of this usually mildly blended pair, in the manner of \citet{ho97}. A copy of the template line was then centered at the wavelengths of each of the [N~II] lines, scaled in width and height (both [N~II] lines were assumed to have the same width, and the flux of the line at 6583 \AA\ was set to 3 times that of the line at 6548 \AA\ in accordance with the ratio of their transition probabilities \citep{agn2}) and subtracted off the spectrum. This process was repeated with different template scale factors until the smoothest broad profile or continuum was attained. While forbidden lines with different critical densities or ionization potentials may have different profiles \citep{der84}, this method worked well overall. The H$\alpha$ profile after [N~II] subtraction for each object in our sample is shown in Appendix 2, with the [O~III] $\lambda$5007 line scaled to match the peak intensity of H$\alpha$ over-plotted as a dashed line for comparison of their widths. \\
\indent From the plots in Appendix 2, we find that many of the Seyferts in our sample, at least at the time when observed by us, do not fit our criteria for classification as 1.8s/1.9s. Our new type classifications for our 34 targets are given in Table~\ref{reddenings}. Eleven objects lack any sign of broad-line emission, and are thus Seyfert 2s. Eleven are type 1.0, 1.2, or 1.5, based upon the definitions of these subclasses by \citet{mai95}. There are eight 1.9s: Mrk 471, Mrk 915, NGC 2622, NGC 3786, NGC 5252, NGC 7314, UGC 7064, and perhaps NGC 1365, though the broad H$\alpha$ is very faint in this object. Seyferts 1.8s are much scarcer than 1.9s in our sample; we found only three true Seyfert 1.8s (Mrk 334, Mrk 1126, and UM 146). The remainder of the objects are LINERS or starburst galaxies. \\
\indent Interestingly, the plots in Appendix 2 show several objects for which the core of the narrow H$\alpha$ line is narrower than the [O~III] $\lambda$5007 line core: Mrk 334, Mrk 516, Mrk 609, Mrk 622, and NGC 7479. This may be due to the addition of a starburst component to the H$\alpha$ profile, as evidenced by the position of most of these objects below the ``Seyfert region'' in Figure~\ref{bpt}. In this case, a larger fraction of the H$\alpha$ emission is coming from the lower ionization gas compared to the [O~III] emission \citep[see][]{bru01}, resulting in a narrower H$\alpha$ profile. \\
\indent Likewise, it is important to note that though we use [O~III] $\lambda$5007 line as the best practical way to estimate the shape of the narrow H$\alpha$, the actual profile of this line is unknown, due to the [N~II] lines on its sides. Therefore, if the narrow H$\alpha$ line has intrinsically wider wings than [O~III], it could still be classified as a 1.9 by our definition even though it does not show true ``broad-line'' emission. This is not likely to be a problem for most objects in our sample, as comparison of the [O~III] line with the [N~II]-subtracted H$\alpha$ usually shows the ``broad'' component of H$\alpha$ to be a few thousand km s$^{-1}$, far broader than expected from the NLR. However, small differences in shape between H$\alpha$ and [O~III] could account for the emission we consider ``broad'' in one case (NGC 1365) in which the H$\alpha$ profile wings are only marginally wider than [O~III]. Thus, because we still classify NGC 1365 as a 1.9 (plus starburst), the number of objects classified as 1.9s in our sample may yet be an upper limit to those that show true broad-line emission. \\
\section{Absorption Measurements}
\indent In this section we discuss how the measurements of the absorption in each object were made from each data set. \\
\subsection{Reddening of the NLR and BLR from Optical Spectra}
\label{findred}
\begin{figure}
\includegraphics[angle=0,scale=0.75]{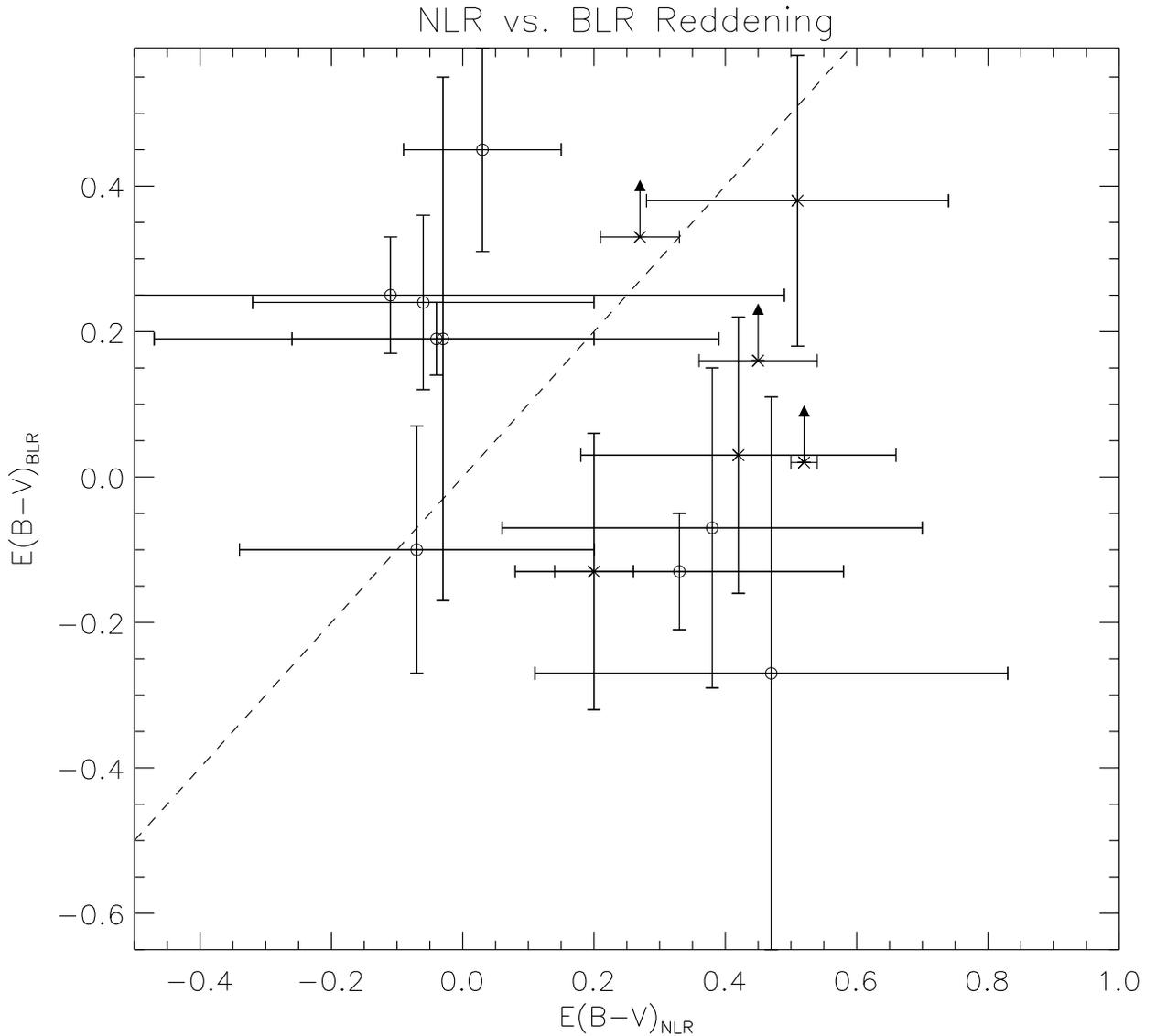}
\caption{The reddening of the BLR vs. the reddening of the NLR for the objects in our sample with both broad H$\alpha$ and H$\beta$, derived assuming an intrinsic flux ratio of H$\alpha$ to H$\beta$ of 2.9 in the NLR and 3.1 in the BLR, and the standard Galactic reddening curve of \citet{sav79}. For the 1.9s, the BLR reddening plotted is the lower limit estimated by the method described in Section~\ref{findred}. Points: Seyfert 1.8/1.9s represented by crosses, Seyfert 1.0/1.2/1.5s represented by open circles. The dashed line represents $E(B-V)_{BLR}=E(B-V)_{NLR}$. \label{eblrvsenlr} }
\end{figure}

\indent In order to measure the broad and narrow Balmer decrements, the narrow emission lines must be separated from the broad lines. We used the same template profile created to subtract the [N~II] lines, scaled in width and height by trial and error until its subtraction from the total blend left a smooth broad-line profile, to remove the narrow Balmer emission. Unfortunately, in objects with strong and irregularly-shaped broad H$\alpha$ (misclassified type 1-1.5 objects that crept into our sample), it is particularly difficult to discern which scale factors give the best fit, and the flux of the template line used to fit the narrow component could be scaled by up to $\pm$40\% and still give a reasonable-looking broad profile. While this lead to some error in the measurement of the flux of the broad components, it lead to a much larger percentage uncertainty in the flux of the narrow components of these objects. However, this method worked well for the true 1.8/1.9s, where the broad lines are reduced in strength and the broad and narrow components easier to differentiate. \\
\indent To determine the the reddening of the NLR in each object, we assumed the intrinsic ratio of the narrow components of H$\alpha$ and H$\beta$ to be equal to the Case B recombination value of f$_{H\alpha, NLR}$/f$_{H\beta, NLR}=2.9$, which is appropriate for the tenuous NLR gas \citep{agn2}. We further assumed the standard Galactic reddening curve $R(\lambda)$ of \citet{sav79} to calculate the reddening using the equation \begin{equation}E(B-V)=\frac{2.5}{R_{H\alpha}-R_{H\beta}}\log\frac{2.9}{\frac{f_{H\alpha}}{f_{H\beta}}}\end{equation}. \\
\indent For the objects that also displayed both broad H$\alpha$ and H$\beta$, the reddening to the BLR was calculated in a similar manner, but with the intrinsic Balmer decrement of the broad components assumed to be slightly higher than for the narrow components (3.1 instead of 2.9) to account for a small amount of collisional enhancement of H$\alpha$ in the higher-density BLR \citep{vei87}. \\
\indent There has been some debate about the validity of assuming a single number for the broad-line Balmer decrement in different objects. Photoionization models predict the ratio to vary widely depending upon the conditions of the BLR \citep{net75, kwa81, kor04}, and \citet{rud88} showed that the steep Balmer decrement of Mrk 609 (in its ``high''optical state observed in 1984) is not due to reddening, based upon the strength of the broad Ly$\alpha$ line in a simultaneous {\it IUE} spectrum. However, a recent observational study by \citet{don08} finds the broad Balmer decrements of a large homogeneous sample of blue (i.e. unreddened) Seyfert 1s/QSOs from SDSS to have a log-Gaussian distribution that peaks at 3.06, with a standard deviation of only 0.03 dex. Additionally, they find no correlation between Balmer decrement and luminosity, accretion rate, or continuum slope. We therefore retain the use of the standard value of f$_{H\alpha, BLR}$/f$_{H\beta, BLR}=3.1$ to calculate the BLR reddening. \\
\indent For the 1.9s, which show only weak broad H$\alpha$, only a lower limit to the reddening of the BLR can be derived. To measure the lower limit to H$\beta$ flux, and to determine if H$\beta$ should be visible in the spectrum in the absence of reddening, we used the following procedure. The broad component of the H$\alpha$ line was cut from the spectrum, scaled in width to make up for the velocity width difference at H$\beta$, and its flux divided by 3.1, to make a template representing the intrinsic emitted H$\beta$ line. This template was then added to the spectrum at the position of H$\beta$. In Mrk 915, NGC 3786, and NGC 7314, the addition of the template made a visible H$\beta$ line, indicating that H$\beta$ emission would be observable in these objects in the absence of reddening. The template was then multiplied by progressively smaller scale factors until the line became indistinguishable. This scale factor times the intrinsic expected H$\beta$ flux was then taken as an upper limit to the amount of  H$\beta$ present in these galaxies, and used to find a lower limit to the broad Balmer decrement and $E(B-V)_{BLR}$ in these objects. However, for most of the 1.9s this procedure revealed that broad H$\beta$ would not be visible even if the BLR were totally unreddened; the expected intrinsic H$\beta$ is indiscernible against the noise and the residuals of the host galaxy subtraction in Mrk 471, NGC 1365, NGC 2622, NGC 5252, and UGC 7064. Thus, for the majority of 1.9s we could not determine if the BLR is reddened based on an optical spectrum alone. \\
\indent The measured NLR and BLR reddening values and 1-$\sigma$ error bars or upper limits are given in Table~\ref{reddenings}. The negative reddening values in this table are consistent with zero within their errors. \\
\indent For the objects that show broad H$\beta$ and H$\alpha$, Figure~\ref{eblrvsenlr} shows the reddening of the NLR plotted against the reddening of the BLR. The Seyferts with types $<$1.8 have reddening points with very large error bars, largely due to the difficulty of separating out the narrow components from the strong broad emission in these objects. It is interesting to note that none of the three Seyfert 1.8s shows evidence for BLR reddening above that of the NLR. Also rather surprisingly, there is not much correlation between the values in Figure~\ref{eblrvsenlr}. About half of the objects are seen to have larger NLR reddening than BLR reddening. We suggest that this is due to the structure of the dust lanes/spirals in the vicinity of the NLR. The plotted values of $E(B-V)$ are averages over the entire emitting regions, and if one of these dusty structures does not happen to pass in front of the BLR, the NLR can have a larger average reddening than the BLR. Furthermore, the fact that the points are distributed equally above and below the dashed line representing $E(B-V)_{BLR}=E(B-V)_{NLR}$ argues against the reddening of the BLR in these objects being in general due to the atmosphere of a central dusty torus, but suggests instead that their reddening is due to the chance obscuration by a dusty structure in the NLR crossing directly over the nucleus. \\
\subsection{N$_H$ from X-ray Spectra}
\begin{figure}
\includegraphics[angle=90,scale=0.8]{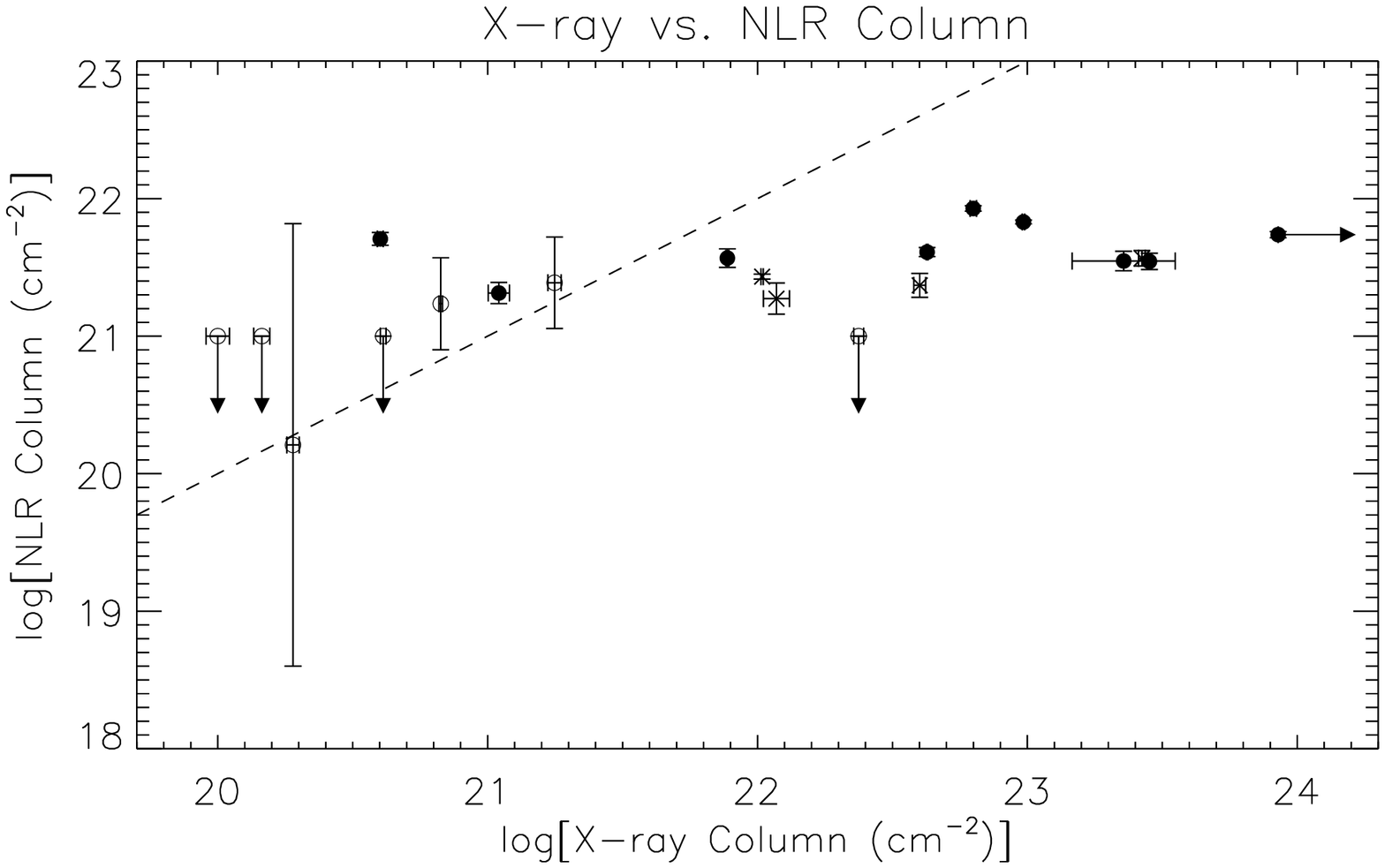}
\caption{ The X-ray column versus the column to the NLR, estimated from the reddening assuming the standard Galactic dust/gas ratio \citep{shu85}. Points: Seyfert 2s represented by filled circles, Seyfert 1.8/1.9s represented by crosses, Seyfert 1.0/1.2/1.5s represented by open circles. The dashed line represents $N_{H, NLR}=N_{H, X-ray}$. \label{nlrvsx} }
\end{figure}

\begin{figure}
\includegraphics[angle=90,scale=0.8]{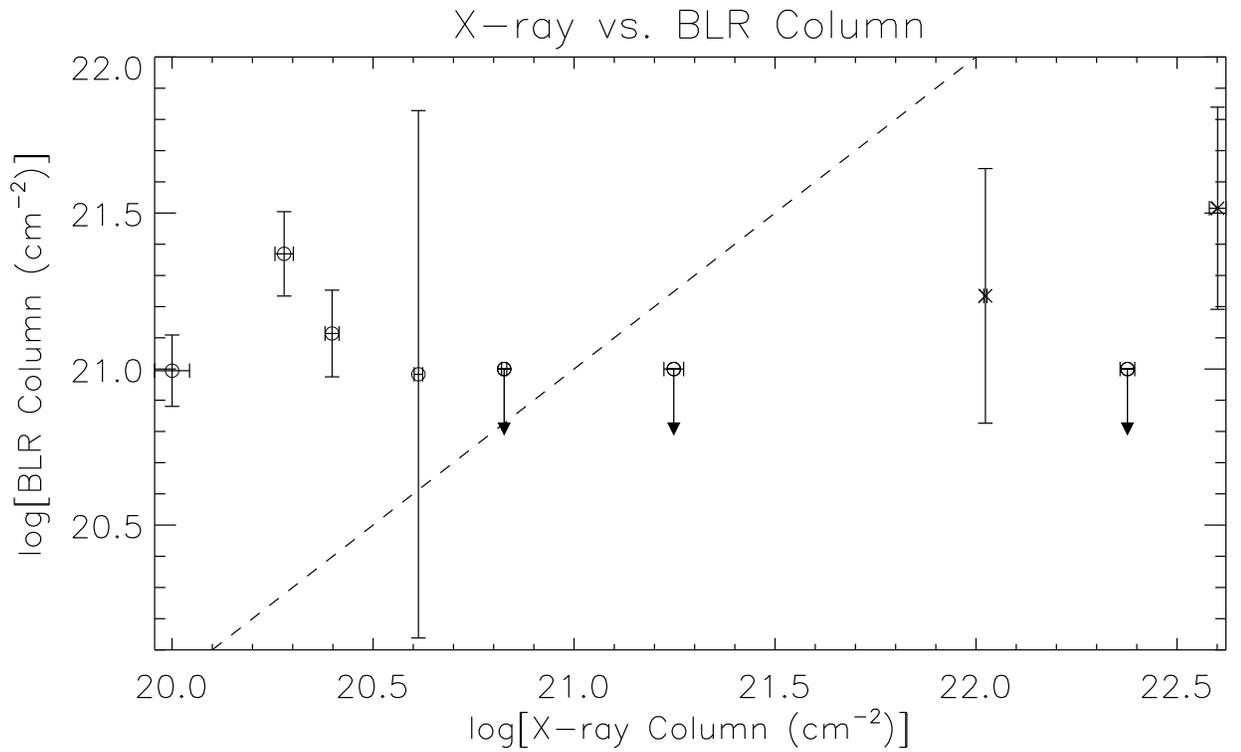}
\caption{ The X-ray column versus the column to the BLR, estimated from the reddening assuming the standard Galactic dust/gas ratio \citep{shu85}.  Points: Seyfert 1.8/1.9s represented by crosses, Seyfert 1.0/1.2/1.5s represented by open circles. The dashed line represents $N_{H, BLR}=N_{H, X-ray}$. \label{blrvsx}}
\end{figure}

\indent To determine the absorbing hydrogen columns (N$_{H}$) of the objects in our sample, the X-ray spectra of the source minus the background in the 0.5 to 10 keV range were fit using XSPEC version 12.3.1. Three models were initially fit to each data set and then compared to determine the best fit. The first was a simple powerlaw modified by Galactic absorption only ($tbabs_{gal}*powerlaw$ in XSPEC, with {\it tbabs} set to the Galactic column), and the second a powerlaw absorbed by a variable cold column ($tbabs_{gal}*tbabs*powerlaw$, with the second {\it tbabs} free). In the third, a second absorbed powerlaw was added, with spectral index tied to that of the first powerlaw; this additional component represents reflected/scattered continuum X-rays or a component of the direct continuum that is not covered by the absorber. Though one of these simple models usually provided a good fit to the data above $\sim$2 keV, many objects showed emission below 2 keV in excess of these preliminary models. Although this ``soft excess'' is a commonly observed feature in the spectra of X-ray obscured AGN \citep{gua05}, its origin is still unknown, though there is increasing evidence that it is due to unresolved emission lines from an extended region of photoionized gas \citep{gua07}. As our main goal was to determine the column density to the central source, we did not attempt detailed models of the soft excess but merely added a thermal plasma emission component ({\it mekal}) or bremsstrahlung emission component ({\it zbremss}) as needed to obtain a good phenomenological fit of the soft end of the spectrum. For the few objects for which the model still did not provide a good fit, we added additional model components as necessary. The final continuum model was selected as the model with the lowest $\chi^{2}$ (see Table~\ref{xmodels}).  We also tested each spectrum for the presence of Fe K$\alpha$ emission by adding a Gaussian profile at 6.4 keV. When the K$\alpha$ line was detected at the 95\%  confidence level, we included it in the spectral model and measured its equivalent width (EW); when the addition of the line did not lead to a significant reduction in $\chi^{2}$,  we only measured the EW upper limit. These values are reported in Table~\ref{xmodels}. Plots of the final unfolded models are presented in Appendix 3. \\
\indent The column densities measured from the spectral fits are presented in Table~\ref{xmodels}. If more than one absorbed powerlaw was used, the column reported in Table~\ref{xmodels} is that of the most heavily absorbed component. Figure~\ref{nlrvsx} presents a comparison of the X-ray column densities with the NLR column densities of all the objects in our sample with X-ray data. The NLR column densities were estimated from the optical NLR reddening measurements, assuming the relation $N_{H}=5.2\times10^{21}E(B-V)$ cm$^{-2}$, derived by \citet{shu85} for the local interstellar medium. As one would expect based on the simplest form of the unified model, most of the objects have greater X-ray columns than NLR columns. In particular, most of the Seyfert 2s and one of the Seyfert 1s have substantially larger X-ray columns than NLR columns, indicating a source of obscuration on a much smaller scale than the NLR. \\
\indent However, two Seyfert 2s in Figure~\ref{nlrvsx} are surprisingly placed: Mrk 609 and Mrk 883 have NLR columns significantly {\it larger} than their X-ray columns suggest (see Figures~1 and 2 in in Appendix 3 for their {\it XMM} spectra). These appear to be ``true'' or ``naked'' Seyfert 2s \citep{pap01, pan02, bia08, pan09}--objects which have apparently unobscured inner nuclei, but which show no evidence of BLRs. These AGN will be discussed in more detail in an upcoming paper (Trippe et al. 2010, in prep.). \\
\indent Figure~\ref{blrvsx} shows the X-ray column vs. the BLR column of those objects with broad H$\beta$, with the BLR columns again estimated by assuming the galactic gas/dust ratio of \citet{shu85}. In three of the Seyfert 1s, Mrk 728, Mrk 1018, and NGC 5033, the BLR-obscuring columns are larger than the X-ray columns. This may indicate that their intrinsic Balmer decrements are slightly greater than the assumed value of 3.1 (their measured H$\alpha$/H$\beta$ ratios are 5.1, 4.1, and 3.8, respectively), causing their derived BLR reddenings to be too high. Or, it may indicate that for some reason the broad lines are more absorbed than the continuum, perhaps by self-absorption. In the rest of the objects, the X-ray columns are consistent with or greater than the BLR columns, in accordance with the expectation there may be a component of X-ray absorbing material inside or originating from within the dust sublimation radius, which does not add appreciably to the extinction in the optical \citep[see][]{wei02}. \\
\subsubsection{Column Densities of Seyfert 1.8/1.9s}
\indent Unfortunately, only a few of the objects that appear as 1.8s/1.9s in our optical spectra had archival {\it XMM} observations. None of the 1.8s had {\it XMM} data available, and only five Seyfert 1.9s did: NGC 1365 (variable N$_{H}$, ranging from $\approx10-27\times10^{22}$ cm$^{-2}$ in our spectra), NGC 2622 (N$_{H}\approx1\times10^{22}$ cm$^{-2}$), NGC 3786 (N$_{H}\approx4\times10^{22}$ cm$^{-2}$), NGC 5252 (N$_{H}\approx27\times10^{22}$ cm$^{-2}$), and NGC 7314 (N$_{H}\approx1.0\times10^{22}$ cm$^{-2}$). Though the 1.9s show significant absorption, their columns are generally lower than that of typical Seyfert 2s. This confirms the results of \citet{ris99}, who found Seyfert 1.8s/1.9s to have a distribution of column densities that falls in between the distributions of Seyfert 1s and Seyfert 2s. \\
\subsection{Cold Dust Column from Spitzer 9.7$\mu$m Absorption}
\indent Mid-IR spectra (see Appendix 4) give information on the temperature of the dust from the shape of the continuum and the amount of extinction due to cold dust in the host galaxy from the depth of the silicate 9.7 $\mu$m feature \citep{deo07}. In our current study, in which we wish to pin down the location of the dust responsible for the reddening in Seyfert 1.8/1.9s, the estimate of the reddening from host galaxy from the mid-IR is useful for comparison with the $E(B-V)$ values of the NLR and BLR.  Unfortunately, the 9.7 $\mu$m line may be surrounded by strong PAH emission lines, which can make placement of the local continuum uncertain and affect line measurement. Nevertheless, we were able to measure the optical depth at the center of the absorption ($\tau_{9.7\mu m}$), and estimate the intrinsic dust extinction of the host galaxy using the relationship for diffuse ISM clouds derived by \citet{roc84}, $A_{V}=18.5\tau_{9.7\mu m}$. The corresponding N$_{H}$ values are listed in Table~\ref{sumtbl}. \\
\section{Notes on Individual Objects}
\label{indiv_comments}
\indent In this section, we give our type classification of each Seyfert in our sample, and summarize its classification history from the literature. From this, we split the objects in our sample into three groups: first, objects that appear as 1.8/1.9s in our optical spectra, second, objects classified as 1.8s/1.9s in the past, that no longer appear as 1.8/1.9s in our spectra, third, objects that were likely originally misclassified as 1.8/1.9s. For the objects in the first two groups, we combine our absorption measurements with previously published data to try to determine if the object appeared as a 1.8/1.9 due to low-flux or internal reddening. \\ 

\noindent {\bf Current 1.8/1.9s} \\

\noindent 1. Mrk 334--Type 1.8, due to low flux \\
\indent The spectrum of Mrk 334 taken at Lowell Observatory in January 2009 shows weak broad H$\alpha$ and very weak broad H$\beta$ after subtraction of the host galaxy spectrum, and thus we classify it as a type 1.8.  The weakness of the broad components makes the broad Balmer decrement difficult to measure, but it seems that the BLR is essentially unreddened. The {\it Spitzer} IRS spectrum of Mrk 334 shows strong PAH emission features due to the star-forming ring in its central kiloparsec \citep{deo07dis}. \\
\indent Mrk 334 was originally classified as a 1.8 by \citet{dah88}, and does not seem to have undergone any type variability since. It was again classified as a 1.8 by \citet{ost93} from a spectrum taken in September 1991, and appears no different in \citet{gal96}. \\
\indent Because the BLR appears to be essentially unreddened and because its position in Figure~\ref{bpt} is slightly below the Seyfert regime, indicating a history of low ionizing flux, we classify Mrk 334 as a low-flux type 1.8.\\

\noindent 2. Mrk 471--Type 1.9, inconclusive cause \\
\indent This object's SDSS spectrum from 2005 shows it to be a Seyfert 1.9 with a highly reddened NLR ($E(B-V)_{NLR}=0.89$). The {\it Spitzer} spectrum of Mrk 471 is heavily contaminated by PAH emission features, making measurement of the 9.7 $\mu$m absorption trough uncertain, but it seems to indicate a reddening of $E(B-V)_{NLR}\approx0.88$, very similar to that of the NLR.\\
\indent There is no evidence for variable broad components in Mrk 471. \\

\noindent 3. Mrk 915--Type 1.9, inconclusive cause \\
\indent Our CTIO spectra from June 2008 show this to be a Seyfert 1.9 galaxy. It is possible that H$\beta$ shows an extremely weak broad component, at the limit of detectability, but not significant enough to classify this as a 1.8. The lower limit of the reddening of the BLR ($E(B-V)_{BLR}>0.33$), derived from the strength of H$\alpha$, is close to the reddening of the NLR ($E(B-V)_{NLR}=0.27$), and originates at least in part in the dust lane seen crossing the central source in its WFPC2 image. \\
\indent Goodrich reports type variability in Mrk 915 between 1984 and 1993, when it appeared to go from a 1.5 to a 1.9. Though he tentatively suggests that the change could be due to a change in reddening, there is not enough data to convincingly favor this scenario.  \\

\noindent 4. Mrk 1126--Type 1.8, due to low flux \\
\indent In our spectrum from CTIO in October 2007, both weak broad $H\alpha$ and very weak broad $H\beta$ are apparent after subtraction of the host galaxy spectrum. The reddening of the BLR and the reddening of the NLR are consistent within their errors, indicating the low-flux scenario for this object. The WFPC2 image of Mrk 1126 shows it to reside in an almost circular face-on spiral galaxy.\\
\indent Our literature search did not provide any evidence for variability; past authors consistently classify it as a 1.8 \citep{bot04, sch90}, with the exception of \citet{ost85} who classified it as a narrow-line Seyfert 1 (NLS1) on the basis that high-ionization iron lines such as [Fe VII] and [Fe X] are present in its spectrum. However, because its Balmer lines are clearly a composite of broad and narrow components, and there there is no measurable [Fe II] emission, we believe that it was wrongly classified at that time, and that 1.8 would have been a better designation. \\

\noindent 5. NGC 1365--Type 1.9 (+starburst), inconclusive cause \\
\indent We classify NGC 1365 as a 1.9, though the broad component of H$\alpha$ in our 2009 January spectrum of this object is extremely faint, if it is present at all. Narrow H$\beta$ appears to be enhanced relative the [O~III] lines, indicating a very strong starburst emission component. This component is also evident in NGC 1365's {\it Spitzer} spectrum, which displays strong PAH emission features. NGC 1365 has been studied extensively in the X-ray, and has been observed to undergo rapid variations in column density \citep{ris05, ris07}. \citet{ris09a} observed changes in X-ray flux that they attribute to an occultation event, and they infer that the X-ray absorption and reflection originates in the BLR clouds. If so, these clouds are well within the dust sublimation radius and the changes seen in the X-ray are therefore probably not related to changes seen in the optical spectra. We base our spectral models of NGC 1365 on the in-depth analysis of \citet{ris09b}; our models show the time-averaged column to be $\approx10\times10^{22}$ cm$^{-2}$ during an observation in January 2004 and $\approx27\times10^{22}$ cm$^{-2}$ during an observation in July 2004. \\
\indent There is some evidence of optical type variability in this object. \citet{sch94} comments on the ``conspicuous'' broad component of H$\alpha$ observed in October 1988, and in \citet{sch99} the authors show a figure of the H$\beta$ profile which appears to have a substantial broad component. Unfortunately, this is still not enough information to establish the cause of this possible variation over time. It seems feasible that, as suggested by \citet{edm82}, most of the BLR absorption occurs in the dust lane which covers the nucleus. \\

\noindent 6. NGC 2622--Type 1.9, due to low flux \\
\indent This object is classified as a type 1.9 based on its SDSS spectrum from 2004. Its {\it XMM} spectrum is well-fit by powerlaw with $\Gamma=2$ attenuated by a column density of $\approx1.2\times10^{22}$ cm$^{-2}$. The available {\it HST} GHRS spectrum shows narrow, but no broad Ly$\alpha$. Conversely, this object's {\it Spitzer} IRS spectrum looks like that of a type 1 AGN, with a continuum rise to lower wavelengths, indicating emission from a hot component. \\
\indent This object has undergone classification changes in the past. \citet{ost83} classed it as a 1.8, but \citet{goo89} observed it as a Seyfert 1, and noted that the changes in the fluxes of the broad lines were consistent with changes in the reddening. He also later showed that the Pa$\beta$ line flux of this object was also consistent with this theory. However, when NGC 2622 faded again in 1993 \citep{goo95}, it did so in a manner inconsistent with a change in reddening. We therefore classify NGC 2622 as a low-flux object, though its nucleus may also be affected by a significant amount of reddening. \\

\noindent 7. NGC 3786--Type 1.9, inconclusive cause \\
\indent We classify NGC 3786 as a Seyfert 1.9 in our Lowell spectrum of this object from January 2009. It is possible that a very faint component to H$\beta$ is also present, but it is at the limit of detectability, even after the correction for the host galaxy contamination. We were not able to tell whether the BLR is more reddened than the NLR; the large uncertainty in the lower limit of reddening of the BLR puts it within the range derived for the NLR. Its {\it Spitzer} IRS spectrum displays the strong PAH emission features indicative of a large starburst, and 9.7 $\mu$m absorption corresponding to a reddening of $E(B-V)=1.74\pm0.22$. NGC 3786's {\it XMM} X-ray spectrum is well-fit by a double powerlaw model, with one component absorbed by a column $N_{H}\approx4\times10^{22}$ cm$^{-2}$, and the other essentially unabsorbed. \\
\indent It is difficult to tell if the broad lines have undergone any variation over time. Past spectra of NGC 3786 certainly look very similar to our own \citep[see][]{huc82, ost83}. \citet{goo83} classify it as a Seyfert 1.8, and measure the broad Balmer decrement to be $\sim8.4$. \citet{kom97} point out that this amount of reddening corresponds to a column of $5.5\times10^{21}$ cm$^{-2}$ to BLR, which closely matches the column they derived by fitting a warm absorber model to its X-ray spectrum from ROSAT. The {\it XMM} spectrum from 2004 is not high enough quality to provide information about the ionization state of the absorber to test this claim. \\

\noindent 8. NGC 5252--Type 1.9, due to reddening \\
\indent The SDSS spectrum from April 2002 shows the spectrum of a Seyfert 1.9 object. It has a reddened NLR, with $E(B-V)_{NLR}\approx0.71$. Based on the weakness of H$\alpha$, H$\beta$ is below the level of detectability in our spectrum, even if the BLR is completely unreddened. Its {\it XMM} spectrum is well-fit by two emission components, one highly absorbed ($N_{H}\approx26.5\times10^{22}$ cm$^{-2}$), and one less absorbed ($N_{H}\approx3.8\times10^{22}$ cm$^{-2}$). \\
\indent There is no evidence for variability in NGC 5252's past observations. However, it is known from polarization studies \citep{you96} that this object has a hidden BLR, and we therefore classify it as a reddened object. \\

\noindent 9. NGC 7314--Type 1.9, due to internal reddening \\
\indent Our CTIO spectra show broad H$\alpha$ but no H$\beta$. We note that though H$\alpha$ is wider than [O~III] $\lambda$5007 (and thus ``broad'' by our definition), it is still relatively narrow, with a FWHM of only $\approx1600$ km s$^{-1}$, similar to NLS1s. Our {\it XMM} spectrum shows it to have a column of $1.1\times10^{22}$ cm$^{-2}$. Interestingly, its {\it Spitzer} IRS spectrum \citep{deo07} looks like a Seyfert 1. \\
\indent NGC 7314 is shown to have a hidden BLR by \citet{lum04}. And, based on the rapid variability in its X-ray spectrum, \citet{dew05} classify it as another ``obscured NLS1''. From this evidence, we expect that NGC 7314 is another case of an internally reddened Seyfert. \\

\noindent 10. UGC 7064--Type 1.9, due to internal reddening \\
\indent There is a very faint broad H$\alpha$ component in our optical spectrum of UGC 7064 from Lowell, but no broad H$\beta$, and thus we classify this object as a type 1.9. The reddening of the NLR was found to be $E(B-V)_{NLR}=0.23$ from this spectrum. The {\it HST} GHRS spectrum of this object shows only narrow Ly$\alpha$. Its {\it Spitzer} IRS spectrum shows a rise in its continuum towards short wavelengths, typical of type 1 AGN. \\
\indent There does not seem to be any evidence for variability in the observed history of UGC 7064. \citet{ost88} and \citet{goo89} classify it as a 1.9, and a blue spectrum from \citet{sal95} shows H$\beta$ that looks the same as in our spectrum. Because a broad H$\alpha$ component was seen in spectropolarimetric observations of UGC 7064 \citep{smi04}, we classify it as a reddened object. \\

\noindent 11. UM 146--Type 1.8, due to low flux \\
\indent Our blue CTIO spectrum of this object shows a very weak broad component to H$\beta$ after the subtraction of the host galaxy spectrum, and thus we classify UM 146 as a type 1.8. The weakness of these components makes measuring the BLR Balmer decrement somewhat uncertain, but it seems that the BLR is not significantly reddened ($E(B-V)_{BLR}=0.03\pm0.19$), while the NLR is reddened by $E(B-V)_{NLR}=0.42\pm0.24$, with the higher reddening of the NLR possibly due to dust spirals/lanes within the NLR. Its {\it Spitzer} IRS spectrum shows clear silicate absorption, indicating that dust in the host galaxy could also be playing a role in the reddening of the NLR. \\
\indent \citet{ost83} classed UM 146 as a 1.5, as do \citet{dah88}. However, neither publication shows the actual spectrum that lead to this classification, and thus we are unable to confirm if the broad components weakened enough for it to have changed type. Because of the low reddening of the BLR, we classify UM 146 as a low-flux object. \\

\noindent {\bf Previous 1.8/1.9s} \\

\noindent 1. Mrk 609--Type 2 (previously a type 1.8, due to low flux) \\
\indent Our CTIO spectrum of Mrk 609 taken in 2007 Oct./Nov. shows it to be a Seyfert 2. It also shows wide [O~III] $\lambda$4959 and $\lambda$5007 lines relative to the other forbidden lines in the spectrum.  Its {\it Spitzer} IRS spectrum shows strong PAH emission features, but little if any 9.7 $\mu$m absorption.  More unusually, both {\it XMM} observations of Mrk 609 show a powerlaw absorbed by only a very small column, consistent with Galactic absorption. This makes Mrk 609 what is known as ``true'' or ``naked'' Seyfert 2 candidate \citep{pan02}, because it appears that we have a clear view to its nucleus, but there is no sign of the broad-line region in the optical, in apparent contradiction of the orientation-dependent unified model. We will explore this possibility in more detail in an upcoming paper. \\
\indent There is evidence that Mrk 609 has undergone type variability since its original classification. Mrk 609 was initially classified as a 1.8 by \citet{ost81} from spectra taken in 1975 and 1976, but he later revised the type to be a 1.5 with better spectra. Goodrich observed it again in November 1986 and also classified it as a 1.5, but noted that it had a large Balmer decrement. \citet{rud88} present a groundbased optical spectrum made simultaneously with an {\it IUE} observation. Broad Balmer components are visible in the optical spectrum, and the {\it IUE} spectrum also shows broad Ly$\alpha$. Interestingly, they found that the ratio $Ly\alpha/H\beta/H\alpha=16/1.0/5.0$ and infer that the BLR is basically unreddened. \\

\noindent 2. Mrk 728--Type 1.2 (previously a type 1.9, due to low flux) \\
\indent The SDSS spectrum from 2004 of Mrk 728 shows it to be a type 1 Seyfert, or, based on the flux ratio requirements of [O~III]$\lambda$5007/H$\beta$ of \citet{mai95}, a type 1.2. The strong broad Ly$\alpha$ line in its GHRS spectrum is consistent with this classification. Its {\it XMM} spectrum shows absorption consistent with the Galactic value. \\
\indent  \citet{ost83} first classified Mrk 728 as a Seyfert 1.9, and it was also classified as a 1.9 when observed by Goodrich in 1986 \citep{goo89}. However, when it was observed by Goodrich again in March 1993 its Balmer lines were much stronger, and the spectrum looks much as it does in our SDSS spectrum \citep{goo95}. Goodrich found the changes in $H\alpha$ and $H\beta$ to be consistent with the same change in reddening, but because $H\beta$ is quite weak to nonexistent in the earlier spectrum (it was classed as a 1.9), this measurement is quite uncertain. Furthermore, Goodrich saw no change in the continuum flux or shape, and therefore states that ``if the lines brightened due to a dust cloud moving out of our line of sight, then the cloud must still cover the continuum source'', which seems unlikely. Because Mrk 728 is not being internally reddened, it seems more likely that the broad line changes seen in Mrk 728 are due to changes in the intrinsic continuum strength. \\

\noindent 3. Mrk 883--Type 2 (previously a type 1.9, due to low flux) \\
\indent The optical spectrum of Mrk 883 from SDSS shows it to be a Seyfert 2. However, its NLR is only reddened by $E(B-V)_{NLR}=0.4\pm0.1$, and its {\it XMM} spectrum shows absorption by a column only slightly in excess of Galactic, attributable to Mrk 883's host galaxy. This makes Mrk 883 another ``true'' Seyfert 2 candidate, similar to Mrk 609. Its {\it Spitzer} spectrum is typical of a Seyfert 2 with a significant starburst component. \\
\indent There are a few small pieces of evidence that Mrk 883 has undergone broad-line variation. \citet{shu81} claimed to see a broad component of $H\alpha$ in early spectra taken between 1975 and 1980. Osterbrock and Dahari later dubbed it a ``marginal 1.9'', due to the weakness of the broad component of $H\alpha$ seen after the subtraction of narrow $H\alpha$ and [N~II] \citep{ost83}. \citet{goo95} saw weak variations in its broad components between 1978 and 1993, inconsistent with a simple change in reddening. \\

\noindent 4. Mrk 993--Type 1.5 (previously a type 1.9, due to internal reddening) \\
\indent The optical spectrum of Mrk 993 from Lowell Observatory in January 2009 shows it to be a type 1.5 Seyfert. The BLR appears unreddened, while the reddening of the NLR is higher ($E(B-V)_{NLR}=0.47$); however, we note both these measurements have large errors due to heavy contamination by the underlying host galaxy spectrum. Mrk 993's {\it XMM-Newton} spectrum shows absorption by only a small column, N$_{H}\approx0.18\times10^{22}$ cm$^{-2}$, attributable to Galactic absorption and likely absorption by its inclined host galaxy (b/a=0.32). The WFPC2 structure map presented in \citet{pog02} confirms these findings, showing an unobscured central nucleus and dust spirals on the scale of the NLR, particularly just to the east of the nucleus, in agreement with the higher NLR column compared with the BLR and X-rays.\\
\indent Mrk 993 has a history of variability, and its transition from its original classification as a 1.9 to a 1.5 has been attributed to variable reddening \citep{tra92}, though it does not appear to be internally reddened in its current state as a 1.5. Its BLR is essentially unreddened in our spectrum, and this lack of reddening confirmed by its low X-ray column. A paper by \citet{cor05} presents an optical spectrum taken simultaneously with the {\it XMM} observation, and this spectrum confirms that it was in the same optical state as in our Lowell observation. \\ 
\indent It is possible that the broad line changes seen by \citet{tra92} are due to both changes in the intrinsic continuum flux and changes in extinction. However, because \citet{tra92} report a reddening of $E(B-V)_{BLR}>0.76$, it seems that it was internally reddened when it appeared as a 1.9. \\

\noindent 5. Mrk 1018--Type 1 (previously a type 1.9, due to low flux) \\
\indent The SDSS spectrum of Mrk 1018 shows it to have been a Seyfert 1 in September 2000. A clear rise in the blue end of the SDSS spectrum not seen in the previous optical spectra from the 1980's indicates that the continuum has gained in apparent brightness since then. \\
\indent This object was a 1.9 when observed by \citet{ost81}. However, it transitioned to a type 1 between this observation and its observation by Cohen et al. in January 1984 \citep{coh86}, and has apparently remained as a type 1 since. \citet{coh86} attribute the observed change in type to an increase in the brightness of the non-stellar continuum, but also note that the increase in flux of broad $H\alpha$ and in broad $H\beta$ are consistent with the same change in reddening. \\

\noindent 6. Mrk 1179--Type 1 (previously a type 1.9, inconclusive cause)\\
\indent Our spectrum of Mrk 1179 from Lowell Observatory in January 2009 shows this to be a type 1 Seyfert galaxy; very strong broad components with FHWM $\approx$ 6,000 km s$^{-1}$ are visible in all of the Balmer lines. Although deconvolution of the components is uncertain due to the overwhelming strength of the broad lines, both the BLR and NLR appear to be essentially unreddened. \\
\indent Interestingly, this object has changed type quite dramatically since it was previously observed, the only new case of type variability in our study. It was classified in the 1980s as a type 1.9 \citep[see][]{ost83,rud85b,goo89}. Unfortunately, these authors give only the classification but no further information about the spectrum, and so the cause of the change, variable flux or variable reddening, cannot be determined. \\

\noindent 7. NGC 2992--Type 2 (previously a type 1.9, due to low flux) \\
\indent We have observed NGC 2992 in the optical many times over the past few years, and because the spectra do not show any evidence for variability, the spectrum presented in Appendix 1 is the average of these spectra. NGC 2992's {\it XMM} spectrum shows absorption by a column density of $\approx0.77\times10^{22}$ cm$^{-2}$. Its {\it Spitzer} spectrum shows strong PAH emission features, and 9.7 $\mu$m absorption corresponding to a reddening of $E(B-V)\approx2.1$. \\
\indent As mentioned in the introduction, NGC 2992 has a history of variability, both in the optical and X-ray regimes \citep[see][for details]{tri08}. Because the changes in its optical classification seem to be correlated with its X-ray brightness, while column density has remained constant, it seems that this object was classified as a 1.9 because it was in a low continuum state. Because the NLR and BLR have similar reddening, the high Balmer decrement observed at the time it appeared as a 1.9 was most likely due to dust in the plane of the host galaxy. \\

\noindent 8. NGC 7603--Type 1 (previously a type 1.9, due to low flux) \\
\indent Our spectrum of NGC 7603 shows a type 1 spectrum with strong broad components. Its {\it XMM} spectrum shows absorption consistent with the Galactic value. NGC 7603's {\it Spitzer} spectrum is also consistent with its classification as a type 1, showing the strong rise towards shorter wavelengths typical of that class. \\
\indent NGC 7603 appeared as a 1.9 in the past, but changed to a type 1, as documented by \citet{toh76}. It went from possessing very strong broad lines in Nov. 1974, to having almost no broad-line emission in Nov. 1975, and then regained in brightness a bit by Feb. 1976. Other campaigns studying the variability of NGC 7603 are presented by \citet{goo89}, \citet{ros94}, and \citet{kol00}. The cause of the variability seems to be intrinsic variability in continuum strength \citep{kol00}. \\

\noindent {\bf Others (Misclassified 1.8/1.9s)} \\

\noindent 1. IRAS 18325-5926--Type 2 \\
\indent Our optical spectra from CTIO show H$\alpha$ to have about the same width as [O~III] $\lambda$5007 (FWHM$\approx$850 km s$^{-1}$), and thus we classify IRAS 18325-5926 as a Seyfert 2. It has a heavily reddened NLR, but its {\it XMM} spectrum shows it to have a total hydrogen column of $N_{H}\approx6.3\times10^{22}$ cm$^{-2}$ several times larger than the equivalent gas column to the NLR, consistent with a Seyfert 2 classification.\\
\indent There is no convincing evidence for variability in this object. Although \citet{iwa95} classified it as a 1.9, their use of Gaussians to fit the [N~II] lines on the sides of H$\alpha$ does not seem to be justified based on the structure of the other forbidden lines in the spectrum, and probably led to a false detection of a broad H$\alpha$ component. \\

\noindent 2. Mrk 423--Type 1.5 \\
\indent Mrk 423's optical spectrum from Lowell Observatory in January 2009 shows that the AGN's  emission spectrum is heavily contaminated by the underlying starlight spectrum from the host galaxy. Although broad H$\alpha$ is clearly visible, a telluric absorption line cuts into its red side and makes measurement of this component uncertain. Further, its unusual continuum shape was difficult to fit, and the spectrum is choppy and noisy even after the template subtraction, making Mrk 423 difficult to classify. Though we formally classify Mrk 423 as a 1.5, based on the definition of this class by \citet{mai95}, it is close in appearance to a 1.8, and its BLR seems to be suffering from slight internal reddening. The NLR is essentially unreddened, and the broad H$\alpha$ component is fairly strong, yet broad H$\beta$ is weak and no broad Ly$\alpha$ is seen in its {\it HST} GHRS (Goddard High Resolution Spectrograph) spectrum, taken in May 1996. \\ 
\indent This galaxy was originally classified as a 1.8 by \citet{ost81}. It is likely that our slightly different classification is due to our new criteria for broad emission, and not real source variability. The spectrum presented by \citet{raf93} looks similar to ours, as does that of \citet{rud85a}. \citet{rud85a} also present a UV spectrum by {\it IUE}, which again shows the Ly$\alpha$ line to be of unresolved width. \\

\noindent 3. Mrk 516--Type 2 \\
\indent The CTIO spectrum of Mrk 516 shows it is a ``transition object''; the ratio of [O~III]/H$\beta$ is only about 1.2 (see Figure~\ref{bpt}). No broad components are detected. Both the [O~III] lines and [N~II] lines have wide, asymmetric profiles. This structure is due to the double nucleus found by \citet{gor95} in {\it HST} images; one is thought to be the AGN and the other is thought to be either the nucleus of another galaxy in the final stages of merging with Mrk 516, or an H~II region triggered by such a merger. \\
\indent There is some evidence that this object has shown weak broad-line variability. \citet{ost81} classified it as a 1.8 based on a spectrum from 1978 which shows a weak broad $H\beta$ component. \citet{goo90} noted that in his August 1986 spectra Mrk 516 had no broad $H\beta$ component and that broad $H\alpha$ also seemed weaker than that reported by Osterbrock. \\

\noindent 4. Mrk 622--Type 2 \\
\indent The SDSS spectrum of Mrk 622 shows its [O~III] lines are stumpy with a double peaked structure, much wider than H$\beta$ or [N~II]. By our criteria, this is a type 2 galaxy. In line with this optical classification, the {\it Spitzer} IRS spectrum shows it to have a typical Seyfert 2 mid-IR spectrum, with the continuum weakening in the 5-15 micron range \citep{deo07} and strong PAH emission features. The low $S/N$ of its X-ray spectrum makes determination of the column very uncertain, but it seems to be absorbed by at least $85\times10^{22}$ cm$^{-2}$, and may even be Compton-thick. \\
\indent \citet{shu81} also classified Mrk 622 as a Seyfert 2. \citet{goo95} includes Mrk 622 in his study of Seyfert 1.8/1.9s but does not show the spectrum, so we do not take this as evidence for type variability. \\

\noindent 5. Mrk 1320--Type 1.5 \\
\indent Our Lowell spectrum of Mrk 1320 from January 2009 show this to be a Seyfert 1.5 object, according to the definition of this class by \citet{mai95}, though it is close in appearance to a Seyfert 1.8. Because the BLR is unreddened, it seems that the weak broad lines in this object are due to a rather low continuum flux. \\
\indent Mrk 1320 was classified by \citet{ost83} as a Seyfert 1.5. He notes that the broad components of the Balmer lines are slightly weaker than other 1.5s, but still easily visible. A blue spectrum displayed in \citet{mar03} shows it to have what seems to be a larger broad H$\beta$ line than that in either our spectrum or that described by Osterbrock, which may be evidence for broad-line variability. \\

\noindent 6. NGC 2639--LINER \\
\indent NGC 2639's nuclear spectrum from Lowell Observatory in January 2009 is heavily diluted by the underlying host galaxy spectrum. It does not show any H$\beta$ emission, broad or narrow, and [O~III] $\lambda$5007 is barely visible even in the host-galaxy subtracted spectrum. Because there is no H$\beta$, we use the criteria of \citet{hec80} ([O~II] $\lambda$3727 $\geq$ [O~III] $\lambda$5007 and [O~I] $\lambda$6300 $\geq$ (1/3) [O~III] $\lambda$5007) to classify NGC 2639 as a LINER. De-blending the [N~II] lines on the sides of H$\alpha$ proved difficult, due to NGC 2639's irregular forbidden-line profiles. The template that best matched the observed [N~II] profiles was a template constructed from the blended [S~II] lines: the blue side of the [S~II] $\lambda$6716 line and the red side of the $\lambda$6731 lines. While this template fit most of the [N~II] profile exactly, the outermost part of the red wing of the $\lambda$6583 line still showed excess flux over the template, leading to the spike of leftover emission seen in the [N~II]-subtracted spectrum of Appendix 2. In spite of this imperfection, it still seems apparent that there is not a broad H$\alpha$ component. The {\it Spitzer} IRS spectrum of NGC 2639 shows strong PAH emission features, 9.7 $\mu$m absorption, and a strong rise in the continuum towards short wavelengths. NGC 2639 was observed by {\it XMM} in April 2005, but due to a very high background level during the observation, we were unable to use it to derive NGC 2639's X-ray column density. \\
\indent Several authors in the past have claimed there to be a broad component to H$\alpha$, leading to its initial classification as a type 1.9 \citep{kee83, huc82}. However, noting (as do \citet{ho97}) the extended wings of the [N II] lines are mostly responsible for the seeming appearance of a broad component to H$\alpha$, we suspect that past decomposition of this blend assuming Gaussian line profiles seriously overestimated the flux of this component. \\

\noindent 7. NGC 3982--Type 2 \\
\indent The SDSS spectrum of this object shows no trace of broad H$\alpha$, showing that this object was a Seyfert 2 when observed in January 2003. In line with this classification, our preliminary model of its {\it XMM} spectrum indicates that the nucleus is heavily attenuated by a column of $N_{H}\approx22.8\times10^{22}$ cm$^{-2}$. This may in fact be a lower limit; the spectrum is equally well-fit by a model employing a reflection spectrum ({\it reflionx}) and thermal emission component ({\it mekal}), indicating that the obscuration may be Compton-thick. Its {\it Spitzer} spectrum is also typical of a Seyfert 2. \\
\indent There is no convincing evidence from the literature that this object has ever changed type. We included it in our study because \citet{qui01} classified it as a ``1.9/2'', but it has apparently always really been a type 2. \\

\noindent 8. NGC 4388--Type 2 \\
\indent The SDSS spectrum of NGC 4388 shows it to be a Seyfert 2. Its {\it XMM} spectrum is absorbed by a column of $28.4\times10^{22}$ cm$^{-2}$, also typical of Seyfert 2s, and shows an excess of soft emission in the 0.5-2 keV range. \\
\indent There is no evidence from the literature for variability from a previous 1.8/1.9 state. Again, this object was included in our study because of its classification in \citet{qui01} as a type 1.9/2, but it appears that it has always really been a Seyfert 2. \\

\noindent 9. NGC 4639--low-luminosity Seyfert 1 \\
\indent Our optical spectrum on NGC 4639 from Lowell Observatory in January 2009 shows that the host galaxy dominates the spectrum. It shows what looks like a broad H$\alpha$ line, but we did not de-blend the [N~II] lines from H$\alpha$ because there are no other forbidden lines in the spectrum to use as a template. Its {\it XMM} spectrum is faint, but appears unabsorbed. Both the {\it XMM} spectrum and the optical spectrum are consistent with its classification by \citet{ho99} as a low-luminosity Seyfert 1. \\
\indent \citet{ho95} display a spectrum in which broad H$\alpha$ looks much more prominent than in our data. Though \citet{ho99} suggest that the BLR is significantly reddened (they measure the ratio of broad H$\gamma$/H$\beta$ to be $\sim0.4$), this measurement is necessarily very uncertain due to the faintness of these lines. \\

\noindent 10. NGC 5033--Seyfert 1.2 \\
\indent The optical spectrum from Lowell Observatory in January 2009 shows NGC 5033 to have the spectrum of a Seyfert 1.2 (according to the definition of this class by \citet{mai95}), heavily contaminated by underlying stellar absorption spectrum from the host galaxy. The H$\beta$ line is apparent only after the subtraction of a host-galaxy template. The BLR seems to be only slightly reddened with $E(B-V)_{BLR}=0.19\pm0.05$, while the NLR seems to be unreddened. NGC 5033's {\it XMM} spectrum is well-fit with an unabsorbed powerlaw, but a narrow Fe K$\alpha$ line is also present, indicating a there may be a reflected component to the spectrum.\\
\indent Broad H$\alpha$ was noted in early spectra of NGC 5033 \citep[see][]{shu80, sta82} and \citet{fil85} note that this broad component of H$\alpha$ seems to be variable from their observations in July 1982 and February 1984. \citet{dah88} classify it as a 1.9, as do \citet{ost93}. \citet{kor95} note the presence of both broad H$\alpha$ and broad H$\beta$ in their spectrum of NGC 5033 from January 1993. It seems likely, however, that many of these differences in classification are due to the difficulty in picking out the H$\beta$ line from the contamination by the host galaxy, and in spite of the weak variations in broad H$\alpha$ noted by \citet{ho95}, it is probable that NGC 5033 has not ever appeared as a bona-fide Seyfert 1.9. \\

\noindent 11. NGC 5273--Type 1.5 \\
\indent Our spectrum from KPNO in June 2008 is rather noisy due to poor seeing conditions, but both broad H$\alpha$ and H$\beta$ are apparent after subtraction of the host galaxy. Neither the NLR or BLR appears to be significantly reddened in this face-on (b/a=0.89) galaxy. However, its X-ray spectrum shows absorption by a significant column, $N_{H}\approx2.4\times10^{22}$ cm$^{-2}$, which may indicate the absorption occurs in a dust-free warm absorber and that the actual column density may be even higher than this value, derived assuming cold absorption. \\
\indent Previous classifications of NGC 5273 are consistent with our current classification \citep{ho95, dah88}, except \citet{ost93}, who classify it as a 1.9. It seems likely that the subtraction of the host galaxy spectrum is the cause of this difference, and therefore we do not take their different classification as evidence for broad-line variability. \\

\noindent 12. NGC 5506--Type 2 \\
\indent From the optical SDSS spectrum of NGC 5506 taken April 2002, we classify it as a Seyfert 2. \citet{nag02} have a near-IR spectrum with permitted lines from the BLR with FWHM$<2,000$ km s$^{-1}$, and therefore claim it to be an ``optically obscured'' NLS1. Some part of the reddening of the nucleus must be due to the dust lane that crosses over the nucleus, seen in its WFPC2 image. \\
\indent We included NGC 5506 in our study because it was listed as a 1.9 in \citet{mai95}; however, it was most likely a type 2 at that time as well \citep{mai94}, at least in the optical part of the spectrum. \\

\noindent 13. NGC 5674--Type 2 \\
\indent NGC 5674 is a Seyfert 2 in our January 2009 spectrum from Lowell. \\
\indent There is no evidence for variability in NGC 5674; it appears it has always been a Seyfert 2. It was first identified as a Seyfert galaxy by \citet{huc82}, who labeled it a type 2. \citet{ost93} list it as a type 1.9, but this was based upon only one unpublished spectrum and therefore is not convincing evidence for broad-line change. \\

\noindent 14. NGC 7479--Type 2 \\
\indent NGC 7479's spectrum from CTIO in August 2008 is dominated by the starlight of the host galaxy. [O~III] $\lambda\lambda$ 4959, 5007 lines are weak, and only H$\beta$ absorption is seen. No broad H$\alpha$ is detected. The narrow lines show an unusual triple-peaked structure. Its {\it XMM} spectrum supports a type 2 classification. We modeled this spectrum with a heavily absorbed powerlaw, and found $N_{H}\approx83\times10^{22}$ cm$^{-2}$. However, a {\it reflionx} model fits the data equally well, which indicates that the obscuration in NGC 7479 may actually be Compton-thick. \\
\indent Our literature search did not find any evidence for variability of NGC 7479. Though it has had different classifications by different authors, this is most likely due to the differences in the criteria used for each type. For example, it was classified as a LINER by \citet{ho93} using the criteria of \citet{hec80}, but as a Seyfert 2 by \citet{mai95} using their own definition. It was also classified as a 1.9 by \citet{ho97}, on the basis that H$\alpha$ might have an extremely faint broad component; however, the authors note that the complex velocity structure of the narrow lines could have played a role in inaccurate subtraction of the [N~II]. \\

\noindent 15. UGC 12138--Type 1.5 \\
\indent Our spectra from CTIO in August 2008 show UGC 12138 to be a Seyfert 1.5.  Its BLR is unreddened, but its NLR is reddened by $E(B-V)_{NLR}=0.33\pm0.25$, apparently due to the dust spirals in the NLR seen in the the WFPC2 structure map of \citet{pog02}. In agreement with its optical appearance as a 1.5, UGC 12138 has an essentially unabsorbed power-law {\it XMM} spectrum. \\
\indent UGC 12138 was first classified as a type 1 by \citet{huc82}. Later spectra of UGC 12138 published by \citet{deg92} and \citet{cru94} also look very similar to ours. \\

\section{Results}
\subsection{Many ``1.8/1.9s'' are not 1.8/1.9s}
\indent The first important consideration brought to light by this work is that many Seyferts alleged to be 1.8s or 1.9s are misclassified (16 of our sample of 34). The most common reason (at least 7 of the 16) for misclassification seems to be a large overestimation of the flux of broad H$\alpha$, due to inaccurate or no [N~II] subtraction. Spectral variability may account for a couple of the type discrepancies, but based on our literature search of previously published spectra, this seems to be unlikely in most cases. It is therefore important for anyone hoping to study the properties of 1.8/1.9s to recognize that optical classifications as such are frequently unreliable.\\
\indent The classification of 1.8 in particular seems to be often misused. Our sample was originally thought to contain nine 1.8s, but only one of these objects actually fit our criteria to be classified as a 1.8. These misidentifications probably result from the fact that a weak broad component to H$\beta$ is very difficult to discern due to contamination by the spectrum of the host galaxy; in the three 1.8s we found in our study, the weak broad H$\beta$ was apparent only after the subtraction of the underlying host galaxy spectrum. It is important to note that classification as a 1.8 depends on spectral $S/N$ and the amount of galaxy contamination. In the case of infinite signal to noise and perfect starlight subtraction, all 1.9s would be classified as 1.8s, as broad H$\beta$ must be present in any object that shows broad H$\alpha$.  We further emphasize that because Seyfert 1.8s seem to be rather rare relative to their purported numbers, the 1.8 classification of many objects should be treated warily by those who wish to study intermediate-type AGN, unless there is firm evidence (i.e., a high $S/N$ host-galaxy subtracted spectrum that shows a weak broad component to H$\beta$) that it truly belongs to this class. Also, because of these misclassifications, previous studies that make general claims about 1.8s (e.g., that they have unusually large broad Balmer decrements) may be unreliable. \\
\subsection{Inventory of 1.8/1.9s}
\indent Are Seyfert 1.8s and 1.9s Seyfert 1 nuclei with reddened BLRs, or are they Seyferts in a temporary low-flux state with weakened broad-line components? From the evidence presented in Section~\ref{indiv_comments}, we conclude that of the 19 objects in our sample that currently or formerly appeared 1.8/1.9s, ten received this designation because they were at least temporarily in low flux states, and four because of reddening. Five do not have evidence that favors one scenario over the other. Sixteen of the 34 were other objects misclassified as 1.8/1.9s. Table~\ref{sumtbl} gives a summary of our conclusions from the last section about the most probable cause of classification for each 1.8/1.9. \\
\indent As can be seen from Table~\ref{sumtbl}, a total of eight objects appeared as 1.9s in our recent spectra. Of these, three have evidence for internal BLR reddening, one for low flux, and four are inconclusive. \\
\indent All three objects that appear as 1.8s in our spectra all seem to be due to low flux states. These objects, Mrk 334, Mrk 1126, and UM 146, have broad Balmer decrements of 2.7, 4.7, and 3.2 respectively. In Mrk 334 and UM 146, the reddening of the NLR is actually higher than the reddening of the BLR, while in Mrk 1126 the BLR and NLR reddenings are consistent with each other within their error bars. Measuring their residual continuum at 5100 \AA\ after starlight subtraction, we estimated their Eddington ratios using the mass-luminosity relation of \citet{pet04} (equation 9), and found that all three have $L_{BOL}/L_{EDD}\approx0.02$. This puts them at the low end of the $L_{BOL}/L_{EDD}$ ratios measured in the \citet{pet04} sample, and supports our idea that they simply objects with a weak ionizing continua. \\
\indent Eight objects previously appeared as 1.8/1.9s in the literature, but changed type such that they were not 1.8/1.9s when observed by us: Mrk 609, Mrk 728, Mrk 883, Mrk 993, Mrk 1018, Mrk 1179, NGC 2992, and NGC 7603. There is evidence that the broad-line changes in four of these objects (Mrk 728, Mrk 1018, NGC 2992, and NGC 7603) were due to an intrinsic change in the amount of ionizing continuum. Only in one, Mrk 993, is there good evidence for variable reddening. Mrk 609 and Mrk 883 both showed previous evidence for broad-line components (weak in the case of Mrk 883) that their current spectra no longer manifest. Both these objects show only small amounts of absorption in their recent X-ray spectra, and therefore it would seem that in these objects the broad-line region itself is now missing. \\
\indent Our overall finding that 10/19 Seyfert 1.8/1.9s received this designation because they were (or are) in a temporarily low flux state implies that samples of 1.8/1.9s are not suitable for use in studies of the gas and dust in the torus. In at least half the objects, their intermediate classification has nothing to do with reddening of the nucleus by the torus; further, in many objects that do show BLR reddening the reddening is likely due to nuclear dust lanes on the scale of the NLR, instead of the torus. These findings also also imply, by extension, that some fraction of optically classified Seyfert 2s may similarly be such low-flux objects, and therefore simply using the ratio of optical type 2 to type 1 AGN without more detailed data on the AGN may lead to inaccurate estimation of the scale height of the torus. \\
\subsection{Variability of 1.8/1.9s}
\indent We were able to obtain multiple spectra of 16 objects in our sample, over baselines in time ranging from 1-5 years. The highest $S/N$ spectrum of each object was used for analysis and is listed in Table~\ref{obslog}, but the repeat spectra (not listed in Table~\ref{obslog}) are in general of comparable quality. Of these 16 Seyferts, none showed any evidence for broad-line variability between our observations . Furthermore, with the exception of Mrk 1179, none of the objects in our sample varied strongly even with respect to their latest published optical spectrum in the literature, which was often a much longer baseline in time, extending back 20 to 30 years for some objects. This lack of variability is clearly at odds with the the previous study of 1.8/1.9s done by \citet{goo95}, in which he found ``nine of the 24 galaxies showed significant spectral variability'' in data taken $\sim$1 year apart. The cause of this discrepancy is unclear. \\
\subsection{Conclusions} 
\indent We have collected evidence from different wavebands to determine the presence and location of dust in a sample of 34 Seyfert 1.8/1.9s. We used optical spectra to determine the reddening of the NLR and BLR, and compared the strengths of the broad lines in these new spectra with previous spectra in the literature to determine if they showed evidence for broad-line variability. We also used X-ray data from {\it XMM} to determine the total hydrogen column to the source. We determined that: \\
1.) The categories of 1.8 and 1.9 are often misused: 16 of our sample of 34 ``1.8 and 1.9s'' were misclassified. This is most commonly due to an overestimation of the flux of broad H$\alpha$, stemming from inaccurate or no subtraction of the adjacent [N~II] emission.\\ 
2.) Seyfert 1.8s are fewer than their reported numbers, due to the difficulty in detecting a weak broad component of the H$\beta$ line. All three of the Seyferts in our sample that appeared as 1.8s (after careful subtraction of the host galaxy spectrum) seem to be low-flux objects that do not suffer from reddening of their BLRs. \\
3.) For most Seyfert 1.9s, an optical spectrum alone cannot provide reddening information; H$\alpha$ is too weak to determine a even lower limit for $E(B-V)_{BLR}$. \\
4.) The reddenings of the BLRs (for those objects with both broad H$\alpha$ and H$\beta$) were not consistently higher than the reddenings of their NLRs. Our study therefore does not provide evidence that these objects are being viewed along a line of sight that grazes the atmosphere of a central dusty torus. Dust spirals on the same size scale as the NLR may instead be responsible for the sometimes-reddened BLRs, as these features randomly cut across the NLR, sometimes blocking the central AGN and sometimes not. \\
5.) Ten of 19 objects that previously appeared as Seyfert 1.8/1.9s received this designation because of their low continuum flux. Four received this designation because of BLR reddening, either by the torus or dust structures in the vicinity of the NLR. \\

{\it Acknowledgments} \\
\indent The spectra of our southern hemisphere targets were taken at the CTIO 1.5-m telescope, operated by the SMARTS Consortium. We would like to thank Todd Henry for helping provide access to the SMARTS time. This work has made use of the NASA/IPAC Extragalactic Database (NED), which is operated by the Jet Propulsion Laboratory, California Institute of Technology, under contract with NASA. \\
\indent Funding for the SDSS and SDSS-II has been provided by the Alfred P. Sloan Foundation, the Participating Institutions, the National Science Foundation, the U.S. Department of Energy, the National Aeronautics and Space Administration, the Japanese Monbukagakusho, the Max Planck Society, and the Higher Education Funding Council for England. The SDSS Web Site is http://www.sdss.org/. The SDSS is managed by the Astrophysical Research Consortium for the Participating Institutions. The Participating Institutions are the American Museum of Natural History, Astrophysical Institute Potsdam, University of Basel, University of Cambridge, Case Western Reserve University, University of Chicago, Drexel University, Fermilab, the Institute for Advanced Study, the Japan Participation Group, Johns Hopkins University, the Joint Institute for Nuclear Astrophysics, the Kavli Institute for Particle Astrophysics and Cosmology, the Korean Scientist Group, the Chinese Academy of Sciences (LAMOST), Los Alamos National Laboratory, the Max-Planck-Institute for Astronomy (MPIA), the Max-Planck-Institute for Astrophysics (MPA), New Mexico State University, Ohio State University, University of Pittsburgh, University of Portsmouth, Princeton University, the United States Naval Observatory, and the University of Washington. \\
\indent Based on observations obtained with XMM-Newton, an ESA science mission with instruments and contributions directly funded by ESA Member States and NASA. \\
\bibliographystyle{apj} 
\bibliography{apj-jour,paper}

\clearpage
\begin{deluxetable}{lrcc}
\tabletypesize{\scriptsize}
\tablewidth{0pt}
\tablecaption{The Sample}
\tablehead{
\colhead{ }       & \colhead{ }       & \colhead{Previous}  & \colhead{Classification}      \\
\colhead{Object } &\colhead{Redshift$^{a}$ }& \colhead{Classification} & \colhead{Reference$^b$}  }
\startdata
IRAS 18325-5926& 0.020231 & 1.8      & 2    \\
Mrk 334        & 0.021945 & 1.8      & 4    \\
Mrk 423        & 0.032268 & 1.9      & 3    \\
Mrk 471        & 0.0340   & 1.8/1.9  & 1    \\
Mrk 516        & 0.028416 & 1.8/1.9  & 1    \\
Mrk 609        & 0.034488 & 1.8/1.9  & 1    \\
Mrk 622        & 0.0230   & 1.8/1.9  & 1    \\
Mrk 728        & 0.0350   & 1.8/1.9  & 1    \\
Mrk 883        & 0.0370   & 1.8/1.9  & 1    \\
Mrk 915        & 0.024109 & 1.8/1.9  & 1    \\
Mrk 993        & 0.015537 & 1.8/1.9  & 1    \\
Mrk 1018       & 0.042436 & 1.9      & 3    \\
Mrk 1126       & 0.010624 & 1.8/1.9  & 1    \\
Mrk 1179       & 0.03760  & 1.8/1.9  & 1    \\
Mrk 1320       & 0.1030   & 1.8/1.9  & 1    \\
NGC 1365       & 0.005457 & 1.8      & 2    \\
NGC 2622       & 0.0280   & 1.8/1.9  & 1    \\
NGC 2639       & 0.011128 & 1.9      & 2    \\
NGC 2992       & 0.007710 & 1.9      & 2    \\
NGC 3786       & 0.008933 & 1.8      & 4    \\
NGC 3982       & 0.003699 & 2/1.9    & 5    \\
NGC 4388       & 0.008419 & 2/1.9    & 5    \\
NGC 4639       & 0.003395 & 1.8      & 2    \\
NGC 5033       & 0.002919 & 1.9      & 4    \\
NGC 5252       & 0.022975 & 1.9      & 4    \\
NGC 5273       & 0.003549 & 1.9      & 4    \\
NGC 5506       & 0.006181 & 1.9      & 2    \\
NGC 5674       & 0.024931 & 1.9      & 4    \\
NGC 7314       & 0.004763 & 1.9      & 2    \\
NGC 7479       & 0.007942 & 2/1.9    & 5    \\
NGC 7603       & 0.029524 & 1.8/1.9  & 1    \\
UGC 7064       & 0.024997 & 1.8/1.9  & 1    \\
UGC 12138      & 0.024974 & 1.8      & 4    \\
UM 146         & 0.017405 & 1.9      & 4    \\   
\enddata
\tablenotetext{a}{Redshift from the NASA/IPAC Extragalactic Database (NED).}
\tablenotetext{b}{1--\citet{goo95}; 2--\citet{mai95}; 3--\citet{ost81}; 4--\citet{ost93}; 5--\citet{qui01}}
\label{sample}
\end{deluxetable}

\clearpage
\begin{deluxetable}{lclcc}
\tabletypesize{\scriptsize}
\tablewidth{0pt}
\tablecaption{Log of Optical Observations}
\tablehead{
\colhead{Object}& \colhead{Observatory} & \colhead{Date} & \colhead{Wavelength Coverage} & \colhead{Slit Width$^{a}$} \\
\colhead{ } &\colhead{ }& \colhead{$($U.T.$)$} & \colhead{$($\AA$)$} & \colhead{$($arcsec$)$}
}
\startdata
IRAS 18325-5926 & CTIO & 2007 Sept. 3 & 3658-5416& 2.0 \\ 
IRAS 18325-5926 & CTIO & 2007 Sept. 10 & 5630-6943& 4.0 \\
Mrk 334 &Lowell&2009 Jan. 31& 3255-6785 & 2.0 \\
Mrk 423 &Lowell&2009 Jan. 28& 3185-6715 & 2.0 \\
Mrk 471 & SDSS & 2005 April 12 & 3800-9200 & 3.0 \\
Mrk 516 & CTIO & 2008 May 10 & 3658-5416 & 2.0 \\ 
Mrk 516 & CTIO & 2008 May 11 & 5534-6844 & 2.0 \\ 
Mrk 609 & CTIO & 2007 Oct. 8 & 3658-5416 & 2.0 \\
Mrk 609 & CTIO & 2007 Nov. 10 & 5630-6943 & 2.0 \\
Mrk 622 & SDSS & 2001 Dec. 11 & 3800-9200 & 3.0 \\
Mrk 728 & SDSS & 2004 April 24 & 3800-9200 & 3.0 \\
Mrk 883 & SDSS & 2004 Aug. 9 & 3800-9200 & 3.0 \\
Mrk 915 & CTIO & 2008 June 27 & 3658-5416 & 4.0 \\
Mrk 915 & CTIO & 2008 June 29 & 5630-6943 & 4.0 \\
Mrk 993 &Lowell& 2009 Jan. 31 & 3255-6785 & 2.0 \\
Mrk 1018 & CTIO & 2007 Oct. 5 & 3658-5416 & 2.0 \\
Mrk 1018 & CTIO & 2007 Oct. 7 & 5630-6943& 2.0 \\
Mrk 1126 & CTIO & 2007 Oct. 5 & 3658-5416 & 2.0 \\ 
Mrk 1126 & CTIO & 2007 Oct. 7 & 5630-6943& 2.0 \\
Mrk 1179 &Lowell& 2009 Jan. 28 & 3185-6715 & 2.0 \\
Mrk 1320 &Lowell& 2009 Jan. 31 & 3255-6785 & 2.0 \\
Mrk 1320 &Lowell &2009 Feb. 2 & 6295-9000 & 2.0 \\
NGC 1365 & CTIO & 2009 Jan. 8 & 3658-5416 & 4.0 \\
NGC 1365 & CTIO & 2009 Jan. 9 & 5630-6943 & 2.0 \\
NGC 2622 & SDSS & 2004 Dec. 8 & 3800-9200 & 3.0 \\
NGC 2639 &Lowell&2009 Jan. 31 & 3255-6785 & 2.0 \\
NGC 2992 & CTIO &average spectrum$^{b}$& 3660-6970 & 2.0-4.0 \\
NGC 3786 &Lowell& 2009 Jan. 28 & 3185-6715 & 2.0 \\
NGC 3982 & SDSS & 2003 Jan. 31 & 3800-9200 & 3.0 \\
NGC 4388 & SDSS & 2004 June 10 & 3800-9200 & 3.0 \\
NGC 4639 &Lowell& 2009 Jan. 28 & 3185-6715 & 2.0 \\
NGC 5033 &Lowell& 2009 Feb. 1 & 3845-7375 & 2.0 \\
NGC 5252 & SDSS & 2002 April 10 & 3800-9200 & 3.0 \\
NGC 5273 & KPNO & 2008 July 2 & 3300-6000 & 2.0 \\
NGC 5273 & KPNO & 2008 July 3 & 5500-7000 & 2.0 \\
NGC 5506 & SDSS & 2002 April 14 & 3800-9200 & 3.0 \\
NGC 5674 &Lowell& 2009 Jan. 31 & 3255-6785 & 2.0 \\
NGC 7314 & CTIO & 2007 Sept. 8 & 3658-5416& 2.0 \\
NGC 7314 & CTIO & 2008 May 9 & 5534-6844 & 2.0 \\
NGC 7479 & CTIO & 2008 Aug. 5 & 3658-5416 & 2.0 \\
NGC 7479 & CTIO & 2008 Aug. 8 & 5630-6943 & 2.0 \\
NGC 7603 & CTIO & 2007 Oct. 5 & 3658-5416 & 2.0 \\
NGC 7603 & CTIO & 2007 Oct. 7 & 5630-6943 & 2.0 \\
UGC 7064 &Lowell& 2009 Jan. 28& 3185-6715 & 2.0 \\
UGC 12138 & CTIO & 2008 Aug. 5 & 3658-5416 & 2.0 \\
UGC 12138 & CTIO & 2008 Aug. 2 & 5630-6943 & 2.0 \\
UM 146 & CTIO & 2007 Oct. 5 & 3658-5416 & 2.0 \\ 
UM 146 & CTIO & 2007 Oct. 7 & 5630-6943 & 2.0 \\
\enddata
\tablenotetext{a}{The SDSS spectra were taken with a 3$''$-diameter circular aperture.}
\tablenotetext{b}{The spectrum of NGC 2992 is the average of many spectra taken from 2006 January to 2007 June \citep[see][]{tri08}.}
\label{obslog}
\end{deluxetable}

\clearpage
\begin{deluxetable}{lcccccc}
\tabletypesize{\scriptsize}
\tablewidth{0pt}
\tablecaption{Measured $E(B-V)$ Values}
\tablehead{ \colhead{} & \colhead{Our} &\colhead{Starlight} &\colhead{Starlight} & \colhead{} & \colhead{} \\
\colhead{Object} &\colhead{Type} & \colhead{Template$^{a}$} & \colhead{Fraction$^{b}$} &\colhead{$E(B-V)_{NLR}^{c}$} & \colhead{$E(B-V)_{BLR}^{d}$}
}
\startdata
IRAS 18325-5926&2  & S0 & 0.72  &1.63$\pm$0.07 & -- \\ 
Mrk 334  & 1.8     & E1 & 0.90  &0.20$\pm$0.06 &-0.13$\pm$0.19 \\
Mrk 423  & 1.5     & E2 & 0.81  &-0.13$\pm$0.17&0.36$\pm$0.22 \\
Mrk 471  & 1.9     & Sa & 0.92  &0.89$\pm$0.23 & -- \\
Mrk 516  & 2       & E1 & 0.72  &0.48$\pm$0.08 & -- \\ 
Mrk 609  & 2       & E1 & 0.85  &0.98$\pm$0.11 & -- \\
Mrk 622  & 2       & E1 & 0.91  &1.03$\pm$0.05 & -- \\
Mrk 728  & 1.2     & E1 & 0.97  &0.03$\pm$0.12 & 0.45$\pm$0.14 \\
Mrk 883  & 2       & E1 & 0.95  &0.40$\pm$0.07 & -- \\
Mrk 915  & 1.9     & E2 & 0.86  &0.27$\pm$0.06 & $>$0.33 \\
Mrk 993  & 1.5     & Sa & 0.93  &0.47$\pm$0.36 &-0.27$\pm$0.38 \\
Mrk 1018 & 1       & S0 & 0.62  &0             & 0.25$\pm$0.08 \\
Mrk 1126 & 1.8     & Sb & 0.78  &0.51$\pm$0.23 & 0.38$\pm$0.20 \\
Mrk 1179 & 1       & Sa & 0.76  &-0.06$\pm$0.26 & 0.24$\pm$0.12 \\
Mrk 1320 & 1.5     & E2 & 0.61  &0.38$\pm$0.32 & -0.07$\pm$0.22 \\
NGC 1365 & 1.9+HII & Sb & 1.00  & 1.30$\pm$0.04 & -- \\
NGC 2622 & 1.9     & Sb & 0.82  & 0.36$\pm$0.09 & -- \\
NGC 2639 &LINER    & E4 & 0.95  & --          & -- \\
NGC 2992 &  2      & Sa & 0.94  & 0.71$\pm$0.11 & -- \\
NGC 3786 & 1.9     & E2 & 0.96  & 0.45$\pm$0.09 & $>$0.16\\
NGC 3982 & 2       & Sb & 0.82  & 0.68$\pm$0.11 & -- \\
NGC 4388 & 2       & Sb & 0.88  & 0.67$\pm$0.09 & -- \\
NGC 4639 &1$^{e}$  & Sa & 0.95  & --           & -- \\
NGC 5033 & 1.2     & Sa & 0.83  & -0.04$\pm$0.43&0.19$\pm$0.05\\
NGC 5252 & 1.9     & S0 & 0.93  & 0.71$\pm$0.09 & -- \\
NGC 5273 & 1.5     & E2 & 0.80  & -0.07$\pm$0.27 & -0.10$\pm$0.17\\ 
NGC 5506 & 2       & Sb & 0.88  & 0.78$\pm$0.06 & -- \\
NGC 5674 & 2       & E2 & 0.95  & 0.14$\pm$0.08 & -- \\
NGC 7314 & 1.9     & E4 & 1.00  & 0.52$\pm$0.02 & $>$0.02 \\
NGC 7479 &2$^{f}$  & E3 & 0.74  & --            & -- \\
NGC 7603 & 1       & E3 & 0.27  & -0.03$\pm$0.23& 0.19$\pm$0.36\\
UGC 7064 & 1.9     & E2 & 0.96  & 0.23$\pm$0.13 & -- \\
UGC 12138& 1.5     & E2 & 0.70  & 0.33$\pm$0.25 & -0.13$\pm$0.08 \\
UM 146   & 1.8     & E2 & 0.90  & 0.42$\pm$0.24 & 0.03$\pm$0.19 \\                                  
\enddata
\tablenotetext{a}{The classification of the inactive galaxy template spectrum used to subtract off the host galaxy contamination from the AGN's spectrum, from \citet{kin96}.}
\tablenotetext{b}{The fraction of the observed continuum at 5,100 \AA\ due to starlight from the host galaxy.}
\tablenotetext{c}{Derived from the ratio of the integrated flux of the narrow component of H$\alpha$ to integrated flux of the narrow component of H$\beta$, assuming the intrinsic value of this ratio to be 2.90, and using the reddening curve of \citet{sav79}. Error bars represent the 1-$\sigma$ uncertainties in the derived values. }
\tablenotetext{d}{Derived from the ratio of the integrated flux of the broad component of H$\alpha$ to the integrated flux of the broad component of H$\beta$, assuming the intrinsic value of this ratio to be 3.10, and using the reddening curve of \citet{sav79}. Error bars represent the 1-$\sigma$ uncertainties in the derived values. }
\tablenotetext{e}{A low-luminosity Seyfert 1 \citep{ho99}. }
\tablenotetext{f}{No H$\beta$ is present in the spectrum, but the spectrum is otherwise characteristic of a type 2 Seyfert.}
\label{reddenings}
\end{deluxetable}

\clearpage
\begin{deluxetable}{llllclcl}
\tabletypesize{\scriptsize}
\rotate
\tablewidth{0pt}
\tablecaption{X-ray Models and Column Densities}
\tablehead{
\colhead{ }    &\colhead{ } &\colhead{ } &\colhead{ } &\colhead{Galactic } &\colhead{Intrinsic} &\colhead{Fe K$\alpha$} &\colhead{Reduced} \\
\colhead{Object} &\colhead{Obs. Date} & \colhead{Obs. ID\#} &\colhead{XSPEC Model Components} &\colhead{Column$^{a}$} &\colhead {Column$^{b}$} &\colhead{EW$^{c}$} &\colhead{$\chi^{2}$(dof)}
}
\startdata
IRAS 18325-5926& 5 March 2001 & 0022940101 &tbabs$_{gal}$*(tbabs*po +tbabs*po)                          &0.066& 6.23$^{+0.22}_{-0.11}$  & $<$0.026                & 1.11(283) \\
Mrk 609 (1)    & 13 Aug. 2002 & 0103861001 &tbabs$_{gal}$*(po +zgauss)                                  &0.044& $\sim$0$^{d}$           & 0.164$_{0.135}^{0.134}$ & 0.90(161) \\
Mrk 609 (2)    & 27 Jan. 2007 & 0402110201 &tbabs$_{gal}$*(po +zgauss)                                  &0.044& $\sim$0$^{d}$           & 0.134$_{0.091}^{0.091}$ & 1.00(290) \\
Mrk 622        & 2 April 2003 & 0138951401 &tbabs$_{gal}$*(tbabs*po +po +zgauss)                        &0.052& $>85$                   &  ---$^{e}$               & 3.67(7)  \\
Mrk 728        & 23 May 2002  & 0103861801 &tbabs$_{gal}$*po                                            &0.019& $\sim$0$^{d}$           & 0.180$^{0.086}_{0.087}$ & 0.92(345) \\
Mrk 883 (1)    & 13 Aug. 2006 & 0302260101 &tbabs$_{gal}$*po                                            &0.040& 0.07$^{+0.01}_{-0.01}$  & $<$0.122                & 1.22(186) \\
Mrk 883 (2)    & 15 Aug. 2006 & 0302260701 &tbabs$_{gal}$*po                                            &0.040& 0.06$^{+0.01}_{-0.01}$  & $<$0.152                & 1.19(229) \\
Mrk 883 (3)    & 21 Aug. 2006 & 0302261001 &tbabs$_{gal}$*po                                            &0.040& 0.07$^{+0.01}_{-0.01}$  & $<$0.147                & 0.81(260) \\
Mrk 993        & 23 Jan. 2004 & 0201090401 &tbabs$_{gal}$*(po +zgauss)                                  &0.057& 0.12$^{+0.01}_{-0.01}$  & 0.137$^{0.104}_{0.104}$ & 1.12(223) \\
Mrk 1018       & 15 Jan. 2005 & 0201090201 &tbabs$_{gal}$*po                                            &0.025& $\sim$0$^{d}$           & $<$0.644                & 1.16(100) \\
NGC 1365 (1)   & 17 Jan. 2004 & 0205590301 &tbabs$_{gal}$*(pcfabs*(po +kdblur*reflionx) +zgauss +zgauss &0.014& 9.67$^{+0.10}_{-0.10}$  & 0.048$_{0.010}^{0.010}$ & 1.36(1528)\\
               &              &            &+zgauss +zgauss +zgauss +pexrav +tbabs*vraymond)$^{f}$      &     &                         &                         &           \\
NGC 1365 (2)   & 24 July 2004 & 0205590401 &tbabs$_{gal}$*(pcfabs*(po +kdblur*reflionx) +zgauss +zgauss &0.014& 27.01$^{+0.44}_{-0.43}$ & 0.091$_{0.015}^{0.015}$ & 1.31(1050)\\
               &              &            &+zgauss +zgauss +zgauss +pexrav +tbabs*vraymond)$^{f}$      &     &                         &                         &           \\
NGC 2622       & 9 April 2005 & 0302260201 &tbabs$_{gal}$*(po +tbabs*po)                                &0.035& 1.14$^{+0.13}_{-0.11}$  & $<$0.095                & 0.95(163) \\
NGC 2992       & 19 May 2003  & 0147920301 &tbabs$_{gal}$*(tbabs*po +tbabs*po +zgauss)                  &0.053& 0.72$^{+0.00}_{-0.00}$  & 0.062$^{0.022}_{0.021}$ & 1.16(1789)\\
NGC 3786       & 24 May 2004  & 0204650301 &tbabs$_{gal}$*(tbabs*po +zgauss +po)                        &0.020& 3.97$^{+0.20}_{-0.17}$  & 0.152$^{0.066}_{0.066}$ & 0.95(262) \\
NGC 3982       & 15 June 2004 & 0204651201 &tbabs$_{gal}$*(tbabs*po +mekal +po +zgauss)                 &0.013& 22.74$^{+13.0}_{-6.9}$  & 0.942$^{0.551}_{0.551}$ & 1.32(26)  \\
NGC 4388       & 12 Dec. 2002 & 0110930701 &tbabs$_{gal}$*(tbabs*po +mekal +po +zgauss)                 &0.026& 28.32$^{+0.52}_{-0.51}$ & 0.207$_{0.066}^{0.061}$ & 1.15(437) \\
NGC 4639       & 16 Dec. 2001 & 0112551001 &tbabs$_{gal}$*po                                            &0.024& $\sim$0$^{d}$           & $<$0.783                & 0.99(90)  \\
NGC 5033       & 18 Dec. 2002 & 0094360501 &tbabs$_{gal}$*(po +gauss)                                   &0.010& $\sim$0$^{d}$           & 0.230$^{0.070}_{0.070}$ & 0.97(468) \\
NGC 5252       & 18 July 2003 & 0152940101 &tbabs$_{gal}$*(tbabs*po +zgauss +tbabs*po +po +mekal)       &0.020& 26.52$^{+0.77}_{-0.74}$ & 0.036$_{0.013}^{0.013}$ & 1.01(1431)\\
NGC 5273       & 14 June 2002 & 0112551701 &tbabs$_{gal}$*(tbabs*po +zgauss +po +mekal)                 &0.009& 2.37$^{+0.10}_{-0.10}$  & 0.510$^{0.250}_{0.250}$ & 1.26(451) \\
NGC 5506 (1)   & 9 Jan. 2002  & 0013140201 &tbabs$_{gal}$*(tbabs*po +zgauss +tbabs*po)                  &0.040& 4.14$^{+0.03}_{-0.03}$  & 0.075$_{0.032}^{0.032}$ & 1.03(1570)\\
NGC 5506 (2)   & 7 Aug. 2004  & 0201830501 &tbabs$_{gal}$*(tbabs*po +zgauss +tbabs*po)                  &0.040& 4.29$^{+0.02}_{-0.02}$  & 0.104$^{0.032}_{0.029}$ & 1.01(1557)\\
NGC 7314       & 2 May 2001   & 0111790101 &tbabs$_{gal}$*(tbabs*po +zgauss +po)                        &0.015& 1.04$^{+0.00}_{-0.00}$  & 0.124$^{0.045}_{0.046}$ & 1.13(1634)\\
NGC 7479       & 19 June 2001 & 0025541001 &tbabs$_{gal}$*(tbabs*po +zgauss +po +mekal)                 &0.049& 82.70$^{+20.2}_{-12.6}$ & 0.721$^{0.368}_{0.368}$ & 1.00(10)  \\
NGC 7603       & 14 June 2006 & 0305600601 &tbabs$_{gal}$*(zbremss +po)                                 &0.041& $\sim$0$^{d}$           & 0.060$^{0.023}_{0.023}$ & 1.06(1011)\\
UGC 12138      & 3 June 2001  & 0103860301 &tbabs$_{gal}$*(zbremss +po)                                 &0.067& $\sim$0$^{d}$           & $<$0.110                & 1.04(438) \\
\enddata
\tablenotetext{a}{Galactic column from \citet{dic90}, in units of 10$^{22}$ cm$^{-2}$. These values were used in the first {\it tbabs} model component.}
\tablenotetext{b}{Intrinsic column (total-Galactic) absorbing the central continuum source, in units 10$^{22}$ cm$^{-2}$.}
\tablenotetext{c}{ Fe K$\alpha$ equivalent width, in keV. Upper limits are given for those lines not detected with significance at the 95\% level.  }
\tablenotetext{d}{Column frozen to Galactic value in model.}
\tablenotetext{e}{ Because the data are very low $S/N$ and our model has $\chi^{2}>2$, we did not attempt to measure Fe K$\alpha$ in the spectrum of Mrk 622. }
\tablenotetext{f}{Components based upon the model presented in \citet{ris09b}.}
\label{xmodels}
\end{deluxetable}

\clearpage
\begin{deluxetable}{lcccccccl}
\tabletypesize{\scriptsize}
\tablewidth{0pt}
\tablecaption{Summary of Basic Data on 1.8/1.9s}
\tablehead{
\colhead{ } &\colhead{Our} &\colhead{ } &\colhead{Spitzer} &\colhead{NLR} & \colhead{BLR} &\colhead{X-ray} &\colhead{Past } &\colhead{Reason for }\\
\colhead{Object} &\colhead{Type} &\colhead{b/a$^{a}$} &\colhead{N$_{H}$$^{b}$} &\colhead{N$_{H}$$^{c}$} & \colhead{N$_{H}$$^{d}$} &\colhead{N$_{H}$$^{e}$} &\colhead{Variability?} &\colhead{Classification as 1.8/1.9}
}
\tablecolumns{9}
\startdata
\multicolumn{9}{c}{Current 1.8/1.9s} \\
\hline
Mrk 334        & 1.8     & 0.70 & 0.57$\pm$0.16 & 0.10$\pm$0.03  & -0.07$\pm$0.10 & --                          & no             & low flux               \\
Mrk 471        & 1.9     & 0.67 & 0.46$\pm$0.09 & 0.46$\pm$0.12  &           --   & --                          & no             & inconclusive           \\
Mrk 915        & 1.9     & 0.30 &  --           & 0.14$\pm$0.03  &$>$0.17         & --                          & yes (1.5-1.9)  & inconclusive           \\
Mrk 1126       & 1.8     & 1.0  &  --           & 0.27$\pm$0.12  & 0.20$\pm$0.10  & --                          & no             & low flux               \\
NGC 1365       & 1.9+HII & 0.55 & 1.21$\pm$0.15 & 0.68$\pm$0.02  &           --   & variable$^{f}$              & ?              & inconclusive           \\ 
NGC 2622       & 1.9     & 0.68 & 0.58$\pm$0.08 & 0.19$\pm$0.05  &           --   & 1.18$^{+0.13}_{-0.11}$      & yes (1-1.9)    & low flux               \\
NGC 3786       & 1.9     & 0.59 & 0.91$\pm$0.11 & 0.23$\pm$0.05  &$>$0.08         & 3.99$^{+0.20}_{-0.17}$      & ?              & inconclusive           \\
NGC 5252       & 1.9     & 0.56 &  --           & 0.37$\pm$0.05  &           --   & 26.54$^{+0.77}_{-0.74}$     & no             & reddened               \\
NGC 7314       & 1.9     & 0.46 &   $\approx$0  & 0.27$\pm$0.01  &$>$0.01         & 1.06$^{+0.00}_{-0.00}$      & no             & internally reddened    \\
UGC 7064       & 1.9     & 1.0  & 0.84$\pm$0.28 & 0.12$\pm$0.07  &           --   & --                          & no             & internally reddened    \\
UM 146         & 1.8     & 0.77 & 0.74$\pm$0.18 & 0.22$\pm$0.13  & 0.02$\pm$0.10  & --                          & ?              & low flux               \\
\hline
\multicolumn{9}{c}{Previous 1.8/1.9s} \\
\hline
Mrk 609        & 2       & 0.90 & 0.33$\pm$0.05 & 0.51$\pm$0.06  &           --   & 0.04$^*$                    & yes (1.5-1.8-2)& low flux               \\
Mrk 728        & 1.2     & 0.65 &  --           & 0.02$\pm$0.06  & 0.23$\pm$0.07  & 0.02$^*$                    & yes (1.9-1.2)  & low flux               \\
Mrk 883        & 2       & 0.61 & 0.66$\pm$0.11 & 0.21$\pm$0.04  &           --   & 0.11$^{+0.01}_{-0.01}$$^{g}$& yes (1.9-2)    & low flux               \\  
Mrk 993        & 1.5     & 0.32 &  --           & 0.24$\pm$0.19  &-0.14$\pm$0.20  & 0.18$^{+0.01}_{-0.01}$      & yes (1.9-1.5)  & internally reddened    \\
Mrk 1018       & 1       & 0.52 &  --           &-0.06$\pm$0.31  & 0.13$\pm$0.04  & 0.03 $^*$                   & yes (1.9-1)    & low flux               \\
Mrk 1179       & 1       & 1.0  &  --           &-0.03$\pm$0.14  & 0.13$\pm$0.06  & --                          & yes (1.9-1)    & inconclusive           \\
NGC 2992       & 2       & 0.31 & 1.09$\pm$0.31 & 0.37$\pm$0.06  &           --   & 0.77$^{+0.00}_{-0.00}$      & yes (1.9-1-2)  & low flux               \\
NGC 7603       & 1       & 0.67 &   $\approx$0  &-0.016$\pm$0.12 & 0.10$\pm$0.19  & 0.04$^*$                    & yes (1.9-1)    & low flux               \\
\enddata
\tablenotetext{a}{Ratio of host galaxy's major axis length to minor axis length, from NED.}
\tablenotetext{b}{Column density estimated from the depth of the silicate 9.7 $\mu$m absorption feature in {\it Spitzer} IRS spectra, in units 10$^{22}$ cm$^{-2}$.}
\tablenotetext{c}{Column density estimated using the reddening of the NLR, assuming the local ISM dust/gas ratio of \citet{shu85}, in units 10$^{22}$ cm$^{-2}$.}
\tablenotetext{d}{Column density estimated using the reddening of the BLR, assuming the local ISM dust/gas ratio of \citet{shu85}, in units 10$^{22}$ cm$^{-2}$.}
\tablenotetext{e}{Hydrogen column density measured from {\it XMM} spectra, in units 10$^{22}$ cm$^{-2}$.}
\tablenotetext{f}{See notes on NGC 1365 in Section~\ref{indiv_comments}.}
\tablenotetext{g}{Average column of multiple {\it XMM} spectra.}
\tablenotetext{*}{X-ray column frozen to Galactic value in model.}
\label{sumtbl}
\end{deluxetable}

\clearpage
\clearpage
\clearpage
\clearpage

\clearpage
\setcounter{figure}{0}
\begin{center}
Appendix 1: Continuum-Subtracted Optical Spectra\\
\end{center}
\indent This section presents the continuum-subtracted optical spectra used to measure the reddenings in Table~\ref{reddenings}. On the first plot, IRAS 18325-5926, the emission lines most commonly referred to in the text are marked for reference.
\begin{figure}[h!]
\begin{center}
\includegraphics[angle=0,scale=1.0]{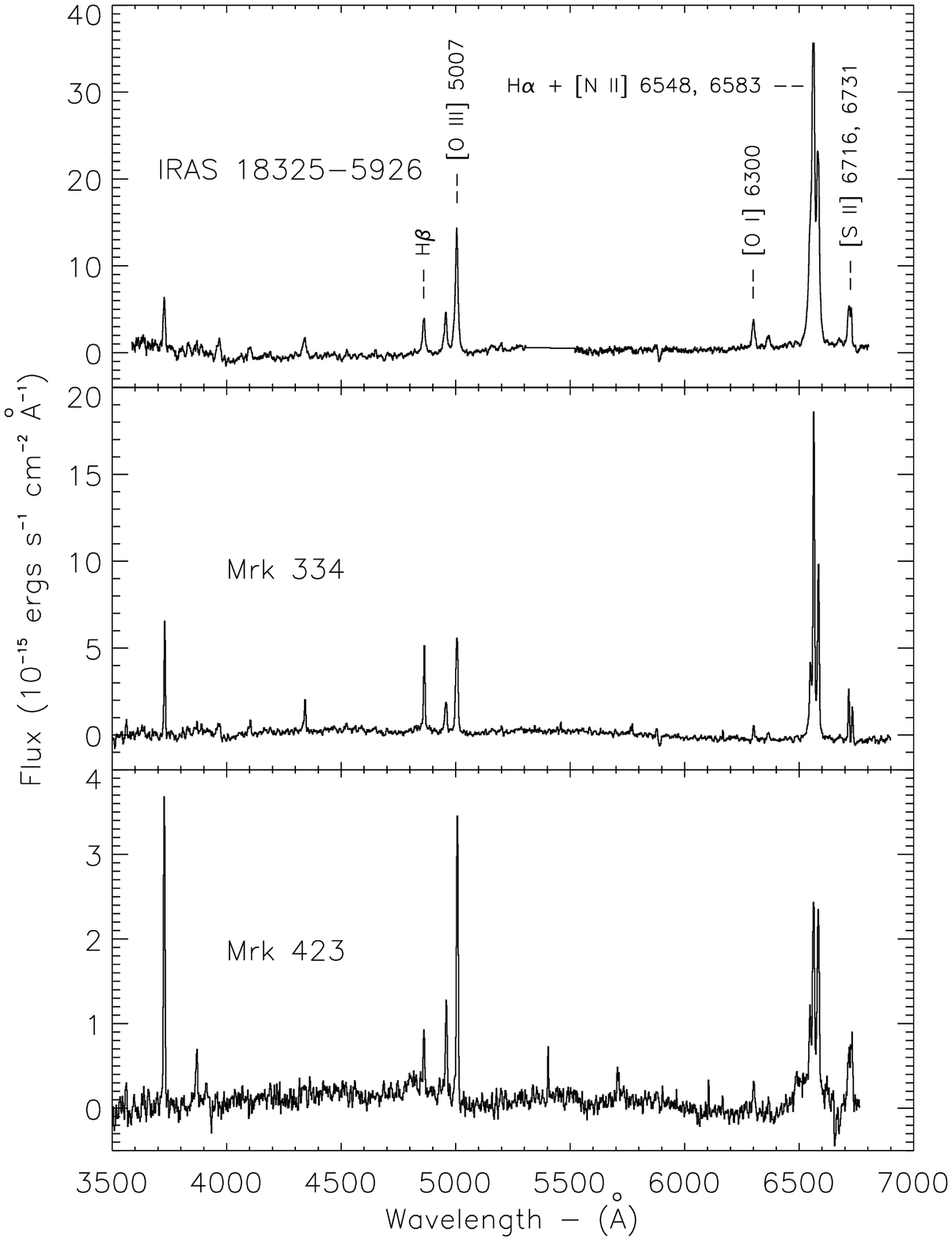}
\end{center}
\end{figure}
\begin{figure}[h!]
\begin{center}
\includegraphics[angle=0,scale=1.0]{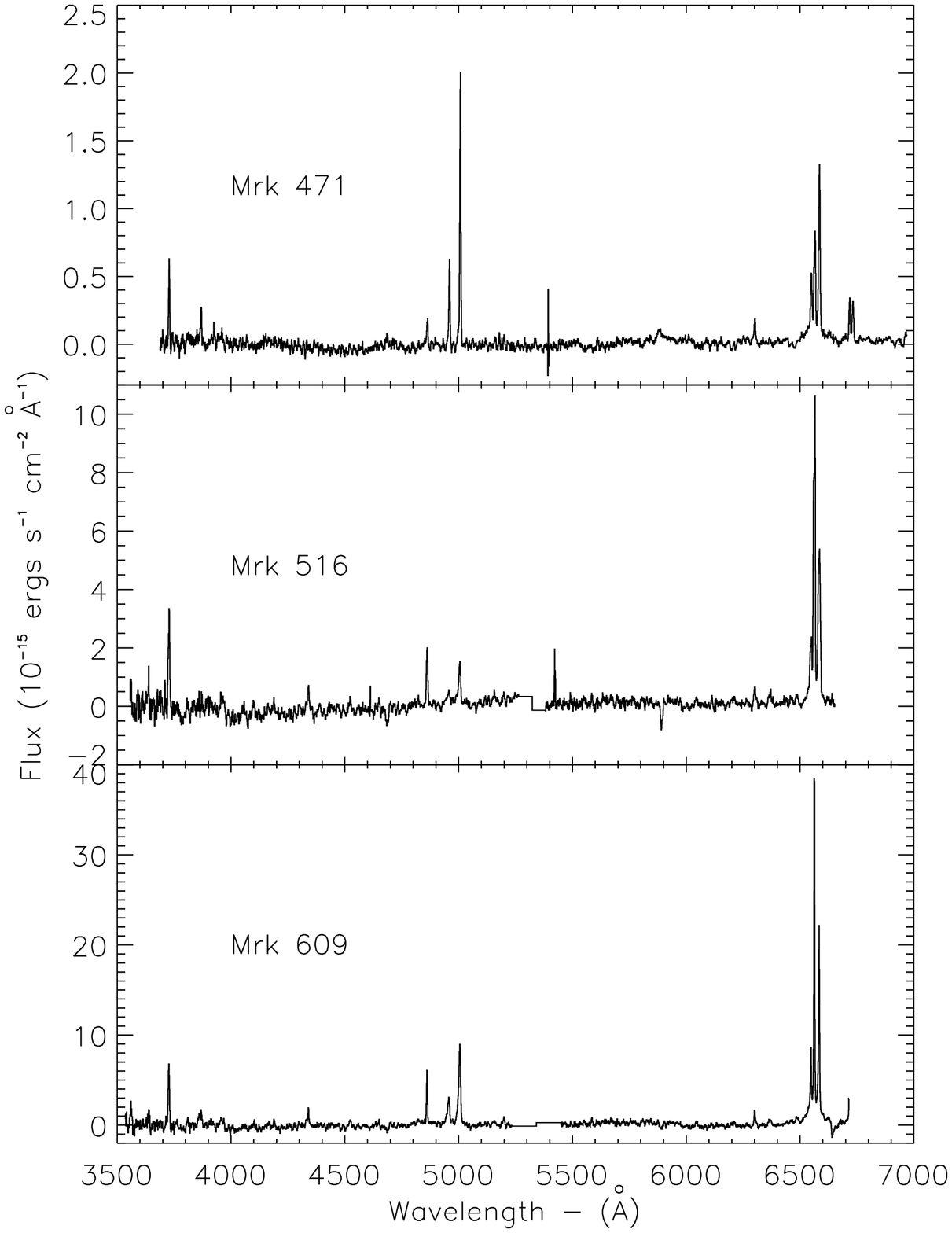}
\end{center}
\end{figure}
\begin{figure}[h!]
\begin{center}
\includegraphics[angle=0,scale=1.0]{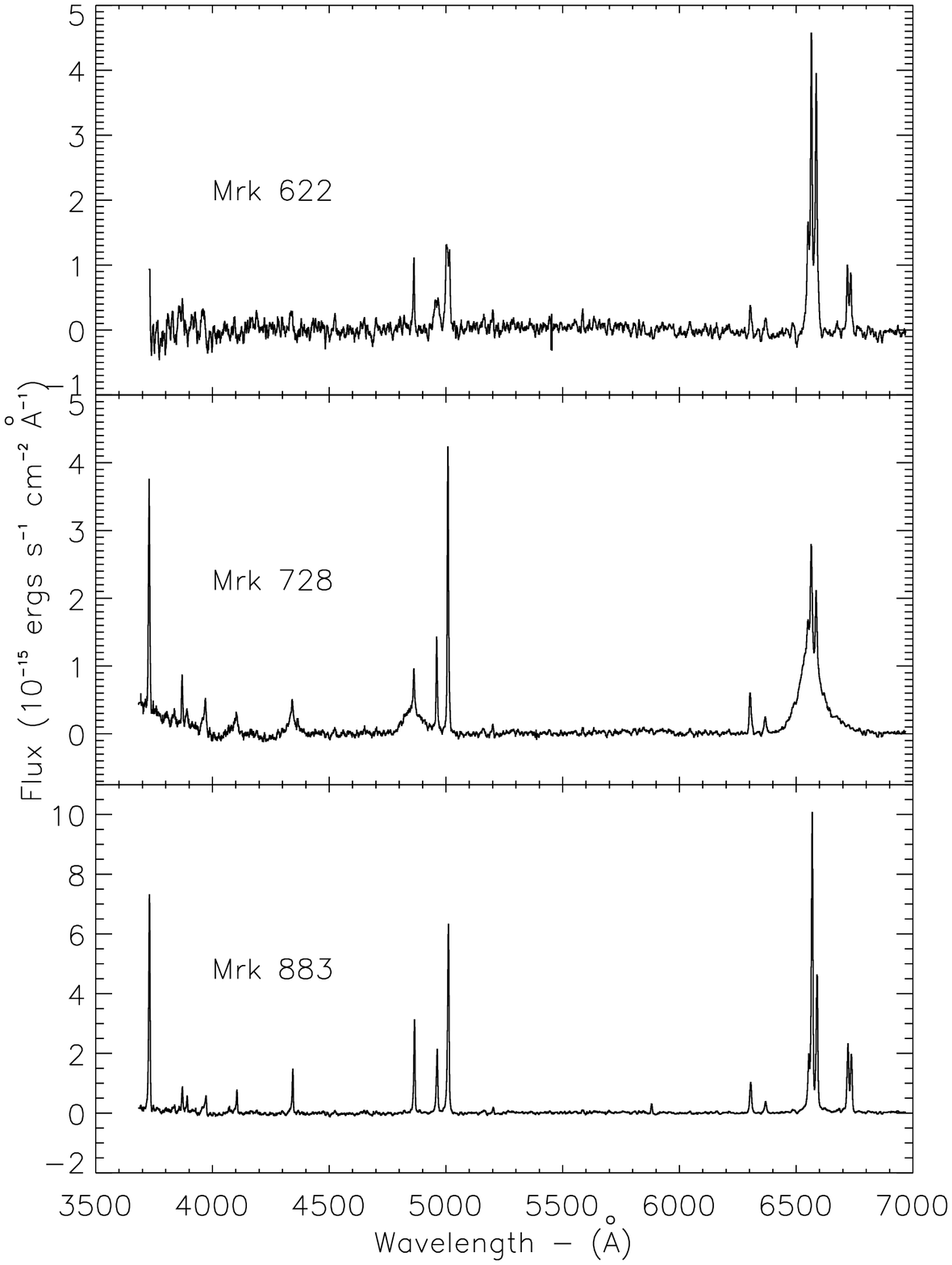}
\end{center}
\end{figure}
\begin{figure}[h!]
\begin{center}
\includegraphics[angle=0,scale=1.0]{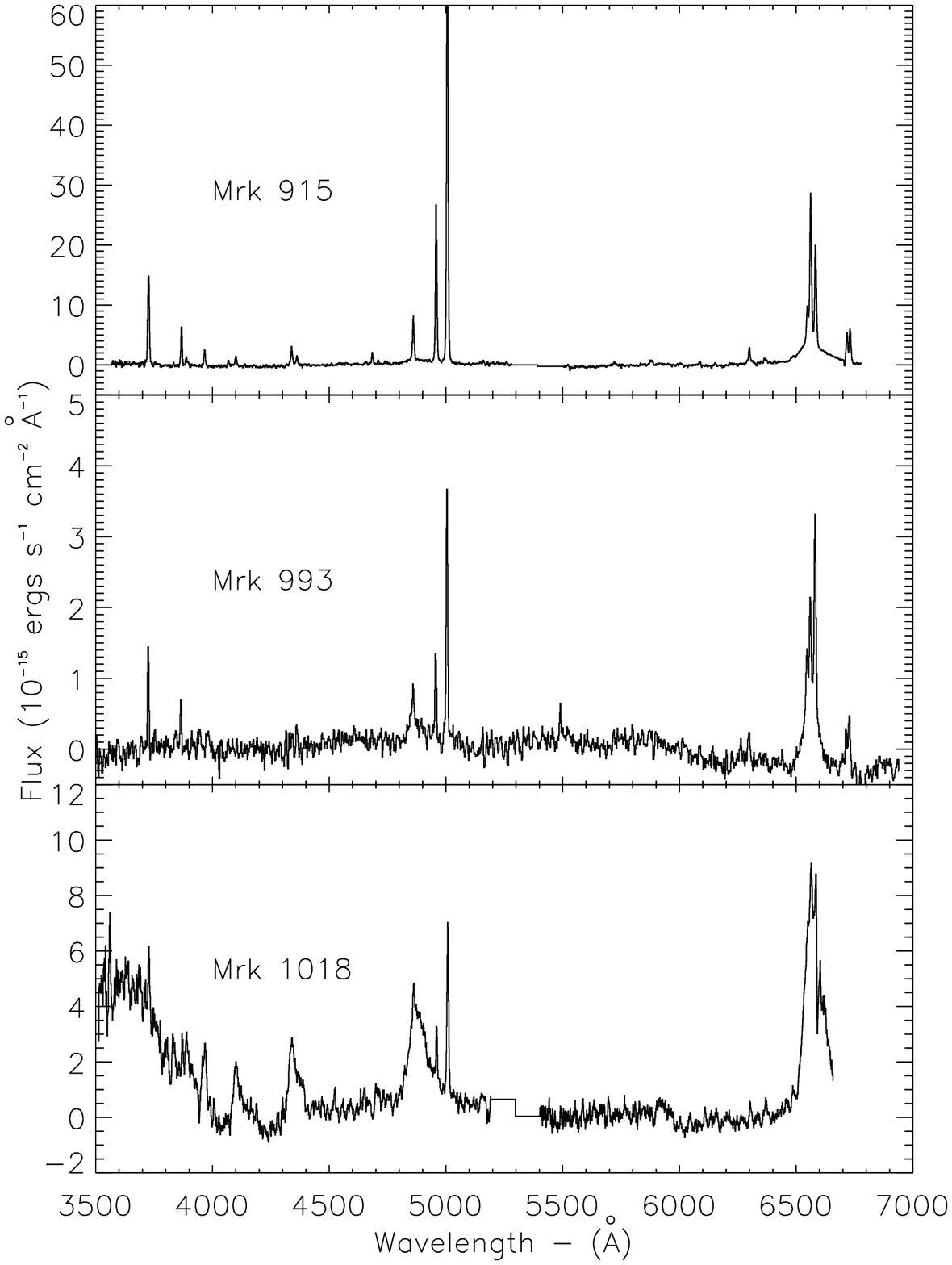}
\end{center}
\end{figure}
\begin{figure}[h!]
\begin{center}
\includegraphics[angle=0,scale=1.0]{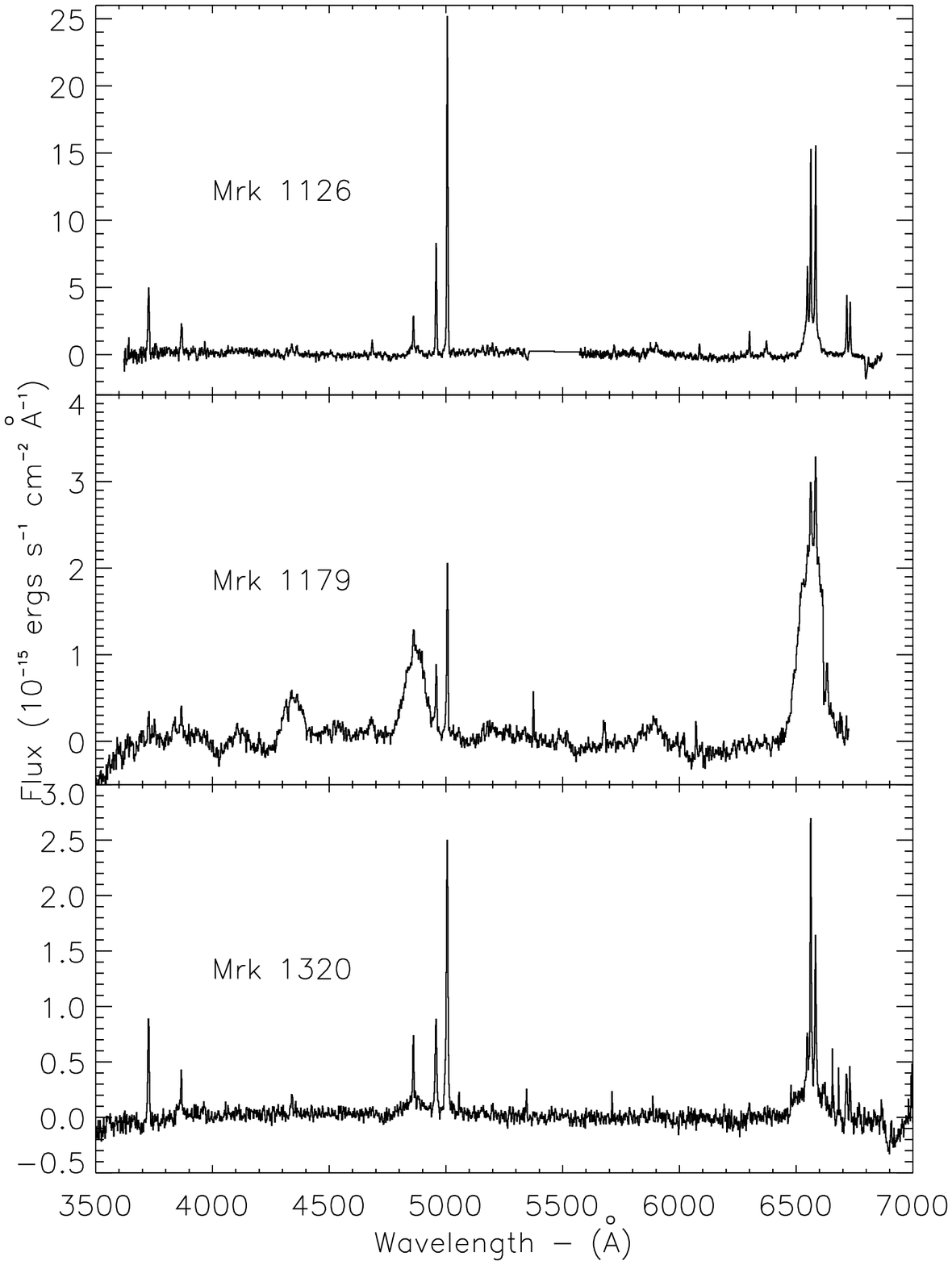}
\end{center}
\end{figure}
\begin{figure}[h!]
\begin{center}
\includegraphics[angle=0,scale=1.0]{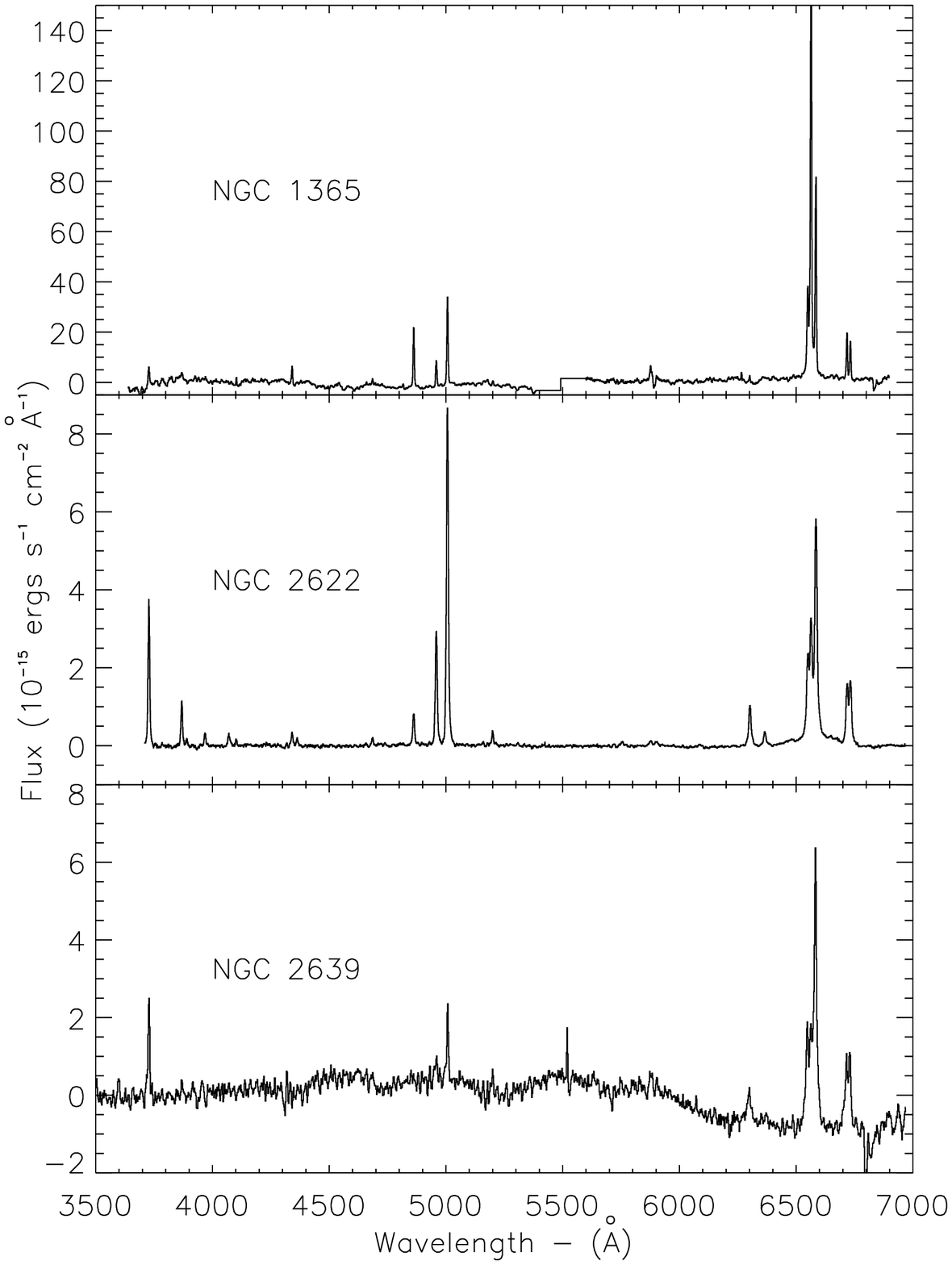}
\end{center}
\end{figure}
\begin{figure}[h!]
\begin{center}
\includegraphics[angle=0,scale=1.0]{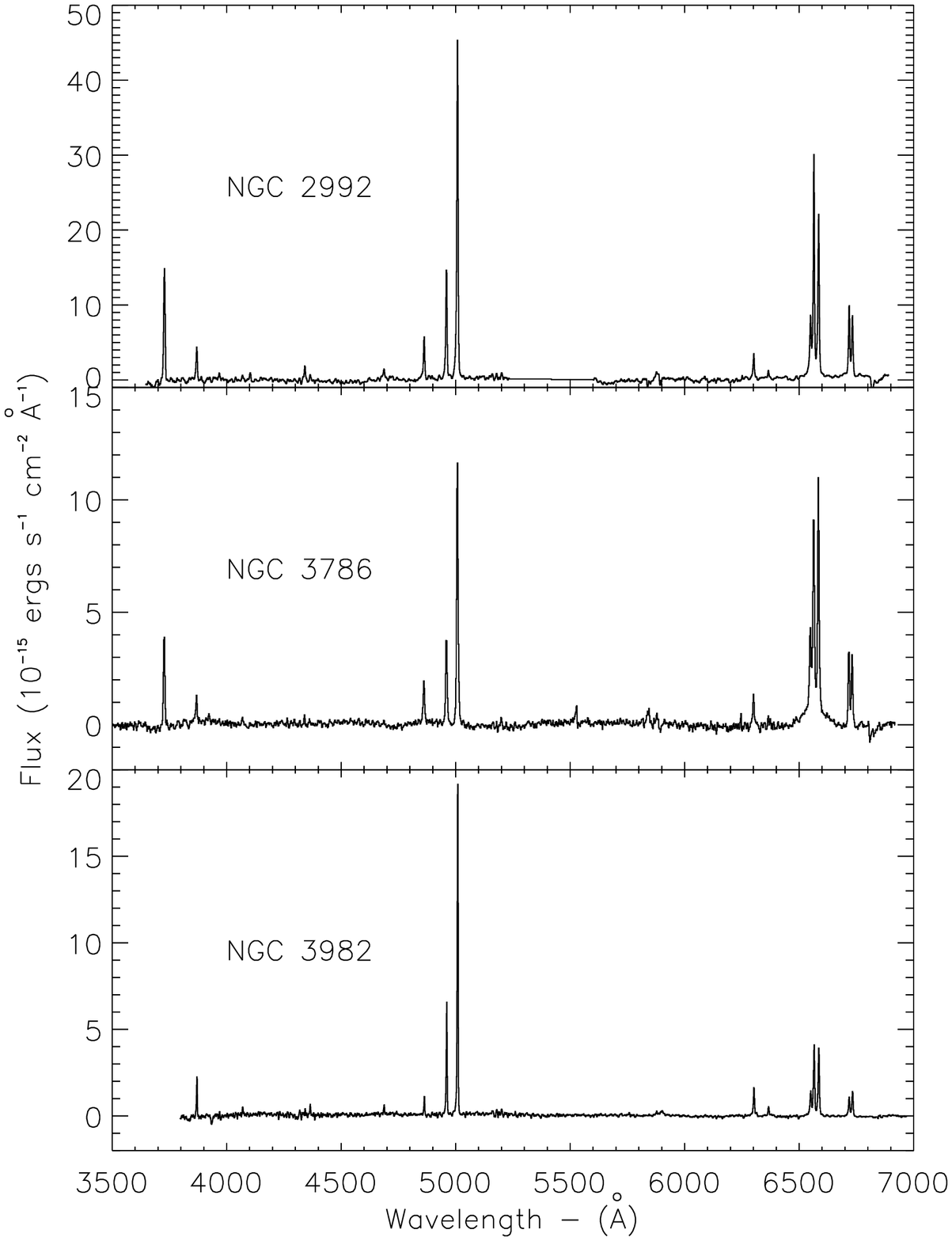}
\end{center}
\end{figure}
\begin{figure}[h!]
\begin{center}
\includegraphics[angle=0,scale=1.0]{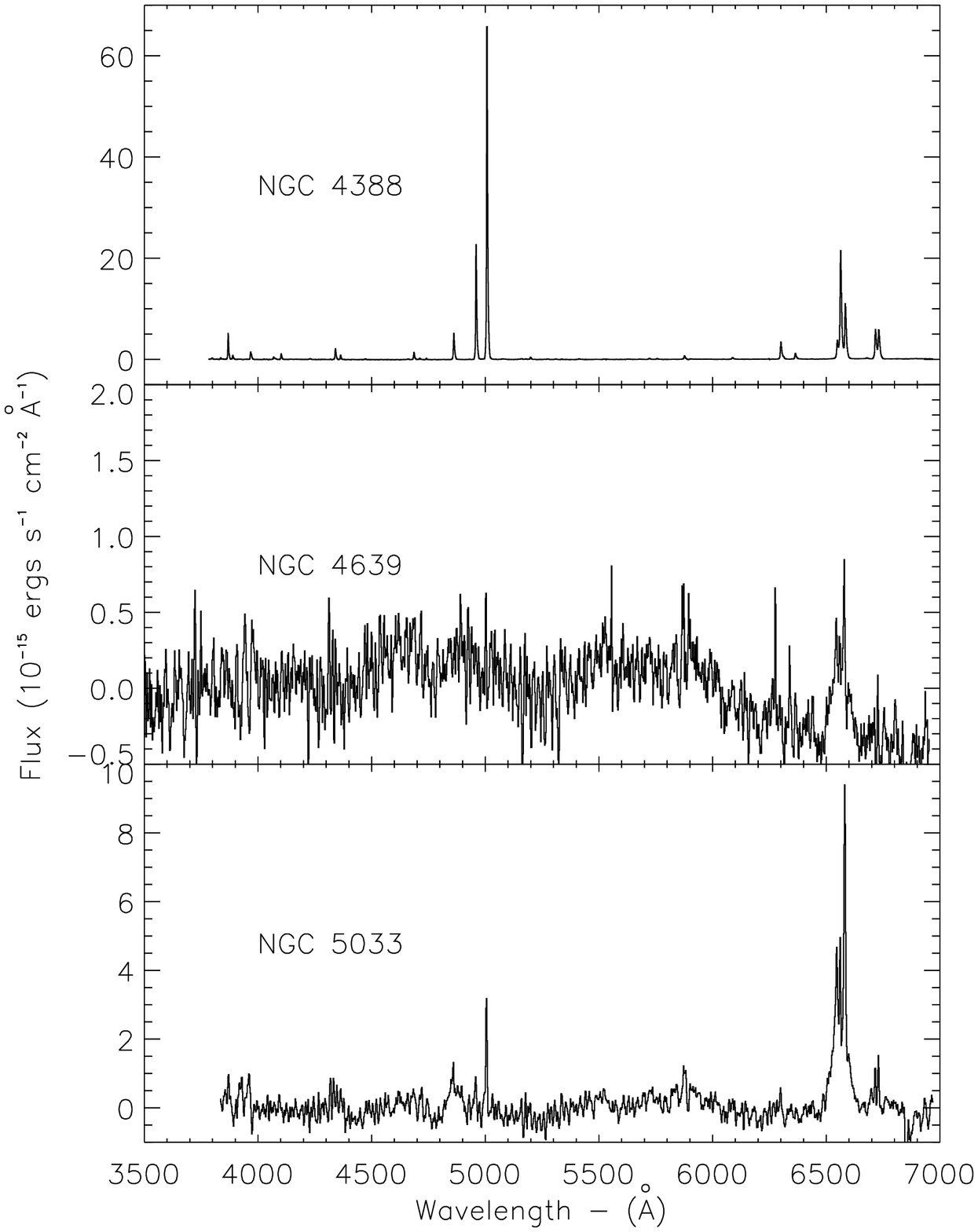}
\end{center}
\end{figure}
\begin{figure}[h!]
\begin{center}
\includegraphics[angle=0,scale=1.0]{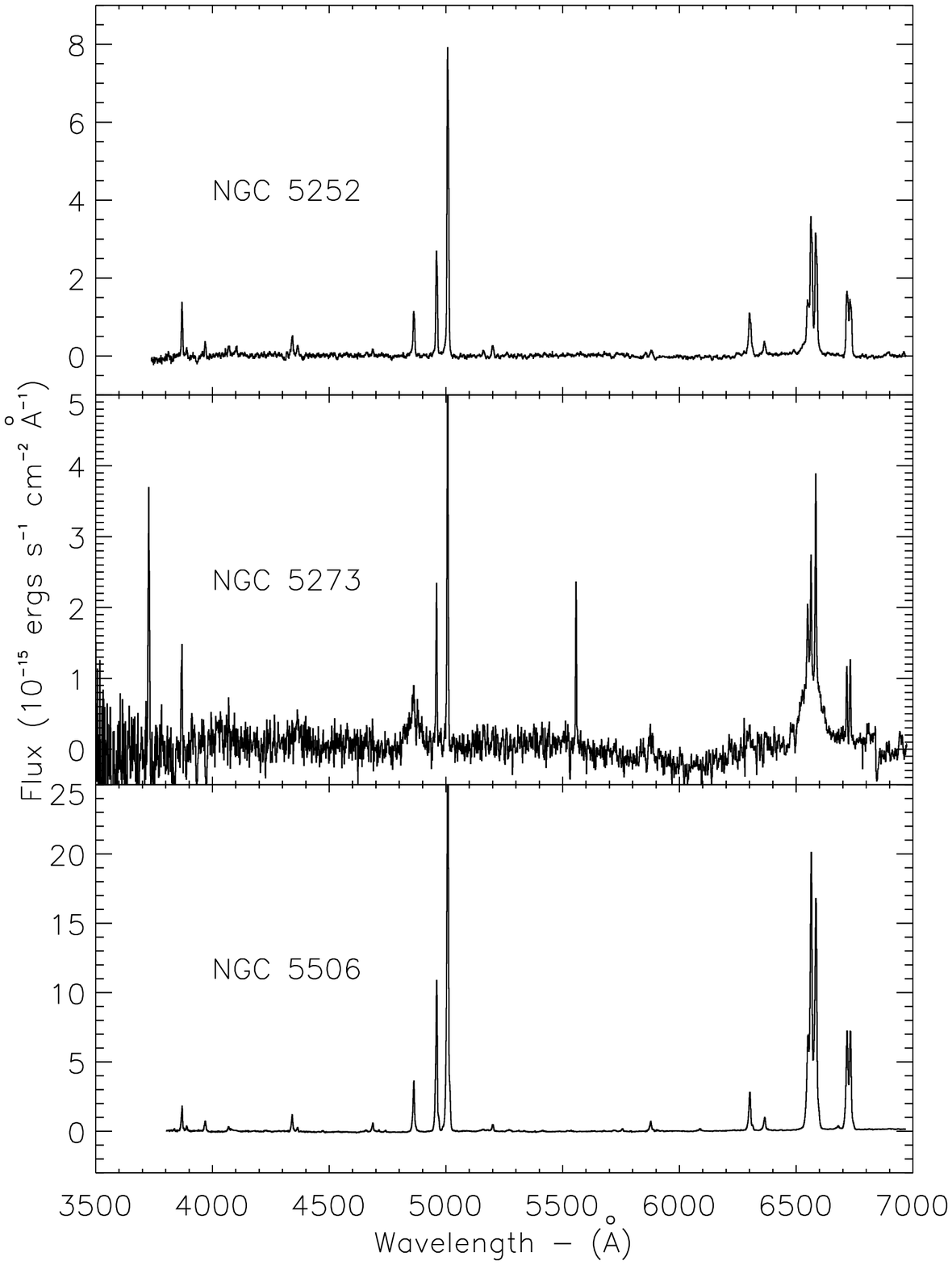}
\end{center}
\end{figure}
\begin{figure}[h!]
\begin{center}
\includegraphics[angle=0,scale=1.0]{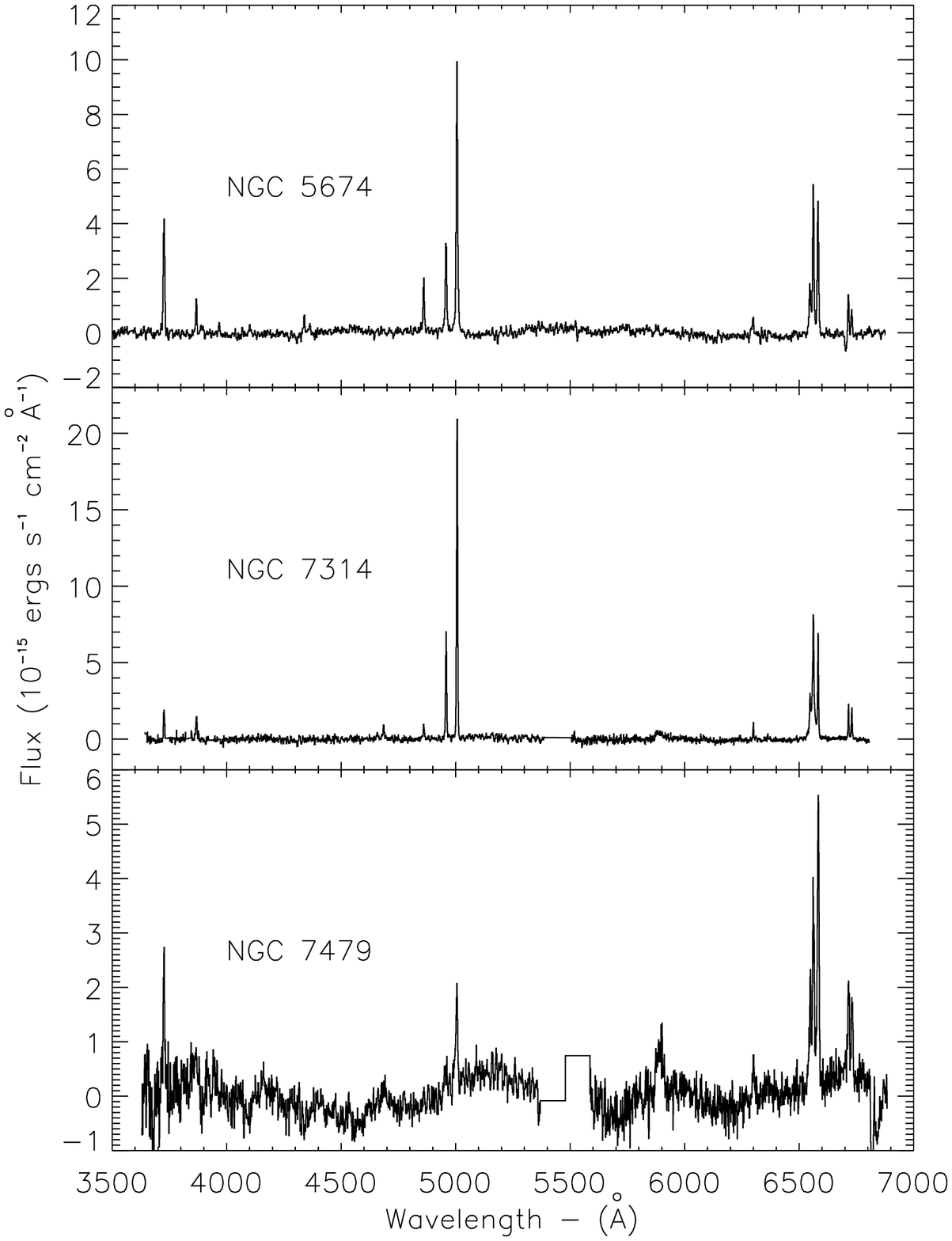}
\end{center}
\end{figure}
\begin{figure}[h!]
\begin{center}
\includegraphics[angle=0,scale=1.0]{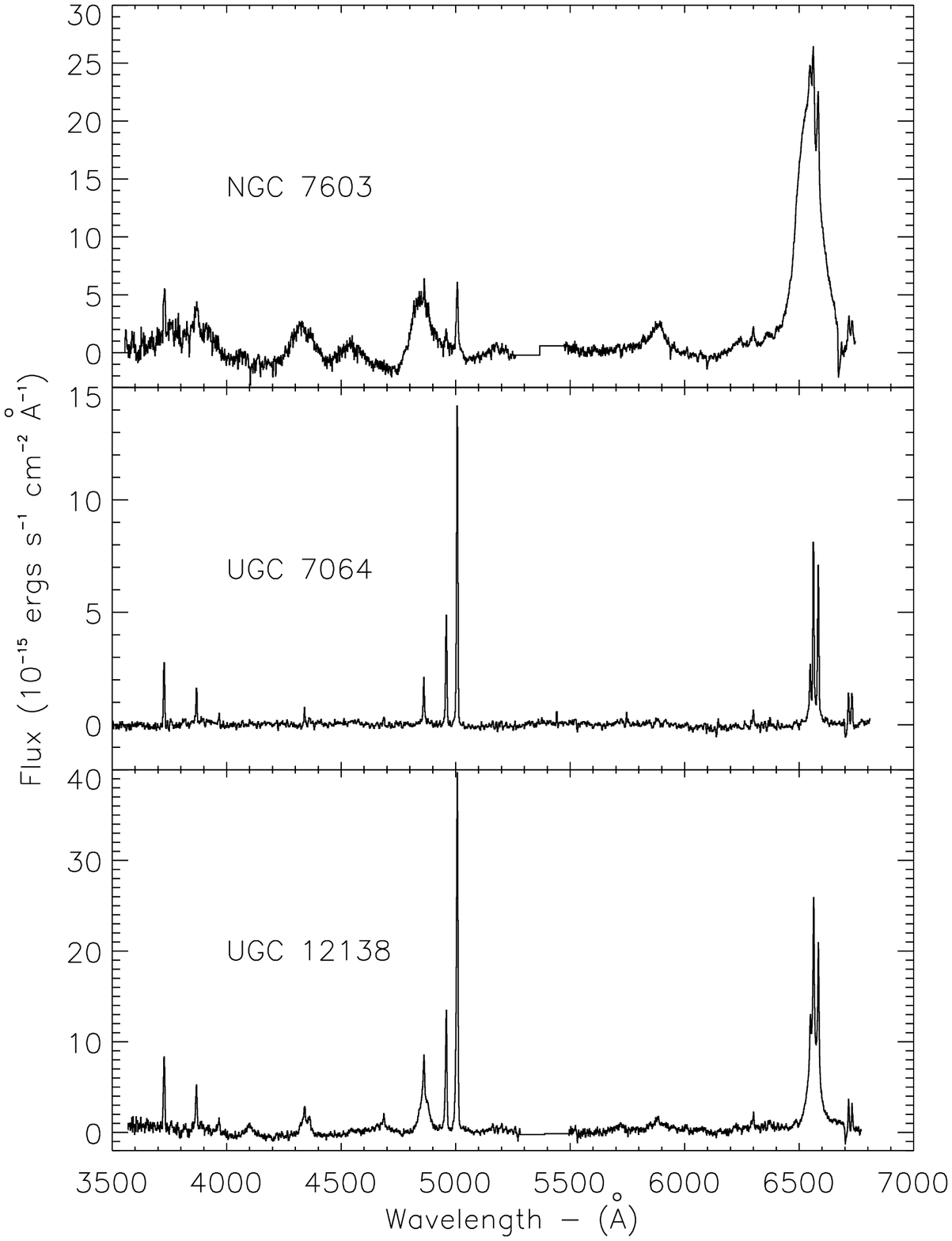}
\end{center}
\end{figure}
\begin{figure}[h!]
\begin{center}
\includegraphics[angle=0,scale=1.0]{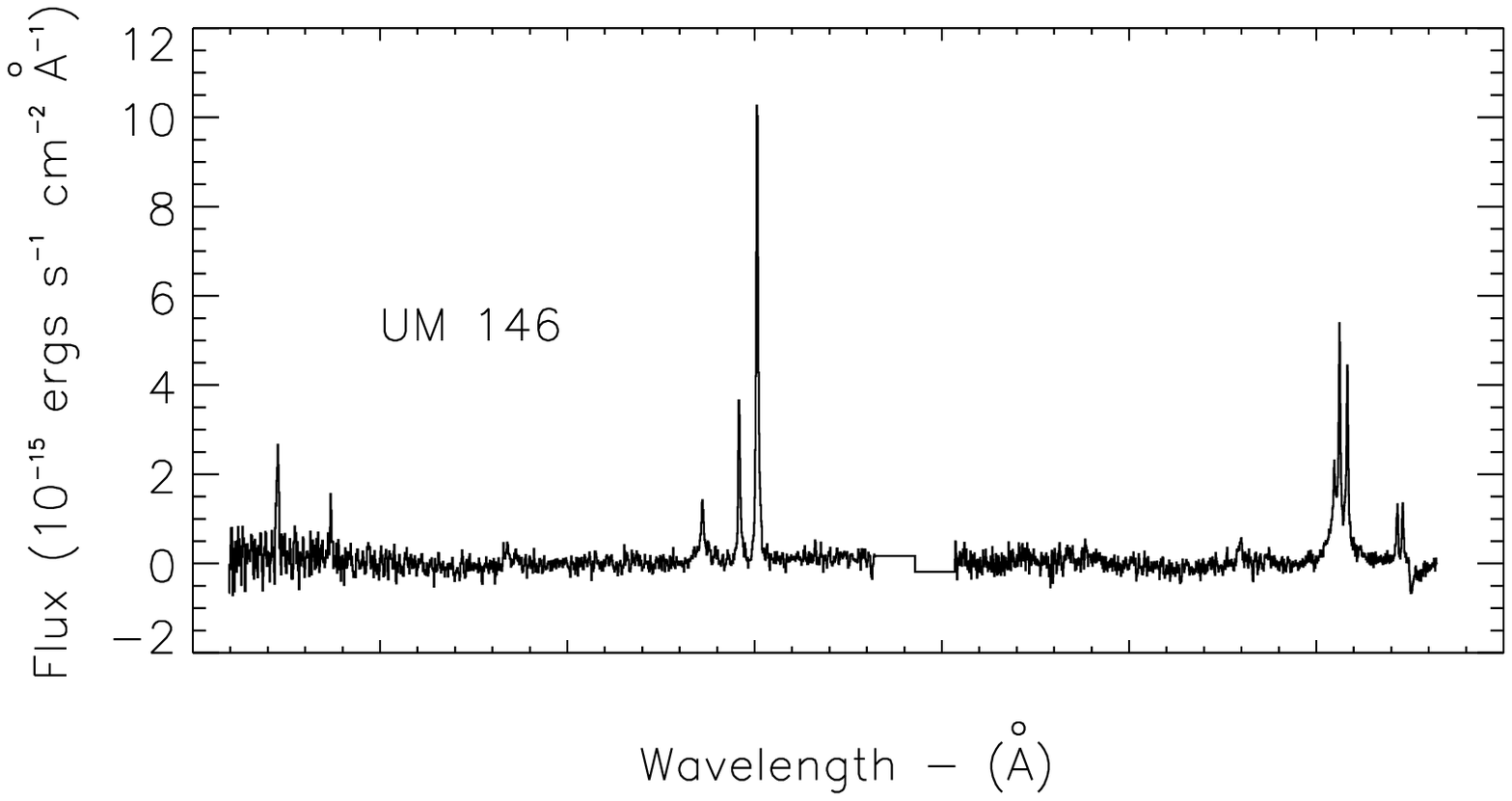}
\end{center}
\end{figure}

\setcounter{figure}{0}
\clearpage
\begin{center}
Appendix 2: H$\alpha$ Profiles\\
\end{center}
\indent This section presents plots of the H$\alpha$ profile after [N~II] subtraction. The [O~III] $\lambda$5007 profile, scaled to match the peak of H$\alpha$ in flux, is over-plotted with a dashed line at the centroid of H$\alpha$ for a comparison of their widths. The H$\beta$ line can be see at the far left end of the plots. \\
\indent The definitions of the different types are from \citet{mai95} ([O~III] refers to the [O~III] $\lambda$5007 emission line in these definitions): \\
\indent $\bullet$ Seyfert 1: Objects showing broad H$\beta$ emission line and with [O~III]/H$\beta$ $<$ 0.3 \\
\indent $\bullet$ Seyfert 1.2: Objects showing broad H$\beta$ and with 0.3 $<$ [O~III]/H$\beta$ $<$ 1 \\
\indent $\bullet$ Seyfert 1.5: Objects showing broad H$\beta$ and with 1 $<$ [O~III]/H$\beta$ $<$ 4 \\
\indent $\bullet$ Seyfert 1.8: Objects showing broad H$\beta$ and with 4 $<$ [O~III]/H$\beta$ \\
\indent $\bullet$ Seyfert 1.9: Objects not showing broad H$\beta$, but having broad H$\alpha$ \\
\indent $\bullet$ Seyfert 2: Objects without broad-line emission that fall in the AGN region in the \citet{vei87} diagnostic diagram.\\
\indent In this paper, we define ``broad'' emission as line emission that is significantly wider than the [O~III] $\lambda$5007 line, {\it in the profile wings}, indicating that the emission is not from the inner NLR, but is generated in a distinct high-density region where [O~III] emission is collisionally suppressed (i.e., the BLR). \\
\begin{figure}[h!]
\begin{center}
\includegraphics[angle=0,scale=1.0]{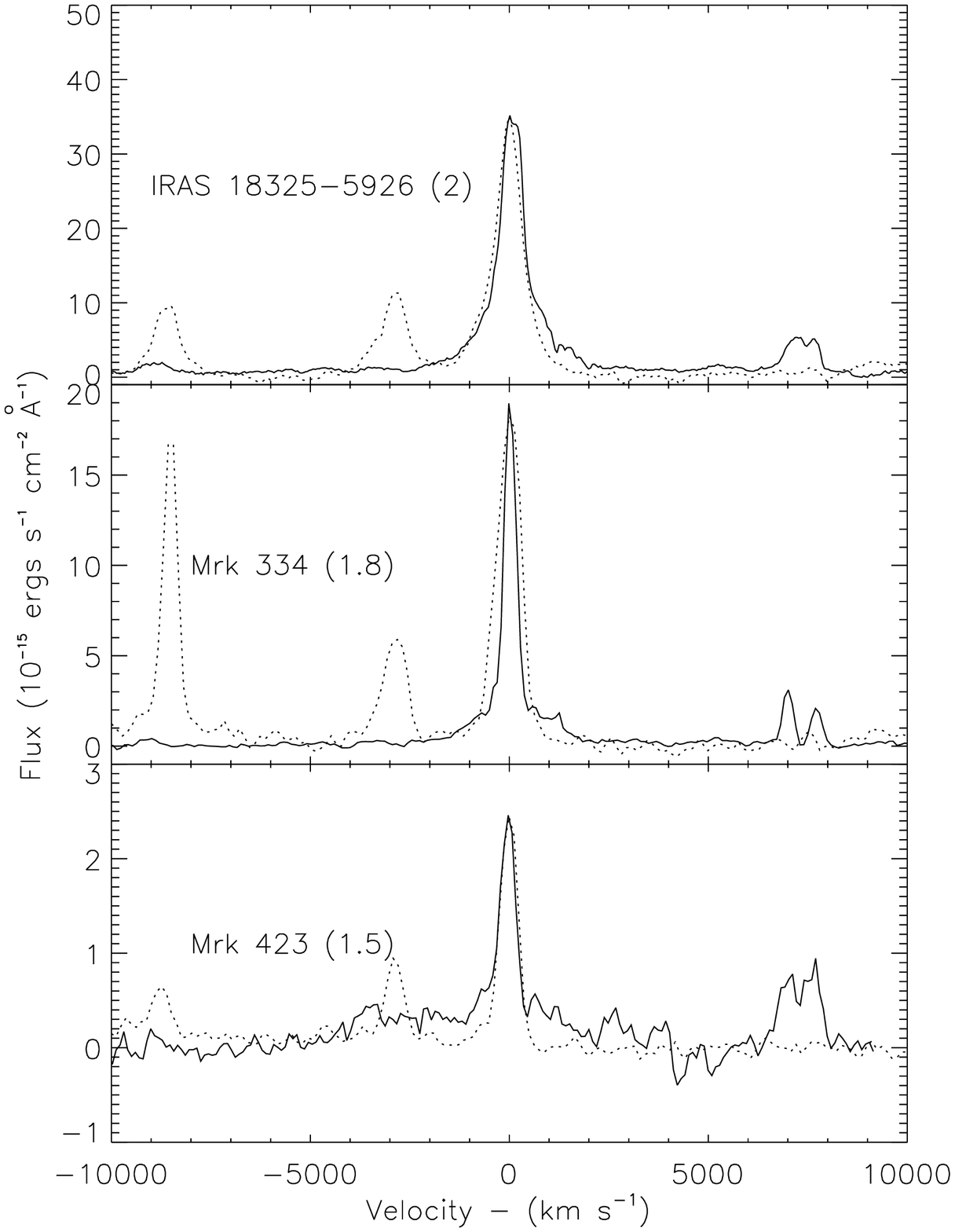}
\end{center}
\end{figure}
\begin{figure}[h!]
\begin{center}
\includegraphics[angle=0,scale=1.0]{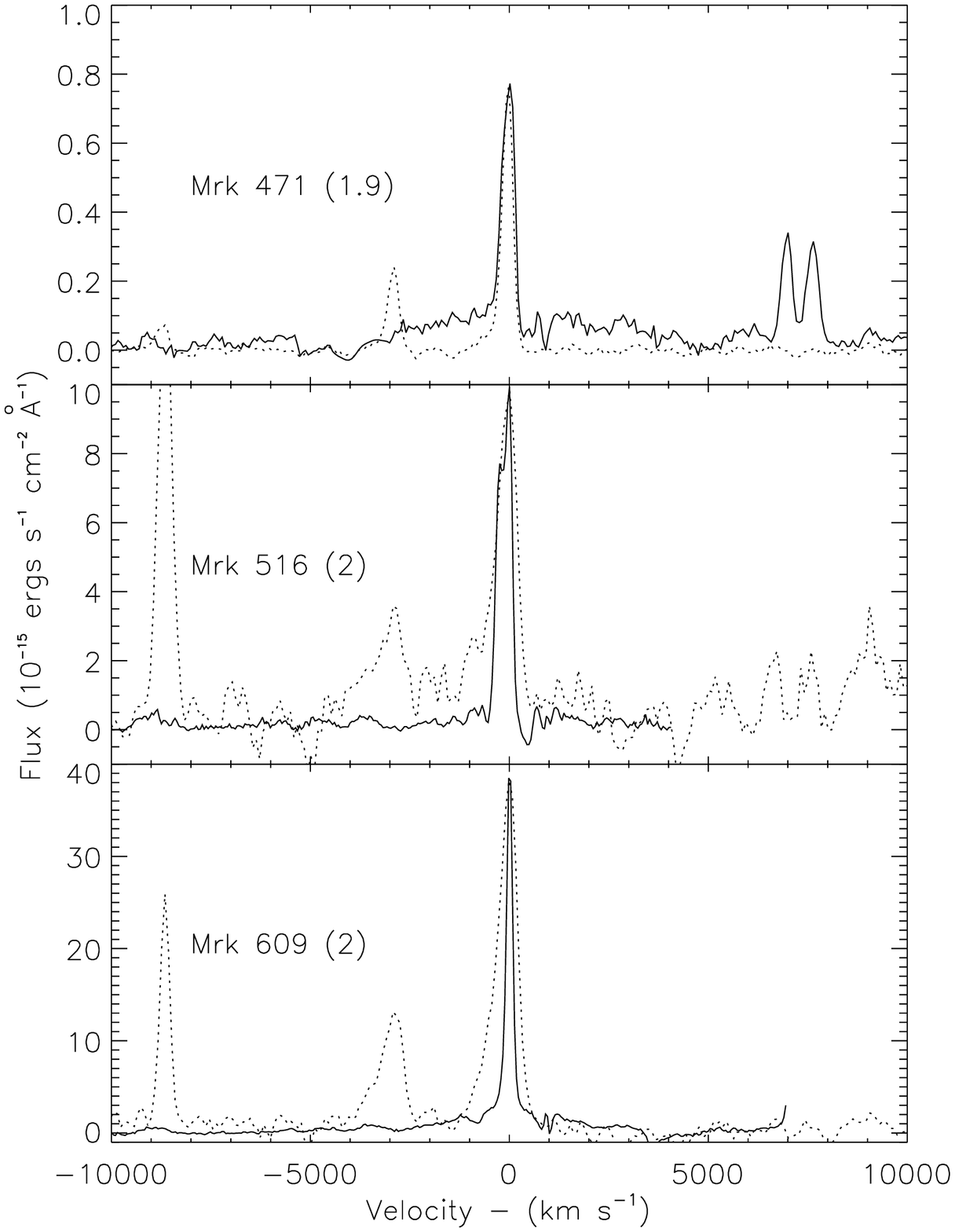}
\end{center}
\end{figure}
\begin{figure}[h!]
\begin{center}
\includegraphics[angle=0,scale=1.0]{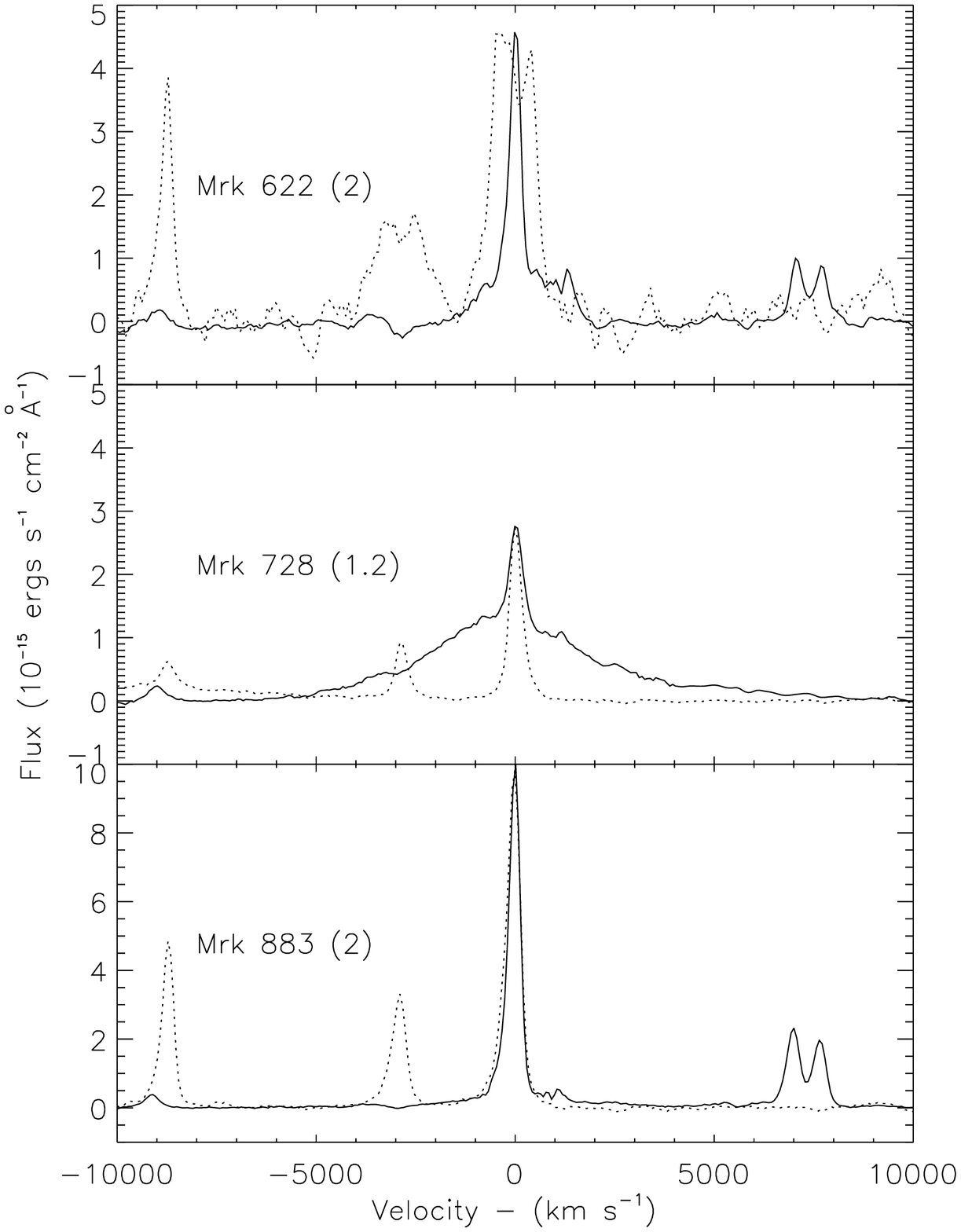}
\end{center}
\end{figure}
\begin{figure}[h!]
\begin{center}
\includegraphics[angle=0,scale=1.0]{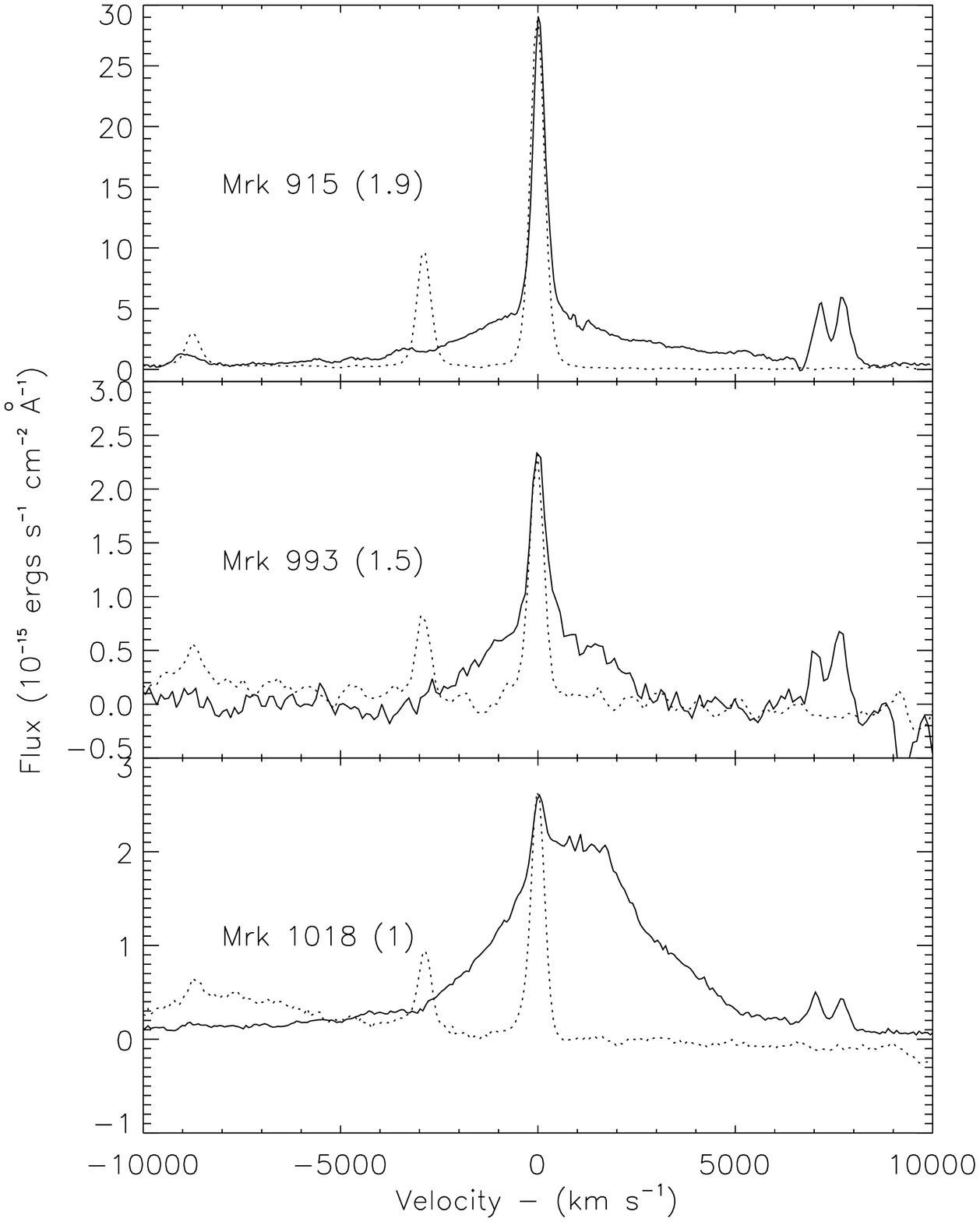}
\end{center}
\end{figure}
\begin{figure}[h!]
\begin{center}
\includegraphics[angle=0,scale=1.0]{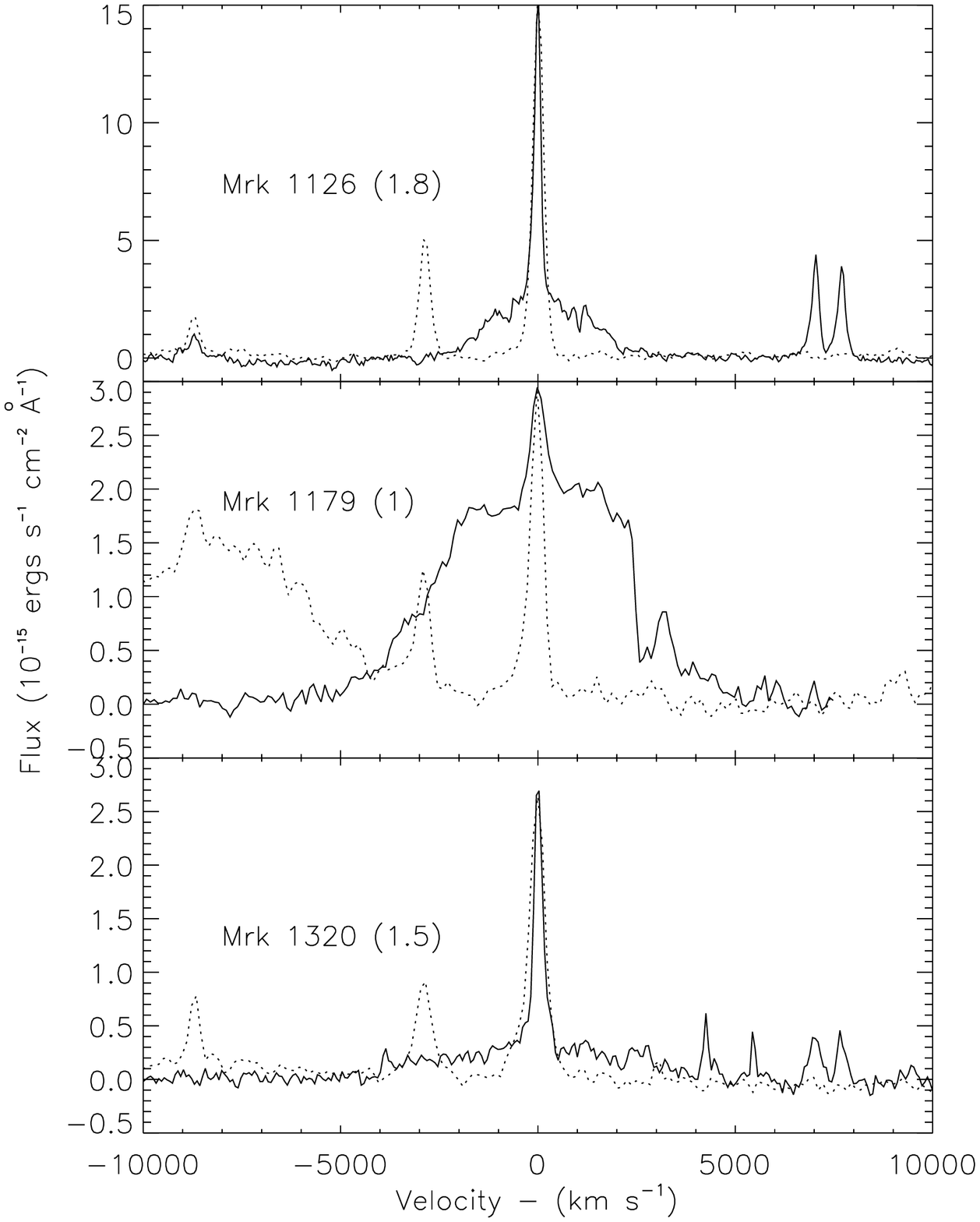}
\end{center}
\end{figure}
\begin{figure}[h!]
\begin{center}
\includegraphics[angle=0,scale=1.0]{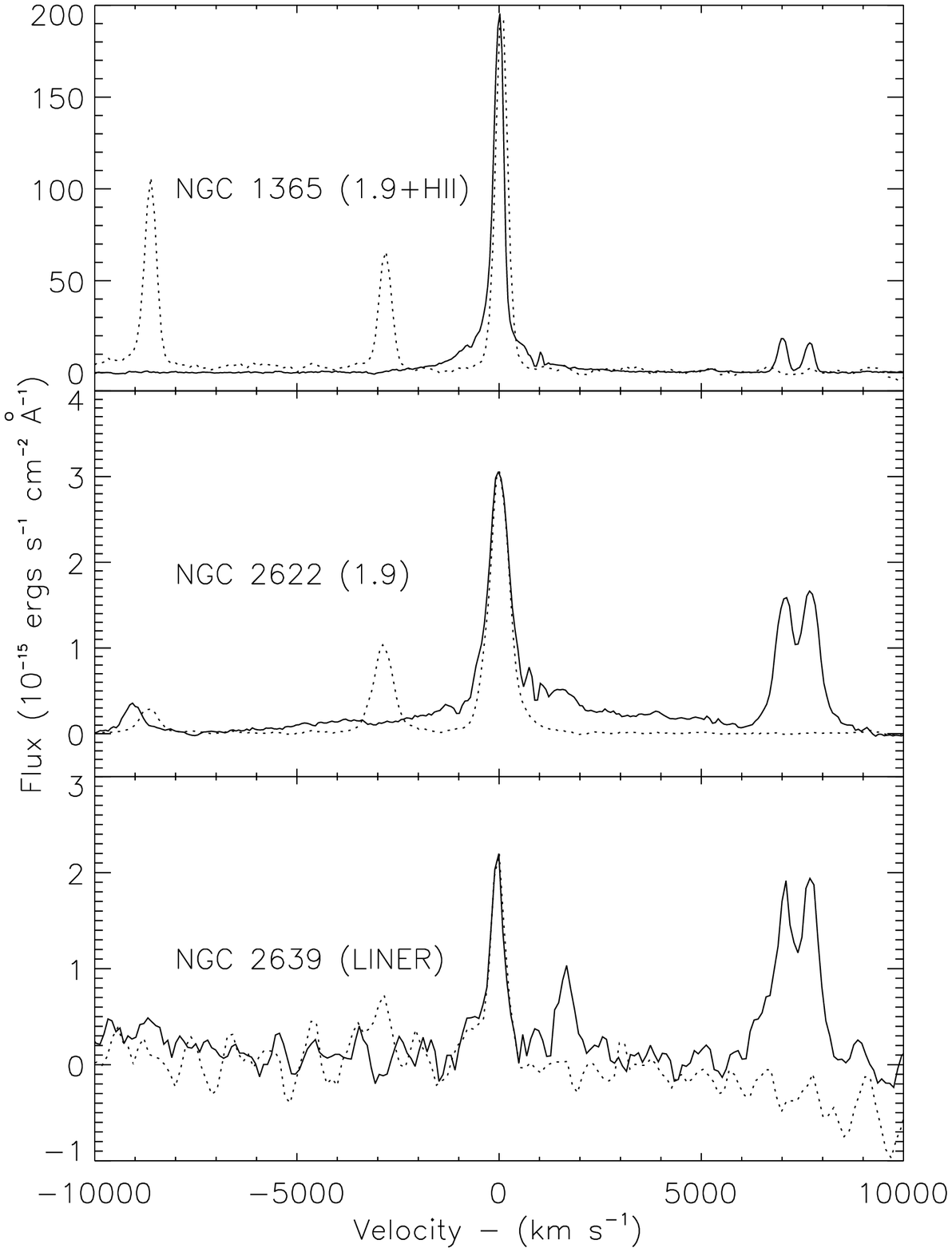}
\end{center}
\end{figure}
\begin{figure}[h!]
\begin{center}
\includegraphics[angle=0,scale=1.0]{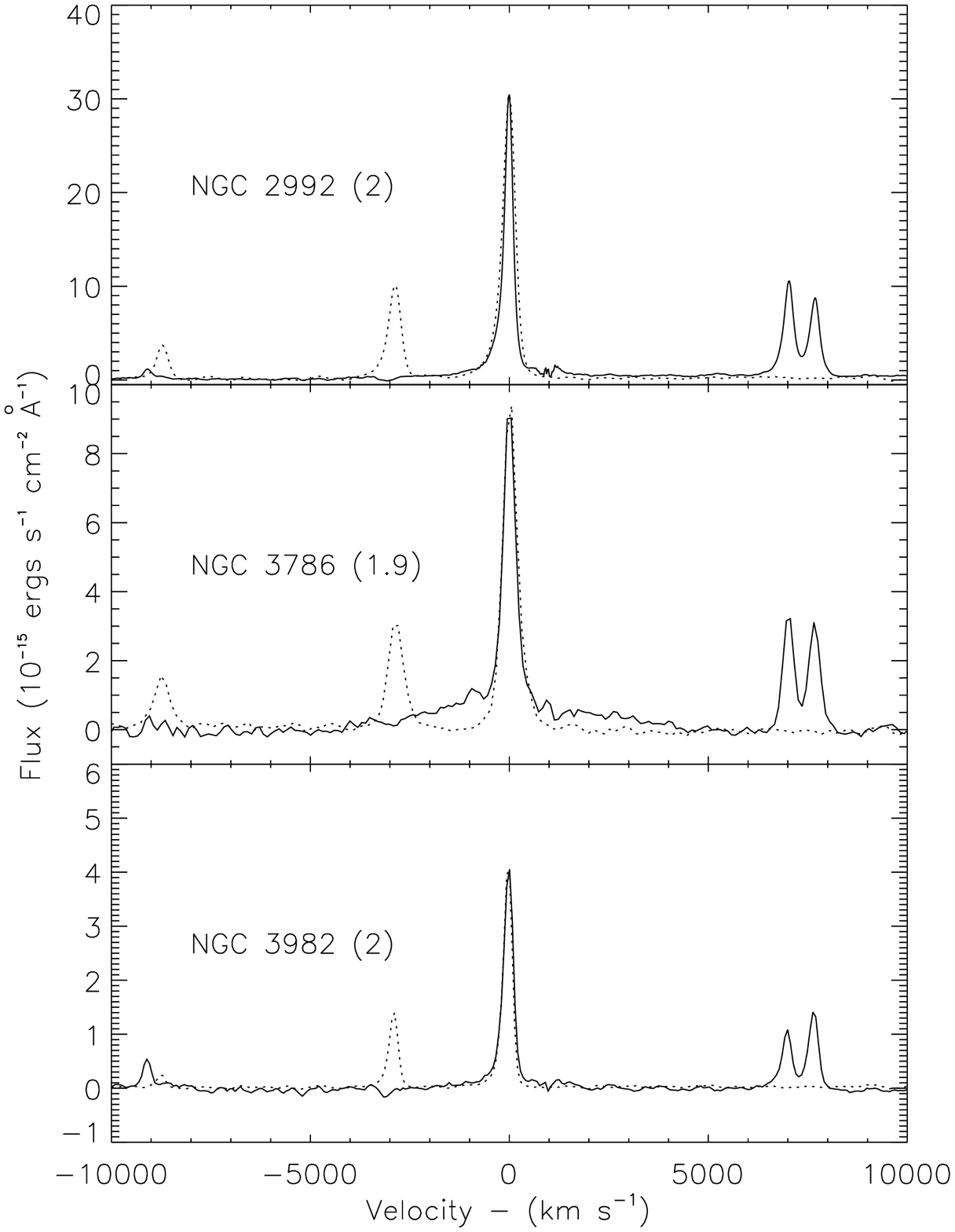}
\end{center}
\end{figure}
\begin{figure}[h!]
\begin{center}
\includegraphics[angle=0,scale=1.0]{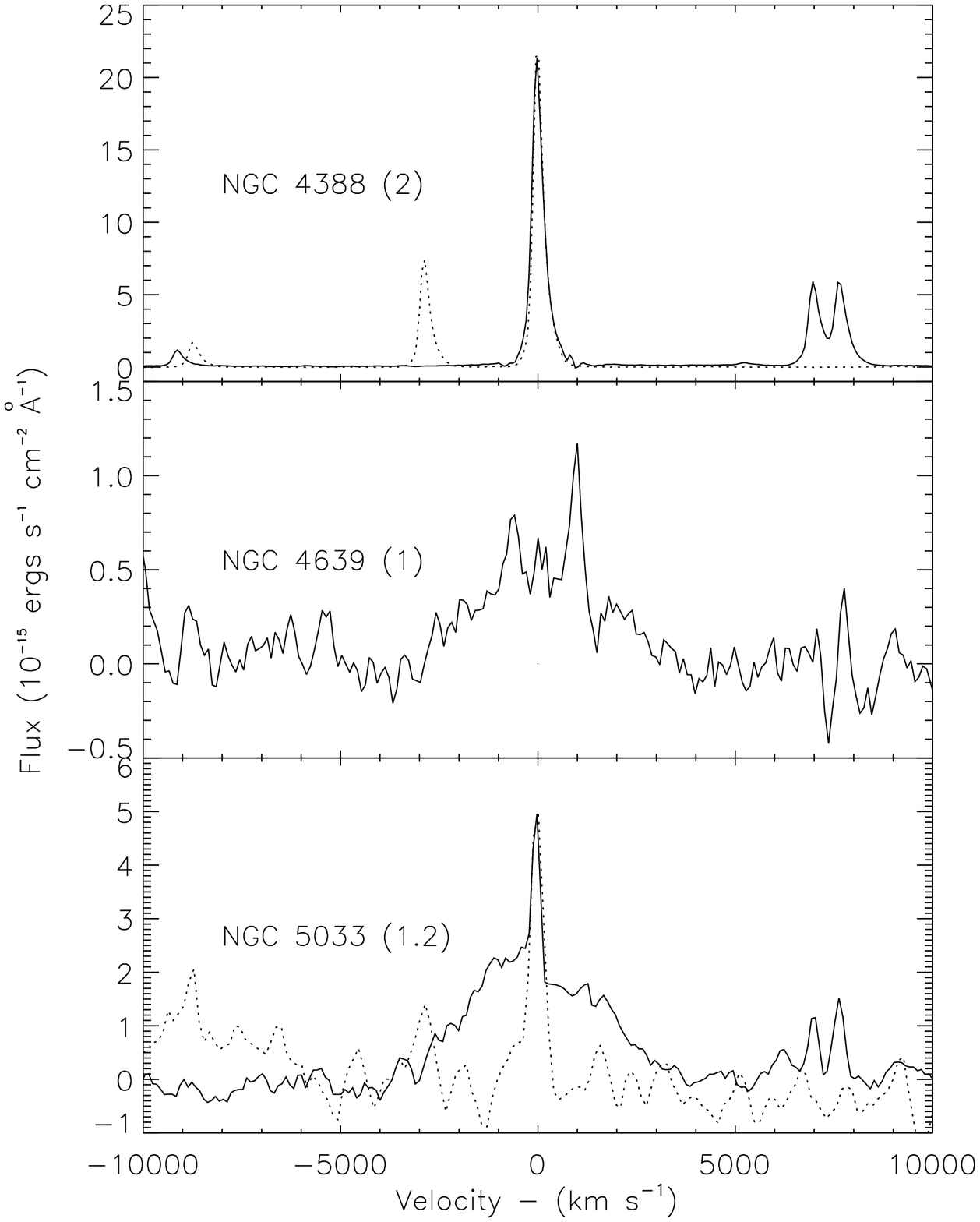}
\end{center}
\end{figure}
\begin{figure}[h!]
\begin{center}
\includegraphics[angle=0,scale=1.0]{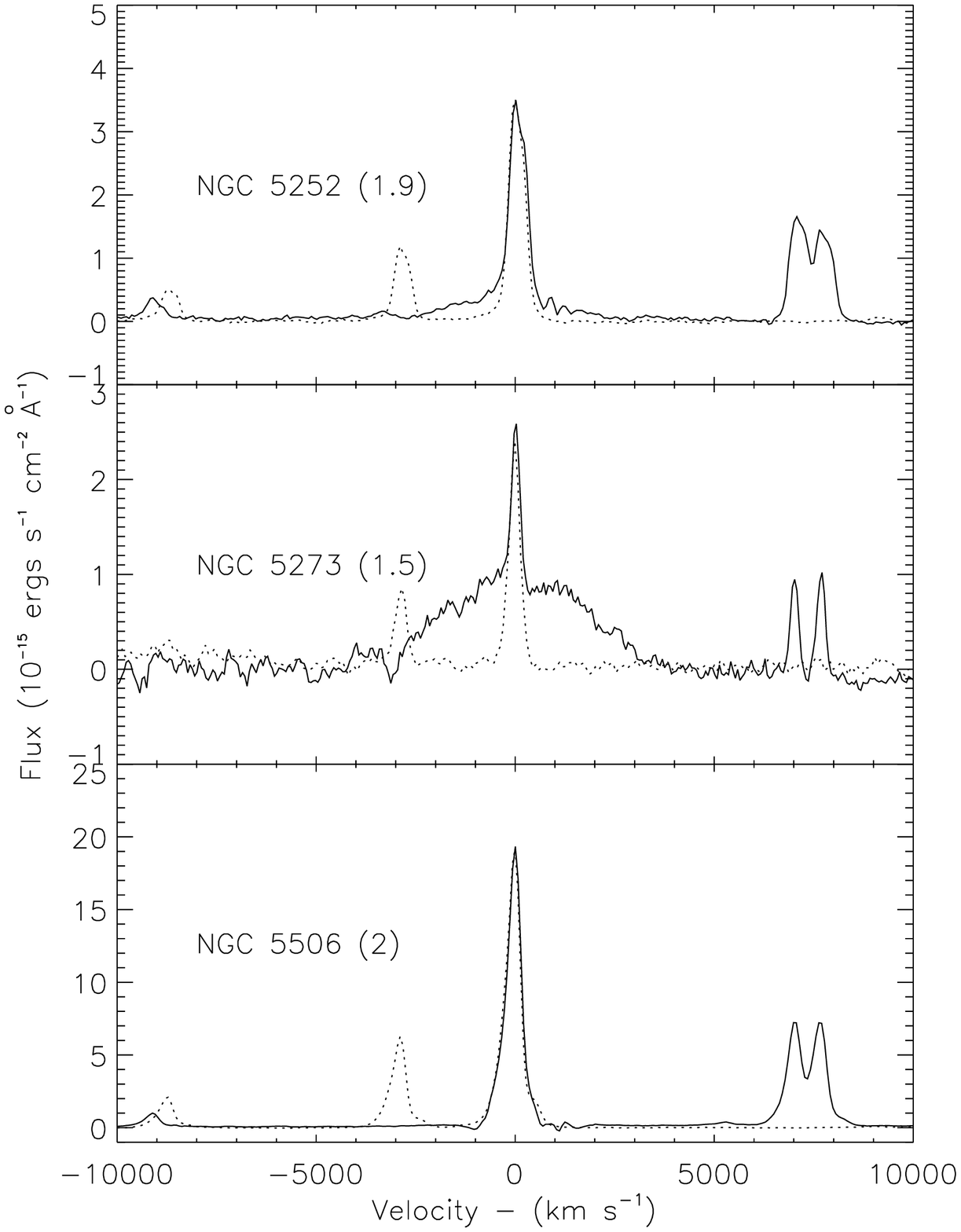}
\end{center}
\end{figure}
\begin{figure}[h!]
\begin{center}
\includegraphics[angle=0,scale=1.0]{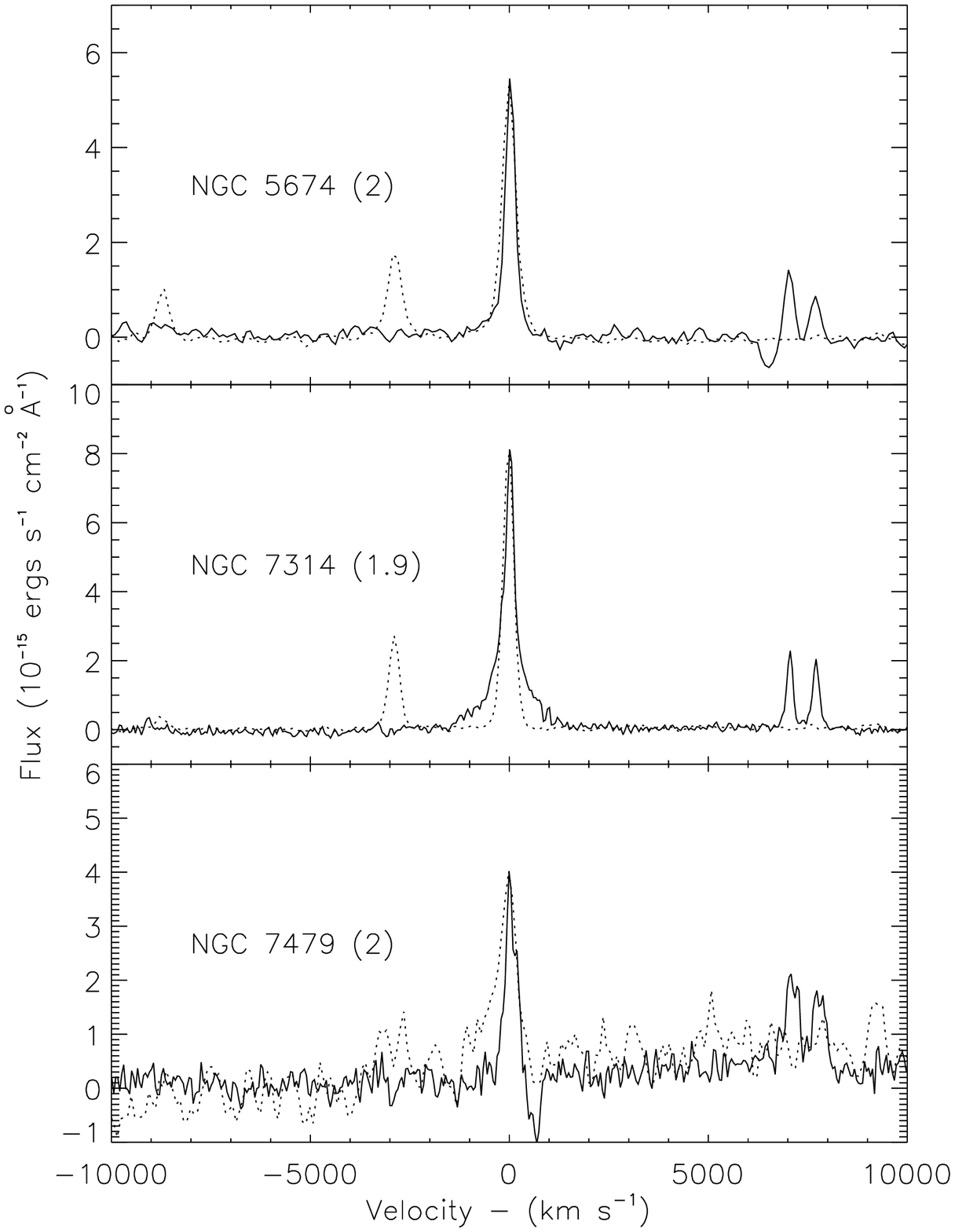}
\end{center}
\end{figure}
\begin{figure}[h!]
\begin{center}
\includegraphics[angle=0,scale=1.0]{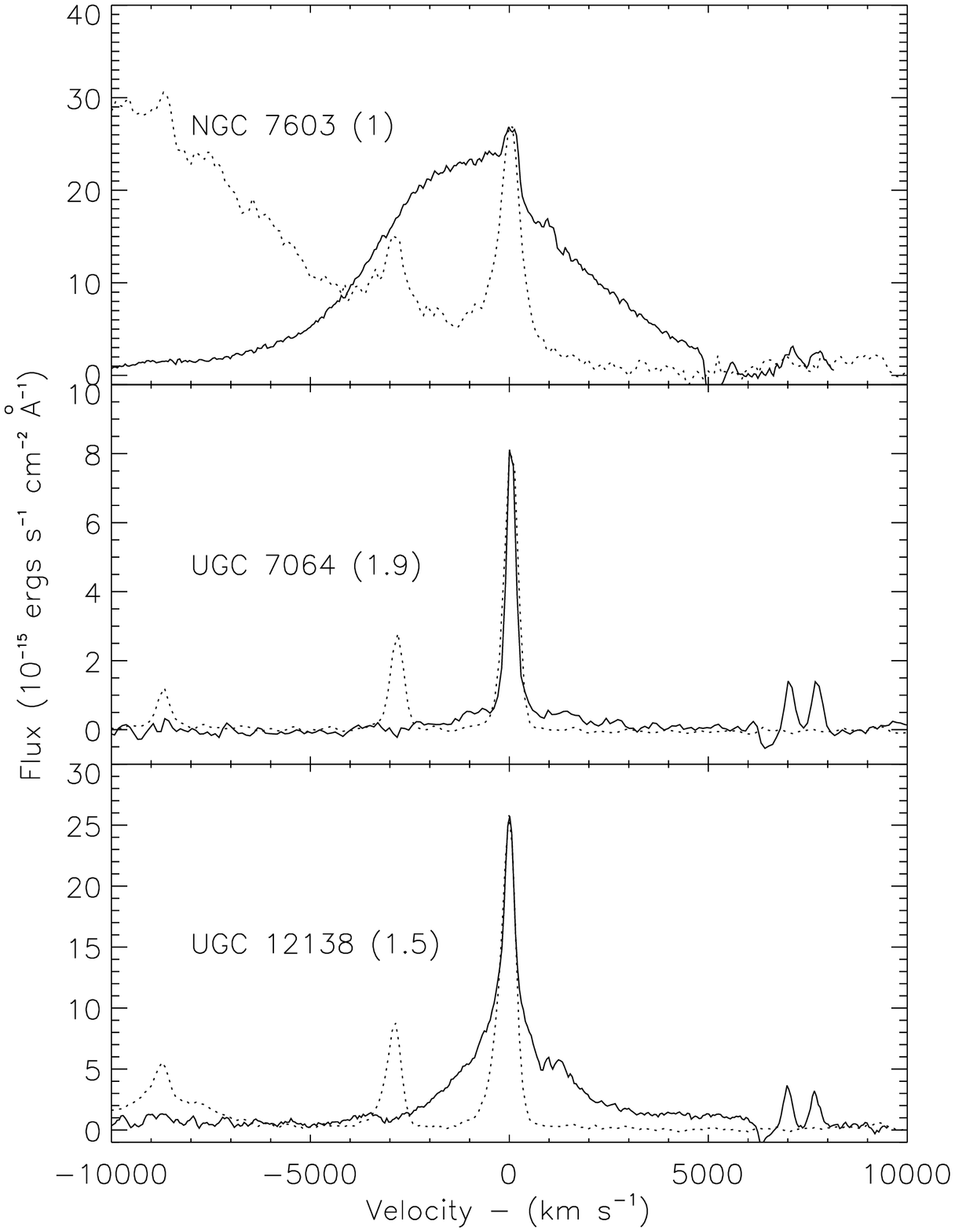}
\end{center}
\end{figure}
\begin{figure}[h!]
\begin{center}
\includegraphics[angle=0,scale=1.0]{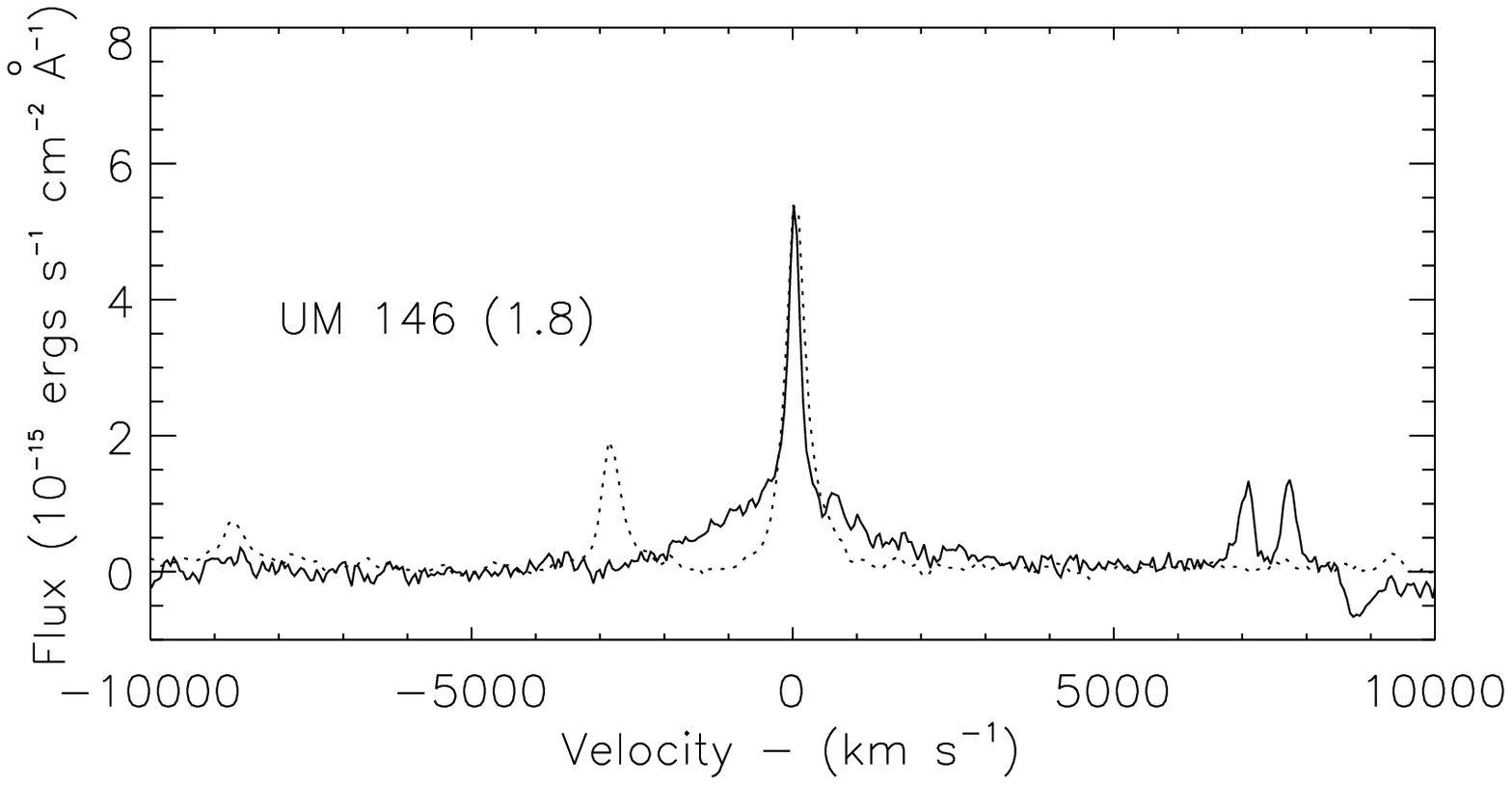}
\end{center}
\end{figure}
\setcounter{figure}{0}
\clearpage
\begin{center}
Appendix 3: {\it XMM} X-ray Spectra \\
\end{center}
\indent This Appendix presents the {\it XMM} spectra. \\
\rotate
\begin{figure}[!h]
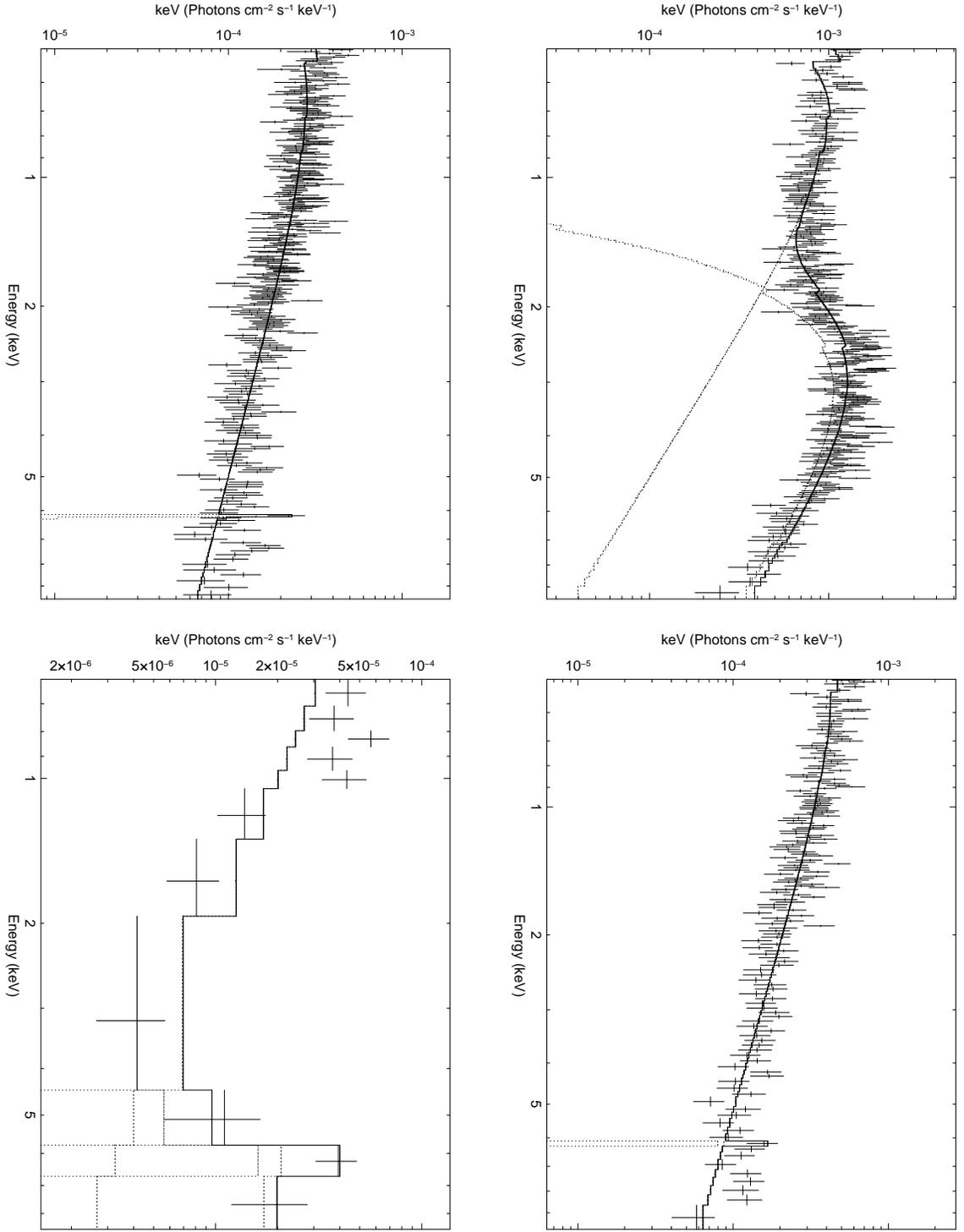

\vspace{-1.0in}
\hspace{-2.0in}
  \centering
  $\begin{array}{l@{\hspace{0.5cm}}r}
    \multicolumn{1}{l}{\mbox{}} &
    \multicolumn{1}{l}{\mbox{}} \\[10pt]
    \hspace{0.5cm}
    \includegraphics[angle=270, trim=0 0 0
    0,totalheight=7cm]
    {iras18325_f.ps} &
    \includegraphics[angle=270, trim=0 0 0
    0, totalheight=7cm]
    {mrk6091_f.ps} \label{mrk609x1}\\[-10pt]
    \multicolumn{1}{l}{\mbox{}} &
    \multicolumn{1}{l}{\mbox{}} \\[10pt]
    \hspace{0.5cm}
    \includegraphics[angle=270, trim=0 0 0
    0,totalheight=7cm]
    {mrk6092_f.ps} \label{mrk609x2}&
    \includegraphics[angle=270, trim=0 0 0
    0,totalheight=7cm]
    {mrk622_f.ps}
  \end{array}$
  \caption{{\it XMM} pn spectra of Seyferts (clockwise from top left): IRAS 18325-5926, Mrk 609 (1), Mrk 609 (2), and Mrk 622}
  \label{xmmfig1}
\end{figure}
\clearpage
\rotate
\begin{figure}[!h]
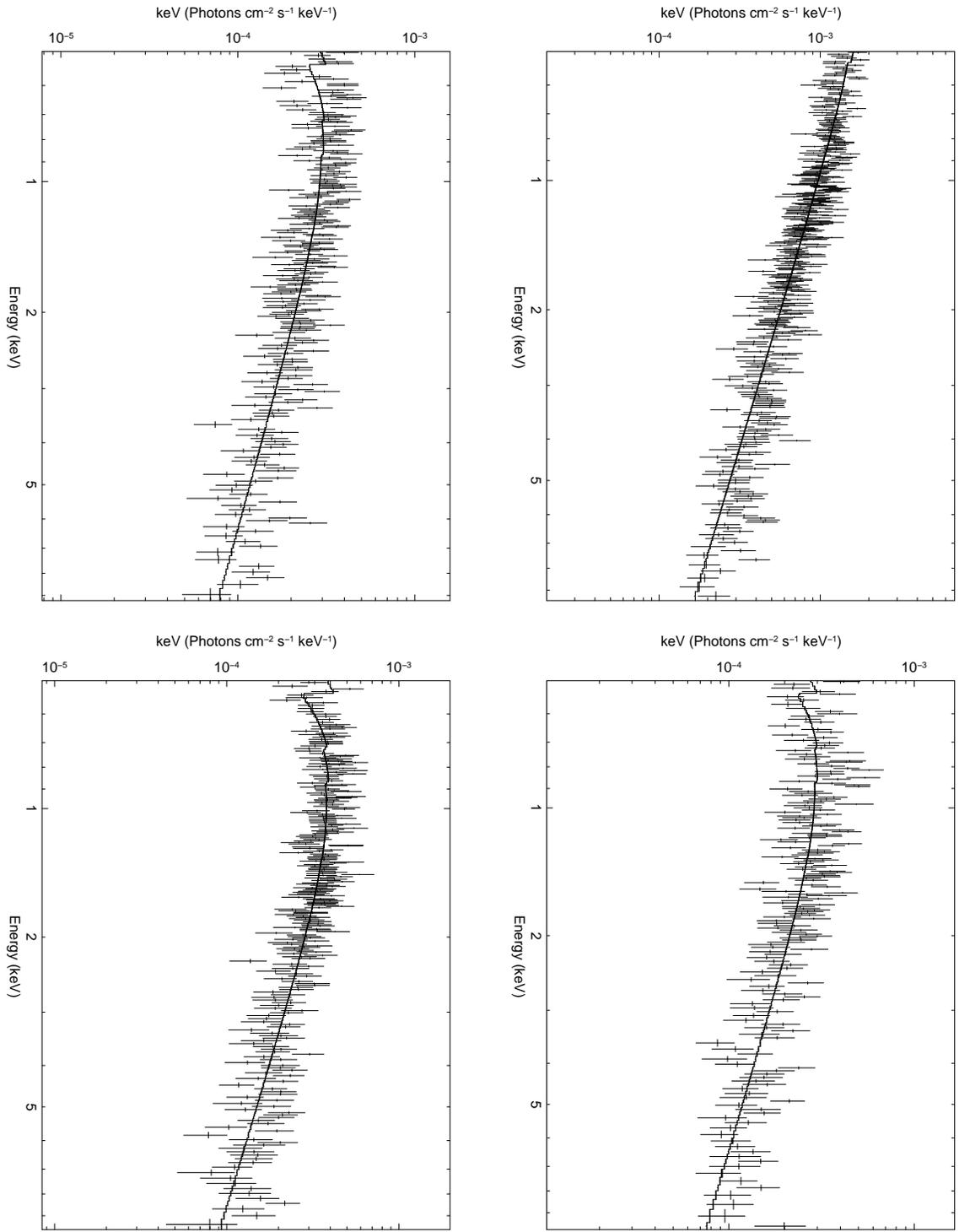

\vspace{-0.5in}
\hspace{-3.0in}
  \centering
  $\begin{array}{l@{\hspace{0.5cm}}r}
    \multicolumn{1}{l}{\mbox{}} &
    \multicolumn{1}{l}{\mbox{}} \\[10pt]
    \hspace{0.5cm}
    \includegraphics[angle=270, trim=0 0 0
    0,totalheight=7cm]
    {mrk728_f.ps} &
    \includegraphics[angle=270, trim=0 0 0
    0, totalheight=7cm]
    {mrk8831_f.ps} \label{mrk883x1} \\[-10pt]
    \multicolumn{1}{l}{\mbox{}} &
    \multicolumn{1}{l}{\mbox{}} \\[10pt]
    \hspace{0.5cm}
    \includegraphics[angle=270, trim=0 0 0
    0,totalheight=7cm]
    {mrk8832_f.ps} \label{mrk883x2} &
    \includegraphics[angle=270, trim=0 0 0
    0,totalheight=7cm]
    {mrk8833_f.ps} \label{mrk883x3}
  \end{array}$
  \caption{{\it XMM} pn spectra of Seyferts (clockwise from top left): Mrk 728, Mrk 883 (1), Mrk 883 (2), and Mrk 883 (3)}
  \label{xmmfig2}
\end{figure}
\clearpage
\rotate
\begin{figure}[!h]
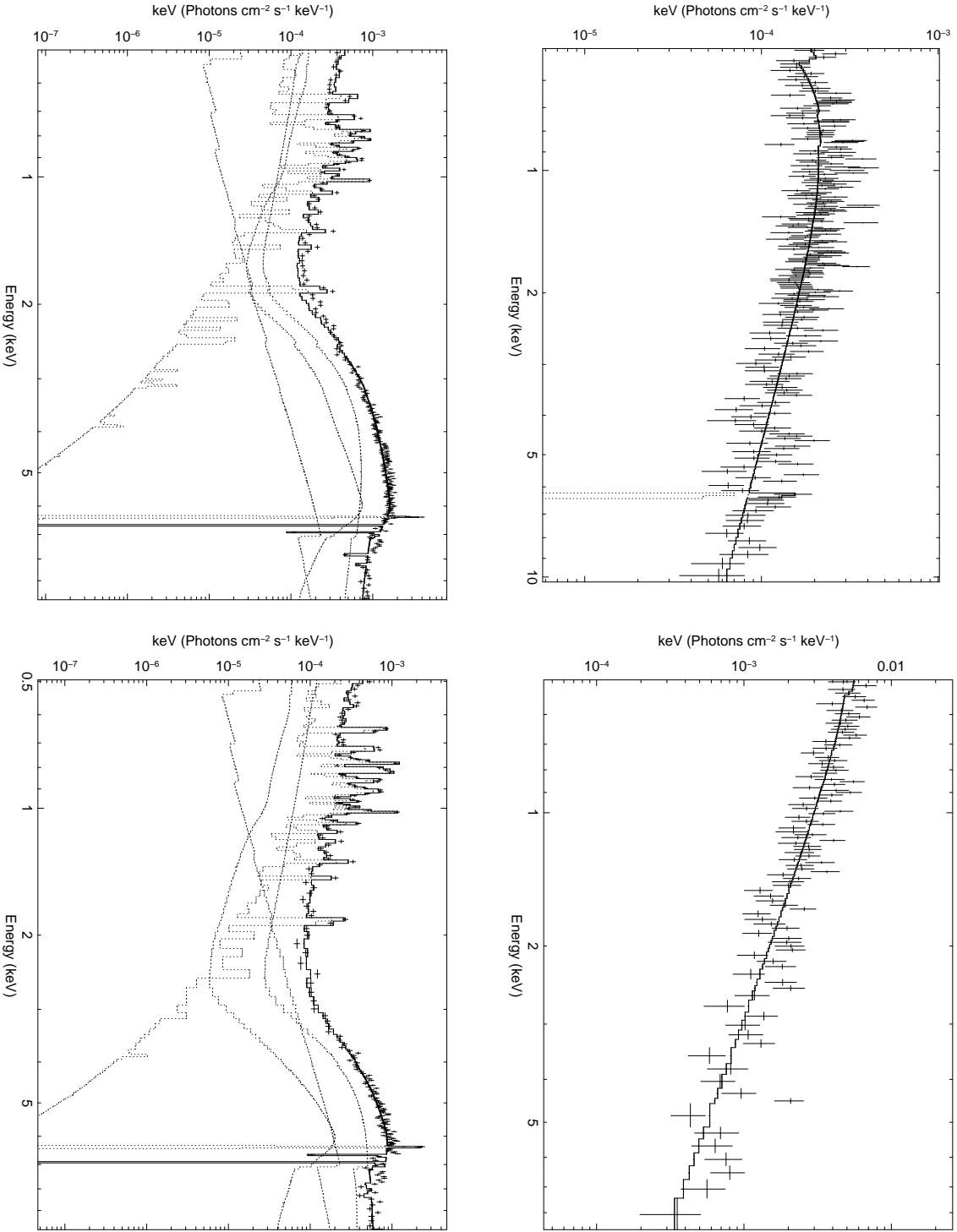

\vspace{-0.5in}
\hspace{-3.0in}
  \centering
  $\begin{array}{l@{\hspace{0.5cm}}r}
    \multicolumn{1}{l}{\mbox{}} &
    \multicolumn{1}{l}{\mbox{}} \\[10pt]
    \hspace{0.5cm}
    \includegraphics[angle=270, trim=0 0 0
    0,totalheight=7cm]
    {mrk993_f.ps} &
    \includegraphics[angle=270, trim=0 0 0
    0, totalheight=7cm]
    {mrk1018_f.ps}\\[-10pt]
    \multicolumn{1}{l}{\mbox{}} &
    \multicolumn{1}{l}{\mbox{}} \\[10pt]
    \hspace{0.5cm}
    \includegraphics[angle=270, trim=0 0 0
    0,totalheight=7cm]
    {ngc13651_f.ps} &
    \includegraphics[angle=270, trim=0 0 0
    0,totalheight=7cm]
    {ngc13652_f.ps}
  \end{array}$
  \caption{{\it XMM} pn spectra of Seyferts (clockwise from top left): Mrk 993, Mrk 1018, NGC 1365 (1), and NGC 1365 (2)}
  \label{xmmfig3}
\end{figure}
\clearpage
\rotate
\begin{figure}[!h]
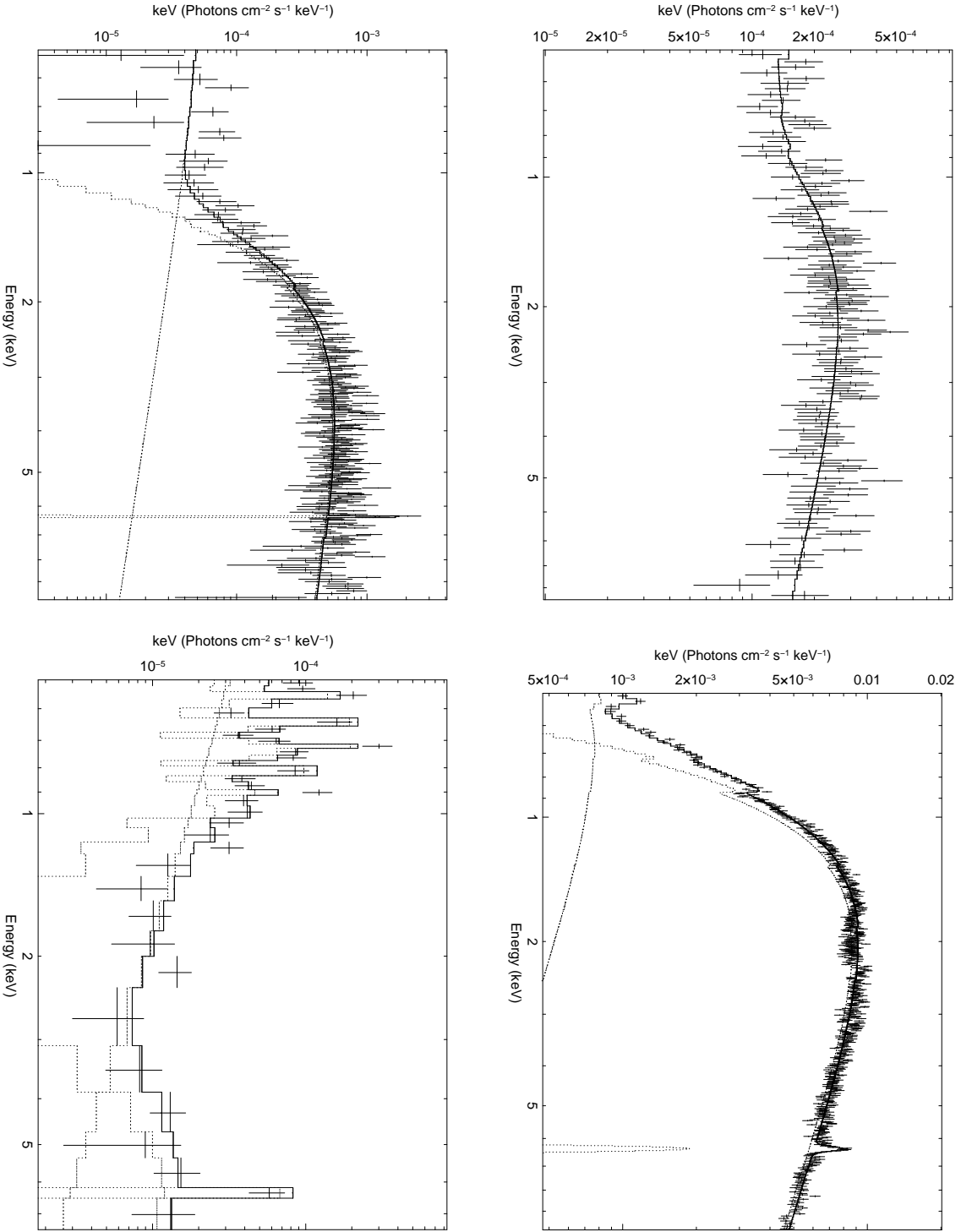

\vspace{-0.5in}
\hspace{-3.0in}
  \centering
  $\begin{array}{l@{\hspace{0.5cm}}r}
    \multicolumn{1}{l}{\mbox{}} &
    \multicolumn{1}{l}{\mbox{}} \\[10pt]
    \hspace{0.5cm}
    \includegraphics[angle=270, trim=0 0 0
    0,totalheight=7cm]
    {ngc2622_f.ps} &
    \includegraphics[angle=270, trim=0 0 0
    0, totalheight=7cm]
    {ngc2992_f.ps}\\[-10pt]
    \multicolumn{1}{l}{\mbox{}} &
    \multicolumn{1}{l}{\mbox{}} \\[10pt]
    \hspace{0.5cm}
    \includegraphics[angle=270, trim=0 0 0
    0,totalheight=7cm]
    {ngc3786_f.ps} &
    \includegraphics[angle=270, trim=0 0 0
    0,totalheight=7cm]
    {ngc3982_f.ps}
  \end{array}$
  \caption{{\it XMM} pn spectra of Seyferts (clockwise from top left): NGC 2622, NGC 2992, NGC 3786, and NGC 3982}
  \label{xmmfig4}
\end{figure}
\clearpage
\rotate
\begin{figure}[!h]
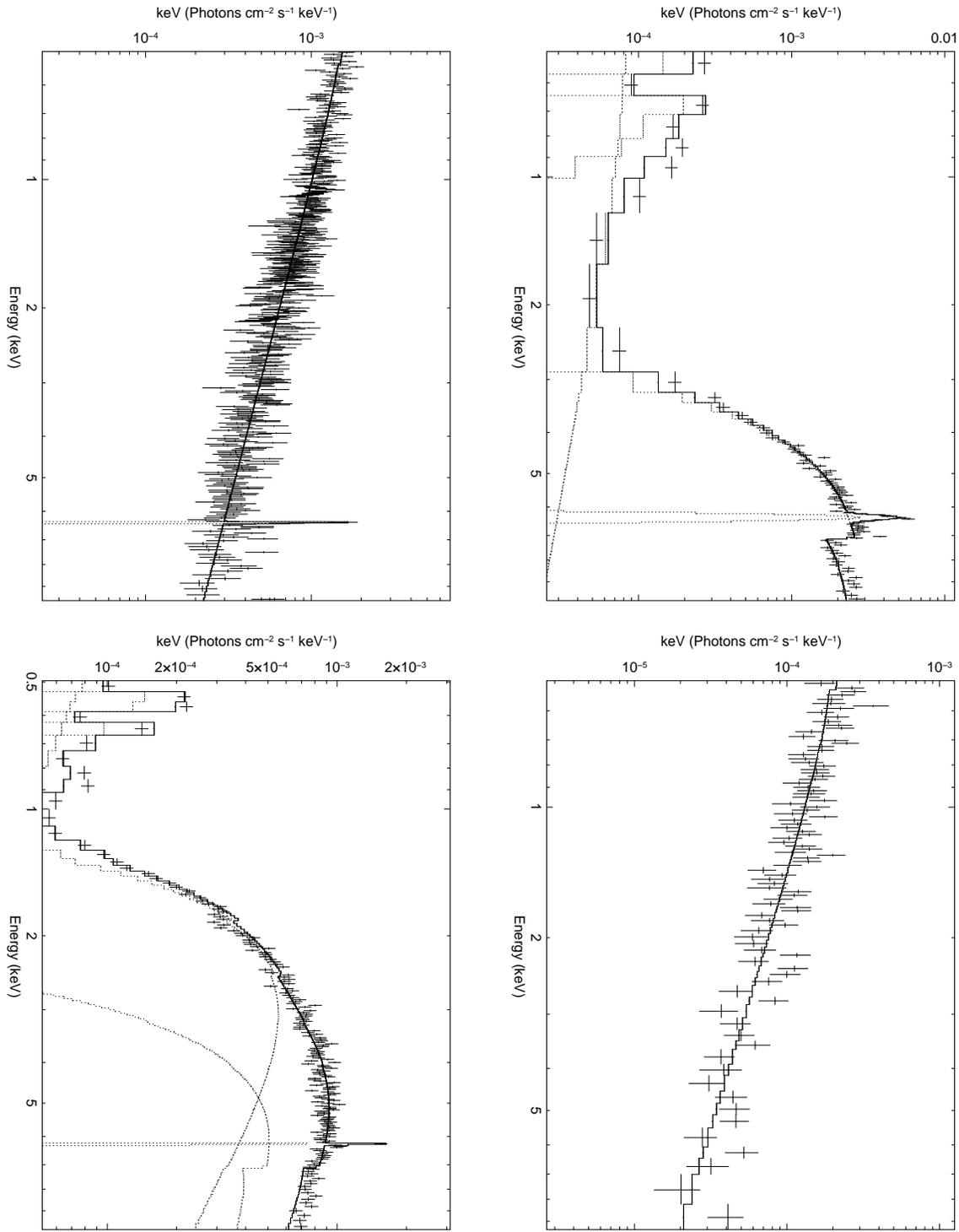

\vspace{-0.5in}
\hspace{-3.0in}
  \centering
  $\begin{array}{l@{\hspace{0.5cm}}r}
    \multicolumn{1}{l}{\mbox{}} &
    \multicolumn{1}{l}{\mbox{}} \\[10pt]
    \hspace{0.5cm}
    \includegraphics[angle=270, trim=0 0 0
    0,totalheight=7cm]
    {ngc4388_f.ps} &
    \includegraphics[angle=270, trim=0 0 0
    0, totalheight=7cm]
    {ngc4639_f.ps}\\[-10pt]
    \multicolumn{1}{l}{\mbox{}} &
    \multicolumn{1}{l}{\mbox{}} \\[10pt]
    \hspace{0.5cm}
    \includegraphics[angle=270, trim=0 0 0
    0,totalheight=7cm]
    {ngc5033_f.ps} &
    \includegraphics[angle=270, trim=0 0 0
    0,totalheight=7cm]
    {ngc5252_f.ps}
  \end{array}$
  \caption{{\it XMM} pn spectra of Seyferts (clockwise from top left): NGC 4388, NGC 4639, NGC 5033, and NGC 5252}
  \label{xmmfig5}
\end{figure}
\clearpage
\rotate
\begin{figure}[!h]
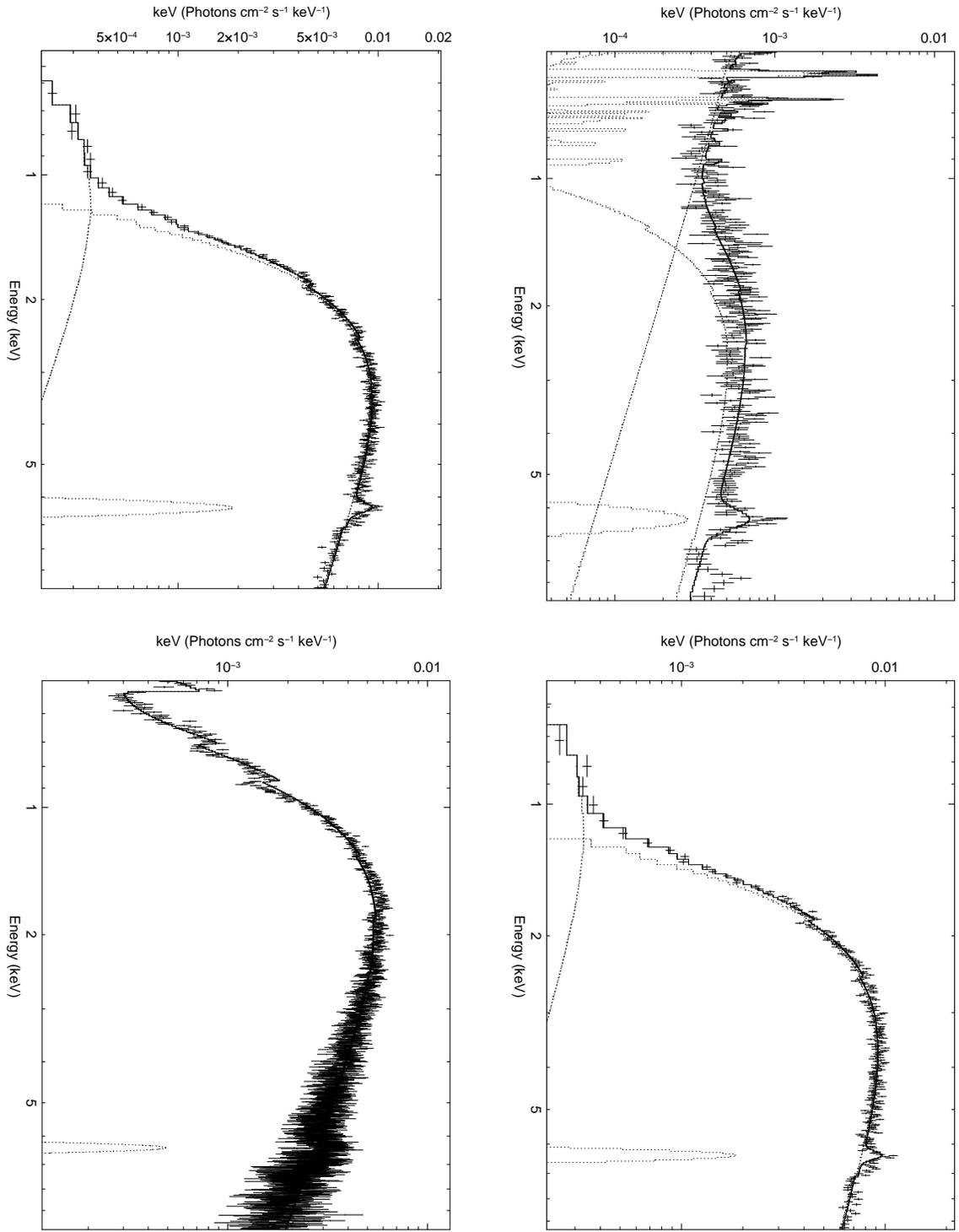

\vspace{-0.5in}
\hspace{-3.0in}
  \centering
  $\begin{array}{l@{\hspace{0.5cm}}r}
    \multicolumn{1}{l}{\mbox{}} &
    \multicolumn{1}{l}{\mbox{}} \\[10pt]
    \hspace{0.5cm}
    \includegraphics[angle=270, trim=0 0 0
    0,totalheight=7cm]
    {ngc5273_f.ps} &
    \includegraphics[angle=270, trim=0 0 0
    0, totalheight=7cm]
    {ngc55062_f.ps}\\[-10pt]
    \multicolumn{1}{l}{\mbox{}} &
    \multicolumn{1}{l}{\mbox{}} \\[10pt]
    \hspace{0.5cm}
    \includegraphics[angle=270, trim=0 0 0
    0,totalheight=7cm]
    {ngc55063_f.ps} &
    \includegraphics[angle=270, trim=0 0 0
    0,totalheight=7cm]
    {ngc7314_f.ps}
  \end{array}$
  \caption{{\it XMM} pn spectra of Seyferts (clockwise from top left): NGC 5273, NGC 5506 (1), NGC 5506 (2), and NGC 7314}
  \label{xmmfig6}
\end{figure}
\clearpage
\rotate
\begin{figure}[!h]
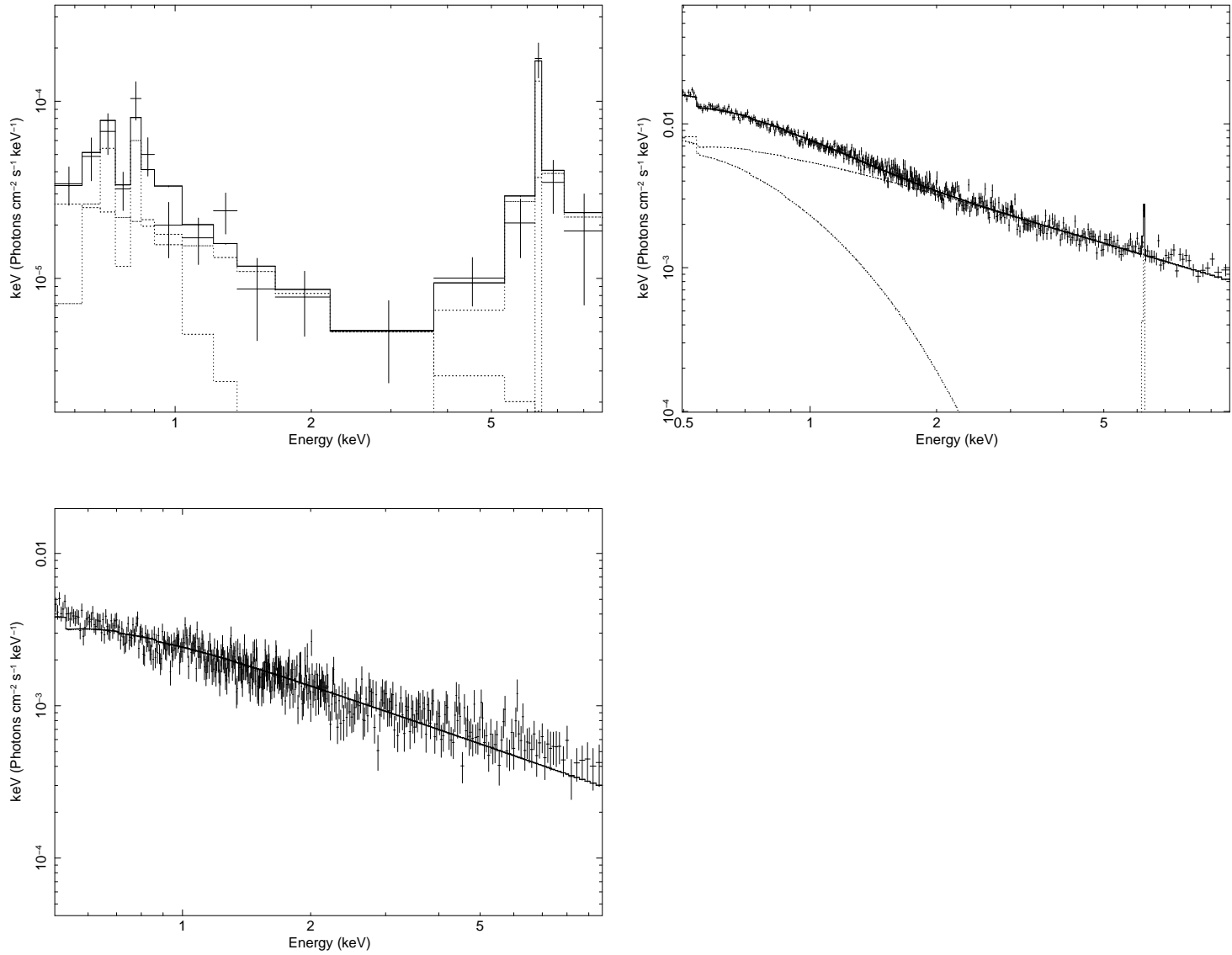

\vspace{-0.5in}
\hspace{-3.0in}
  \centering
  $\begin{array}{l@{\hspace{0.5cm}}r}
    \multicolumn{1}{l}{\mbox{}} &
    \multicolumn{1}{l}{\mbox{}} \\[10pt]
    \hspace{0.5cm}
    \includegraphics[angle=270, trim=0 0 0
    0,totalheight=7cm]
    {ngc7479_f.ps} &
    \includegraphics[angle=270, trim=0 0 0
    0, totalheight=7cm]
    {ngc7603_f.ps}\\[-10pt]
    \multicolumn{1}{l}{\mbox{}} &
    \multicolumn{1}{l}{\mbox{}} \\[10pt]
    \hspace{0.5cm}
    \includegraphics[angle=270, trim=0 0 0
    0,totalheight=7cm]
    {ugc12138_f.ps} &
  \end{array}$
  \caption{{\it XMM} pn spectra of Seyferts (clockwise from top left): NGC 7479, NGC 7603, and UGC 12138}
  \label{xmmfig7}
\end{figure}

\setcounter{figure}{0}
\clearpage
\begin{center}
Appendix 4: {\it Spitzer} IRS Spectra\\
\end{center}
\indent This Appendix presents the {\it Spitzer} IRS spectra discussed in Section~\ref{spitzer_obs}. \\
\rotate
\begin{figure}[!h]
\vspace{-1.0in}
  \centering
  $\begin{array}{l@{\hspace{0.5cm}}r}
    \multicolumn{1}{l}{\mbox{}} &
    \multicolumn{1}{l}{\mbox{}} \\[-10pt]
    \hspace{-0.6cm}
    \includegraphics[angle=90, trim=5 5 5
    5,clip,width=0.5\textwidth,height=0.4\textheight]
    {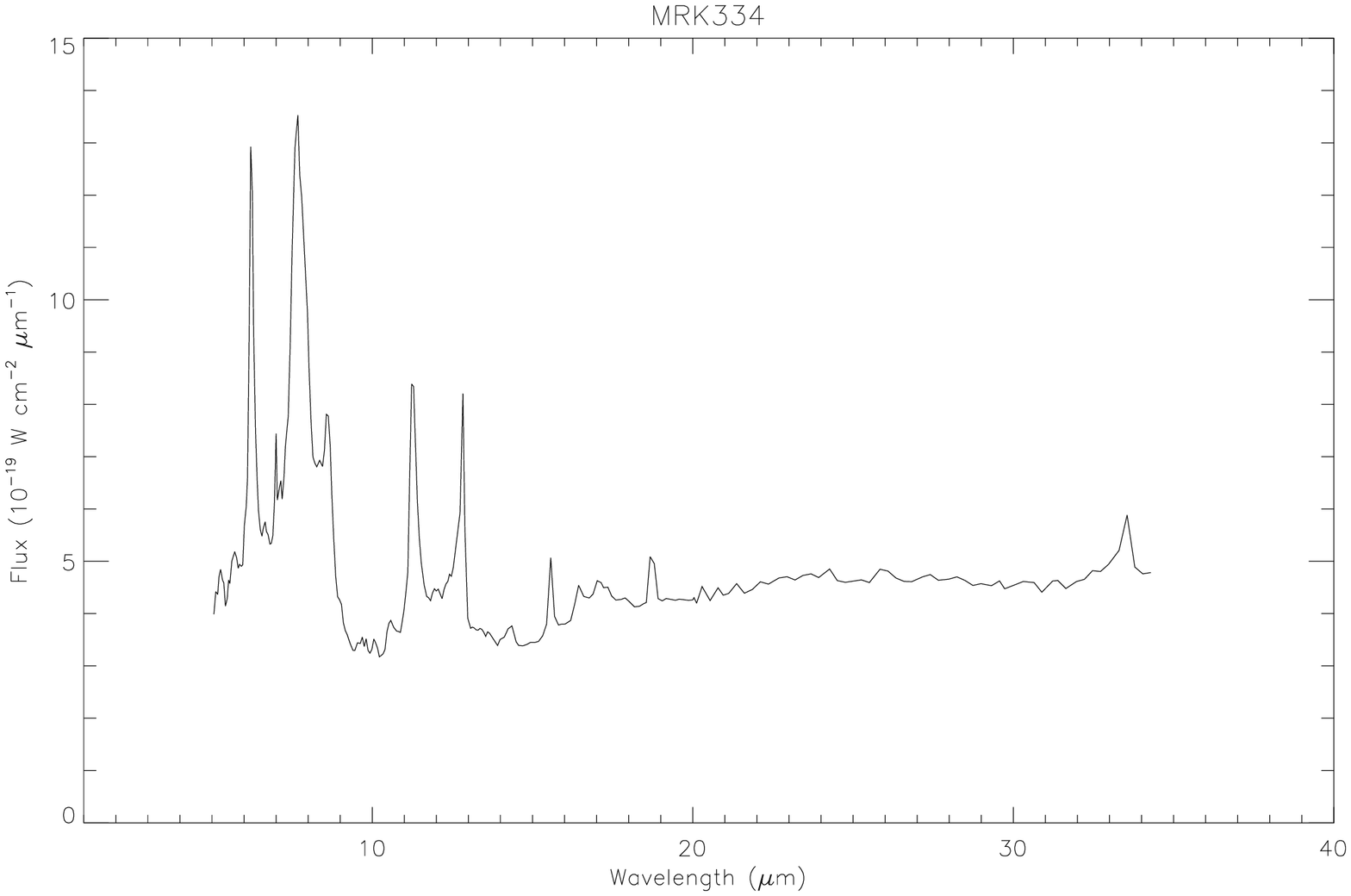} &
    \includegraphics[angle=90, trim=5 5 5
    5,clip,width=0.5\textwidth,height=0.4\textheight]
    {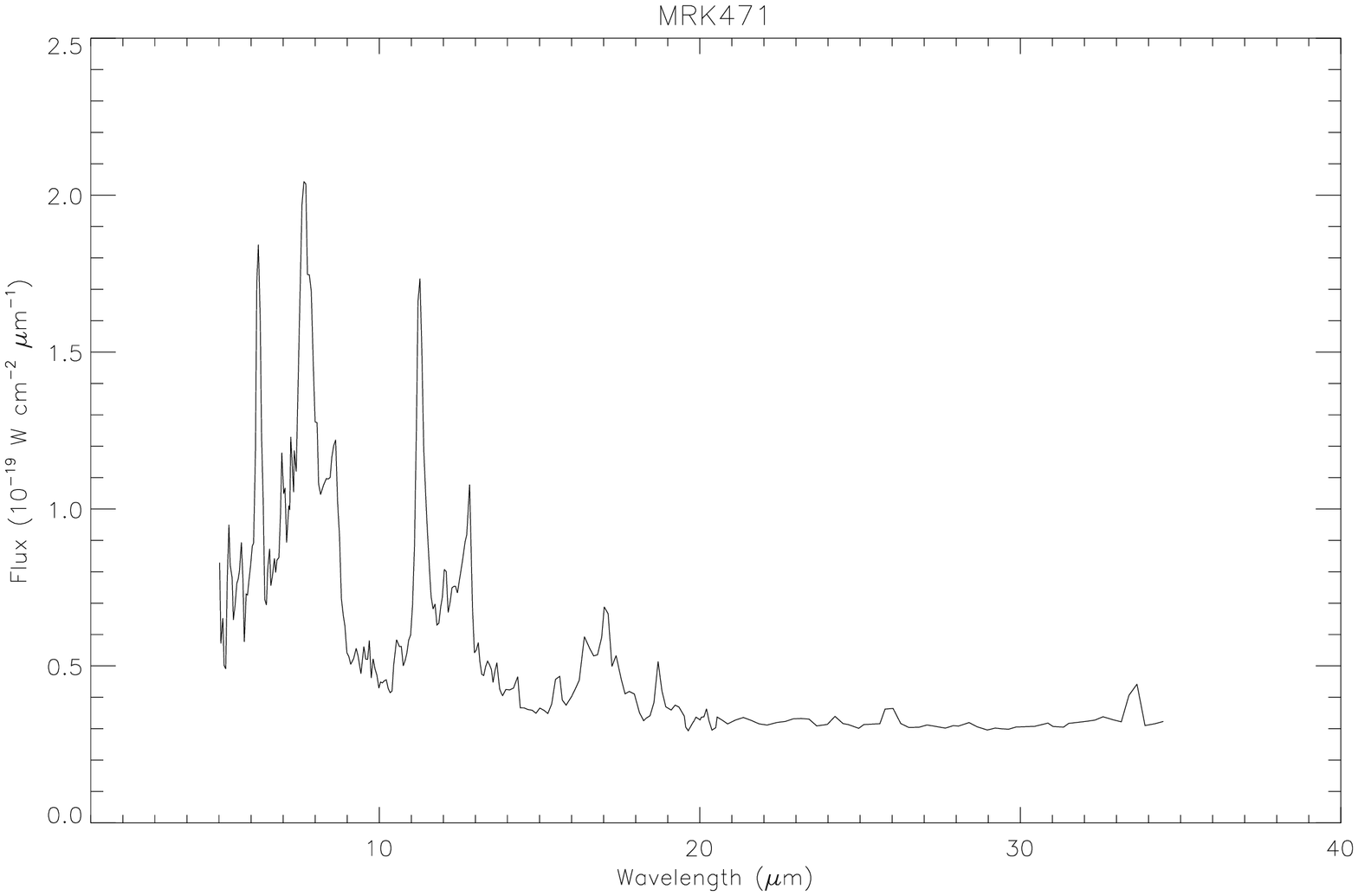}\\[-10pt]
    \multicolumn{1}{l}{\mbox{}} &
    \multicolumn{1}{l}{\mbox{}} \\[-10pt]
    \hspace{-0.6cm}
    \includegraphics[angle=90, trim=5 5 5
    5,clip,width=0.5\textwidth,height=0.4\textheight]
    {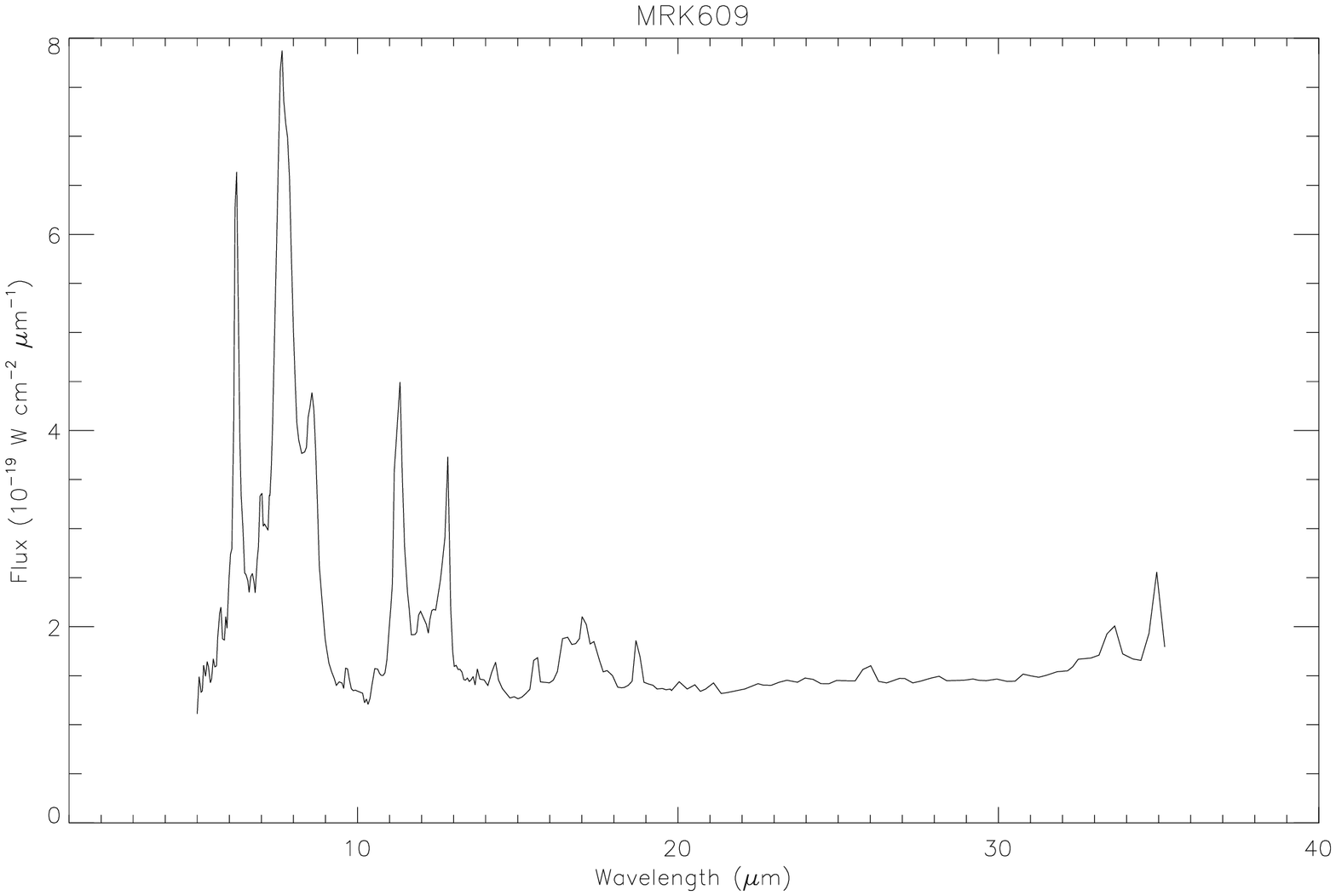} &
    \includegraphics[angle=90, trim=5 5 5
    5,clip,width=0.5\textwidth,height=0.4\textheight]
    {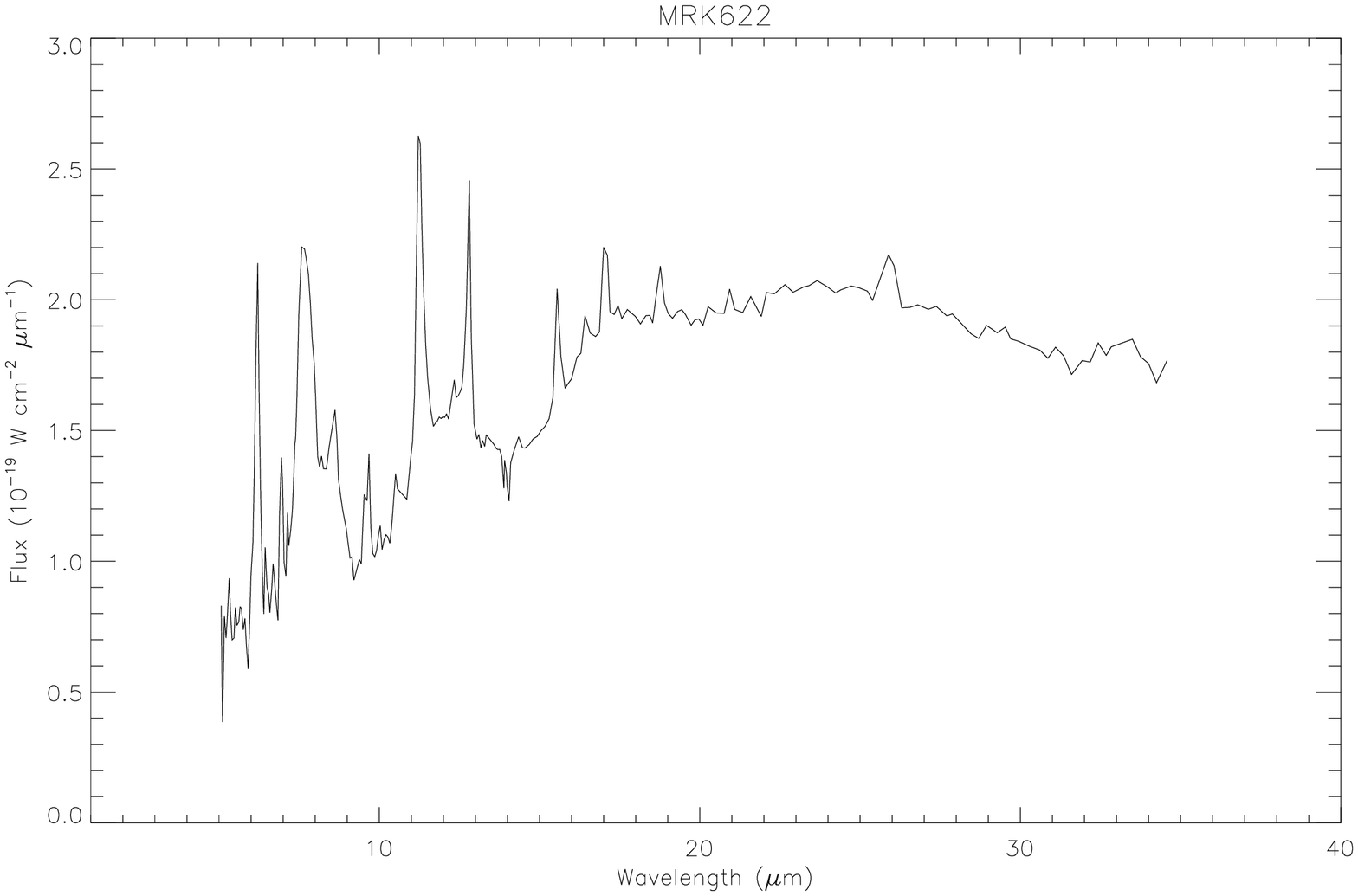}
  \end{array}$
  \caption{{\it Spitzer} IRS spectra of Seyferts (clockwise from top left): Mrk 334, Mrk 471, Mrk 609, and Mrk 622}
  \label{spitzerfig1}
\end{figure}
\clearpage
\rotate
\begin{figure}[!h]
\vspace{-1.0in}
  \centering
  $\begin{array}{l@{\hspace{0.5cm}}r}
    \multicolumn{1}{l}{\mbox{}} &
    \multicolumn{1}{l}{\mbox{}} \\[-10pt]
    \hspace{-0.6cm}
    \includegraphics[angle=90, trim=5 5 5
    5,clip,width=0.50\textwidth,height=0.4\textheight]
    {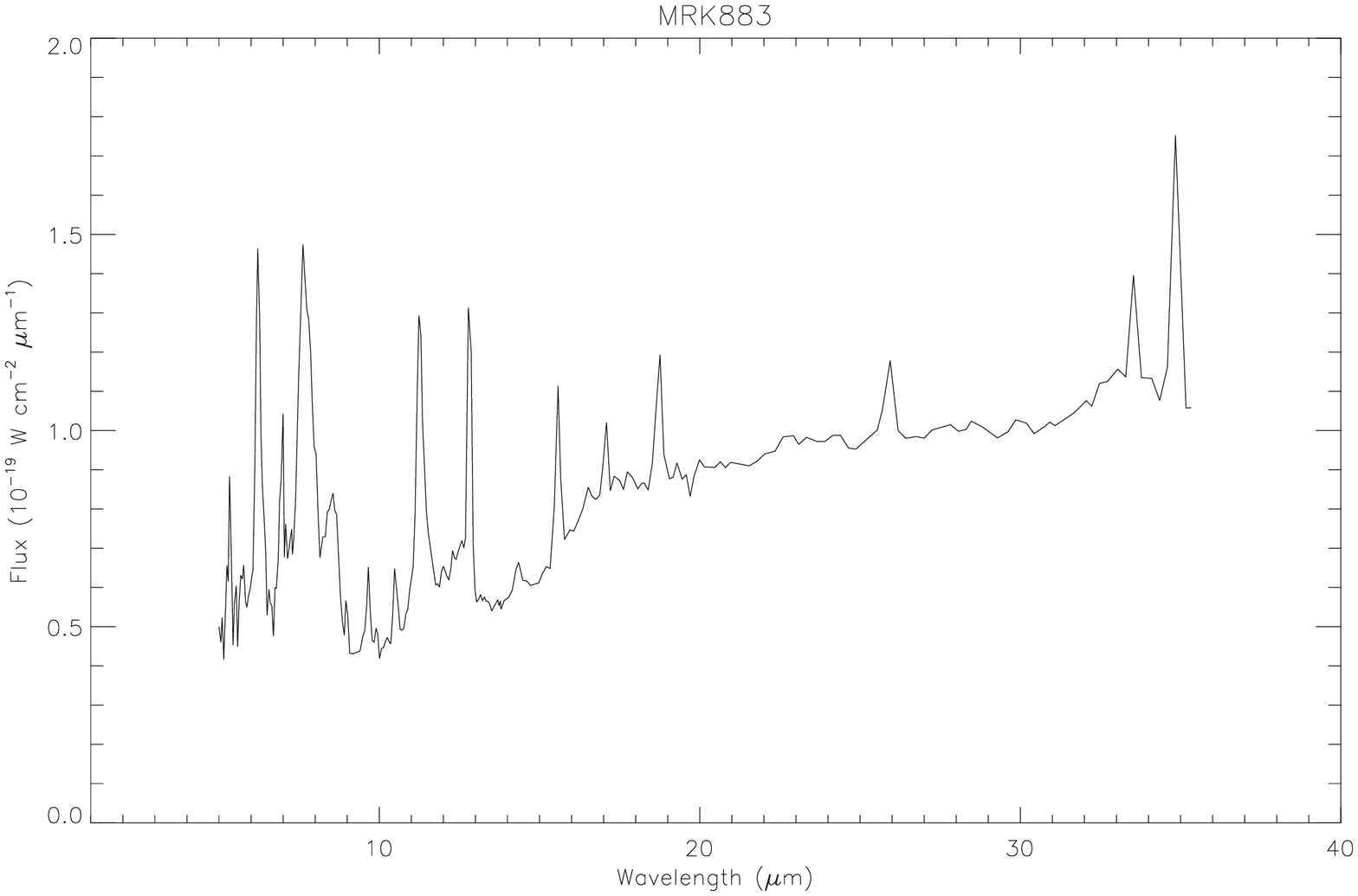} &
    \includegraphics[angle=90, trim=5 5 5
    5,clip,width=0.50\textwidth,height=0.4\textheight]
    {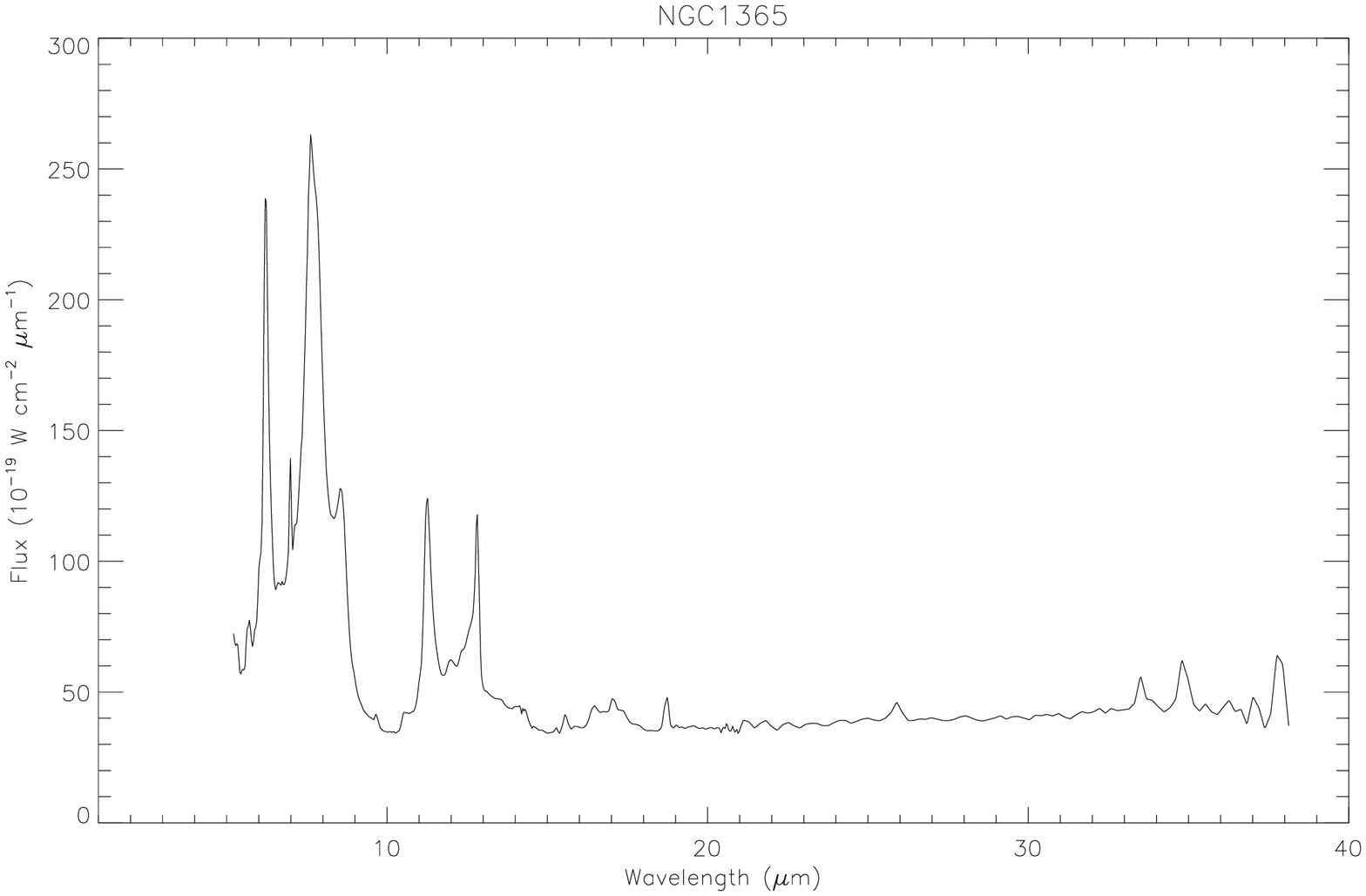}\\[-10pt]
    \multicolumn{1}{l}{\mbox{}} &
    \multicolumn{1}{l}{\mbox{}} \\[-10pt]
    \hspace{-0.6cm}
    \includegraphics[angle=90, trim=5 5 5
    5,clip,width=0.50\textwidth,height=0.4\textheight]
    {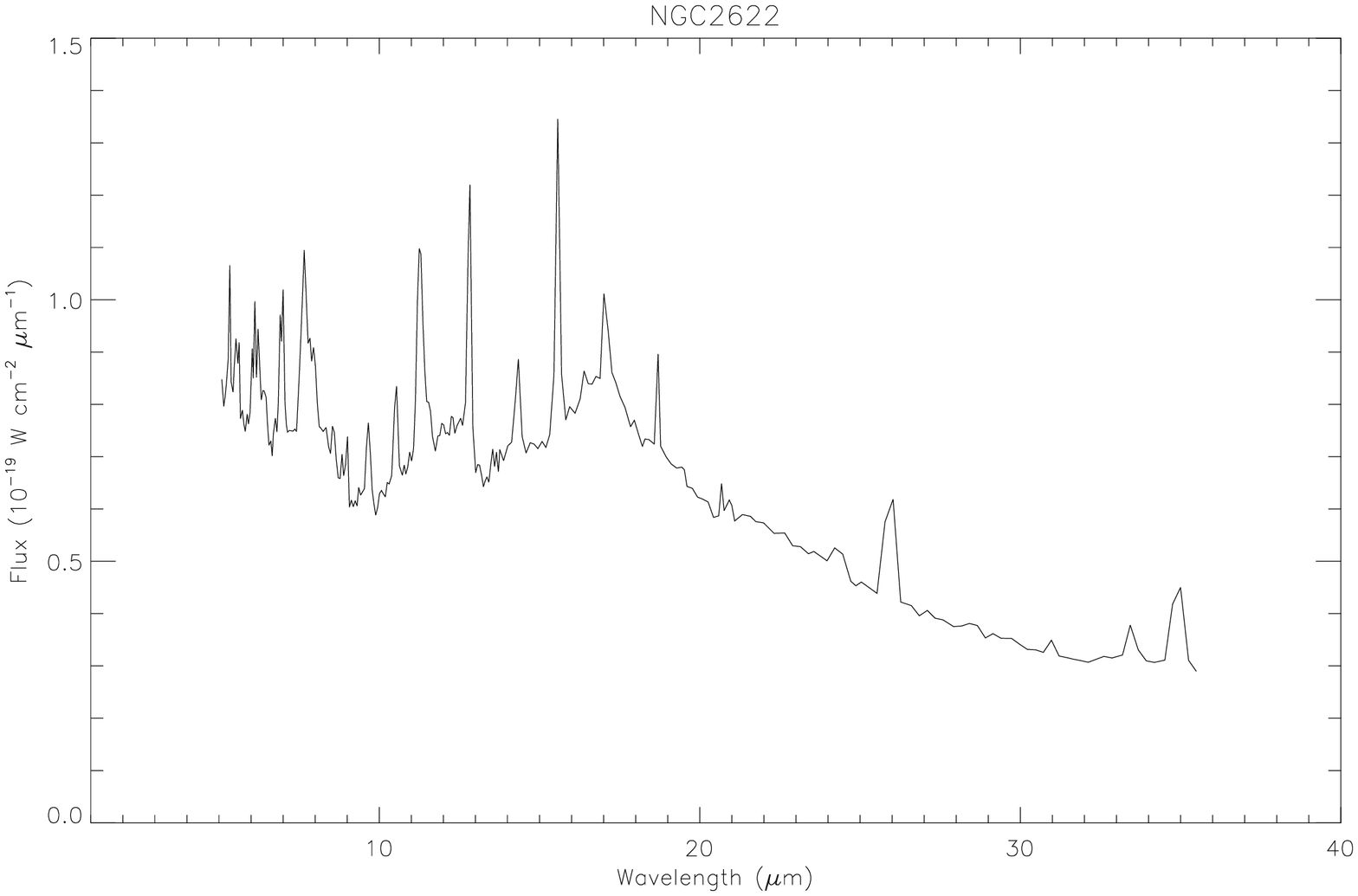} &
    \includegraphics[angle=90, trim=5 5 5
    5,clip,width=0.50\textwidth,height=0.4\textheight]
    {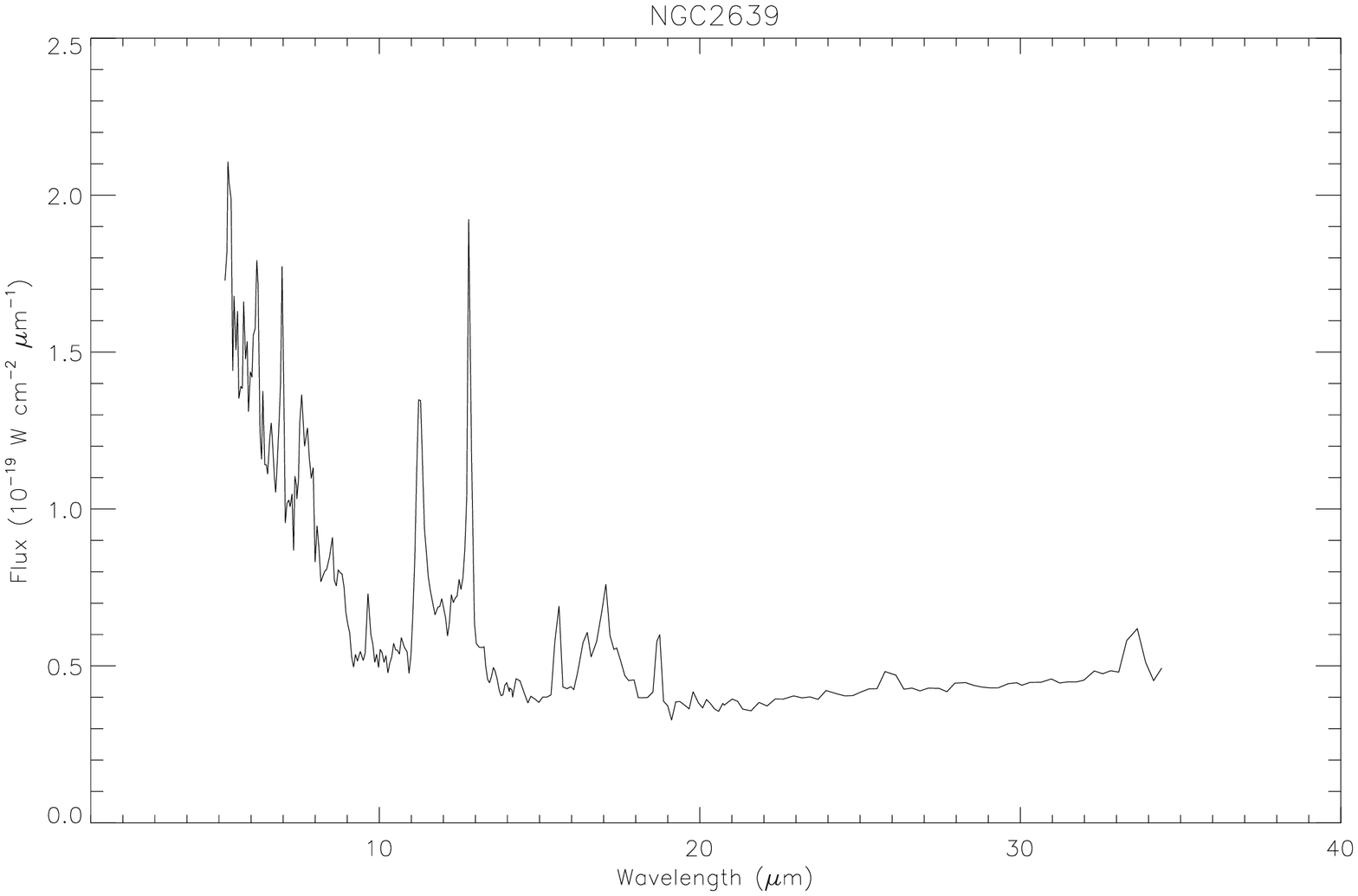}
  \end{array}$
  \caption{{\it Spitzer} IRS spectra of Seyferts (clockwise from top left): Mrk 883, NGC 1365, NGC 2622, and NGC 2639}
  \label{spitzerfig2}
\end{figure}
\clearpage
\rotate
\begin{figure}[!h]
\vspace{-1.0in}
  \centering
  $\begin{array}{l@{\hspace{0.5cm}}r}
    \multicolumn{1}{l}{\mbox{}} &
    \multicolumn{1}{l}{\mbox{}} \\[-10pt]
    \hspace{-0.6cm}
    \includegraphics[angle=90, trim=5 5 5
    5,clip,width=0.50\textwidth,height=0.4\textheight]
    {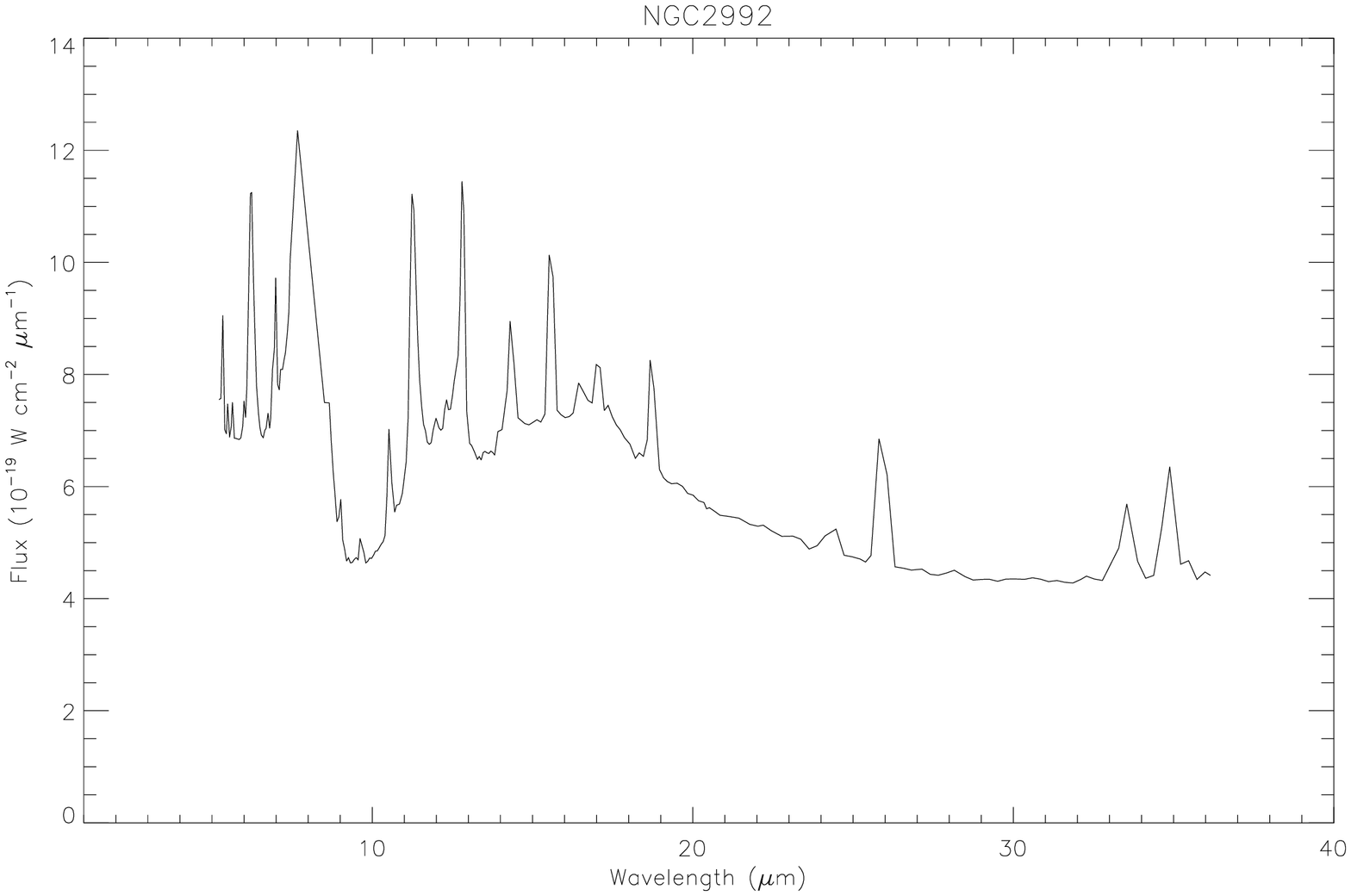} &
    \includegraphics[angle=90, trim=5 5 5
    5,clip,width=0.50\textwidth,height=0.4\textheight]
    {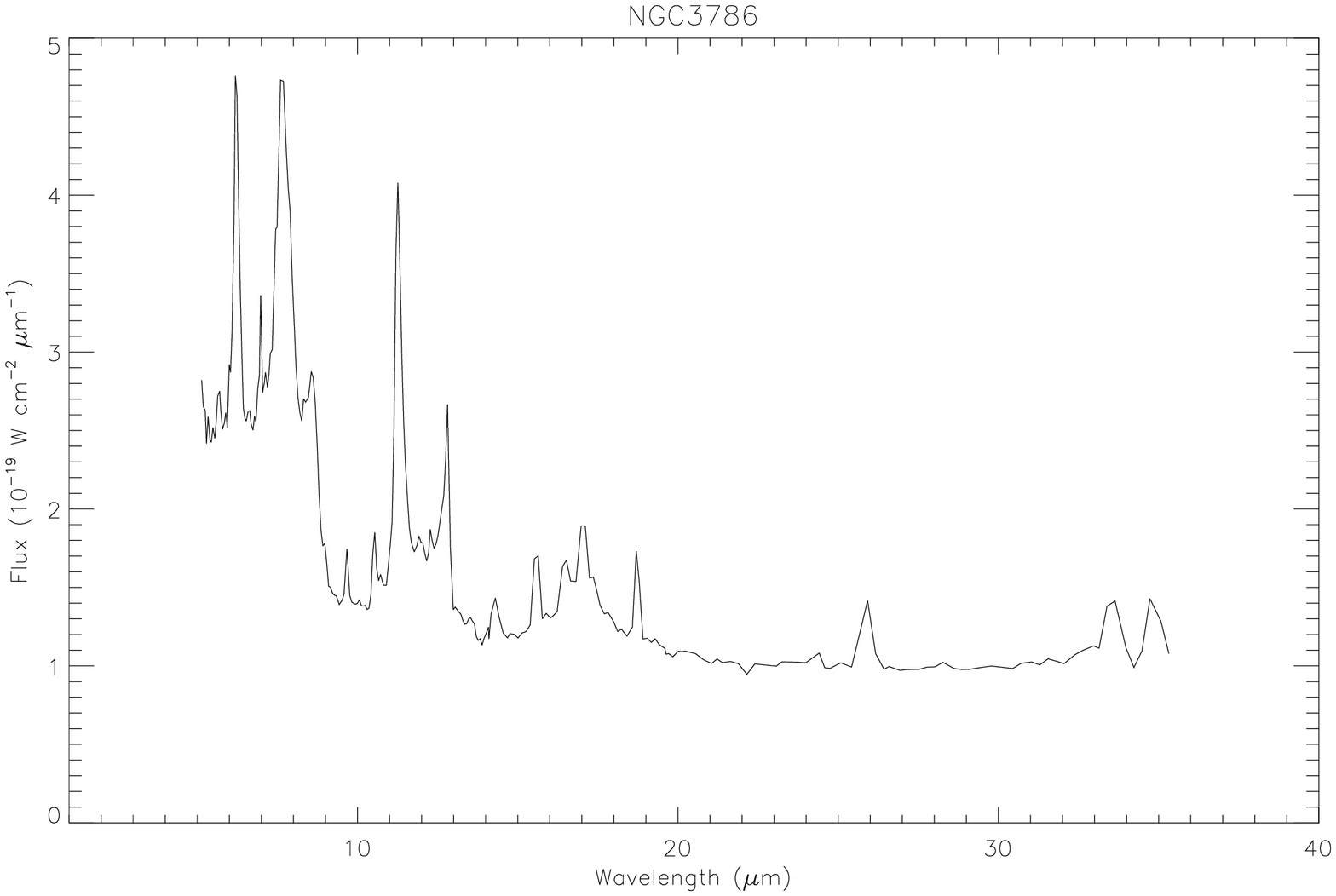}\\[-10pt]
    \multicolumn{1}{l}{\mbox{}} &
    \multicolumn{1}{l}{\mbox{}} \\[-10pt]
    \hspace{-0.6cm}
    \includegraphics[angle=90, trim=5 5 5
    5,clip,width=0.50\textwidth,height=0.4\textheight]
    {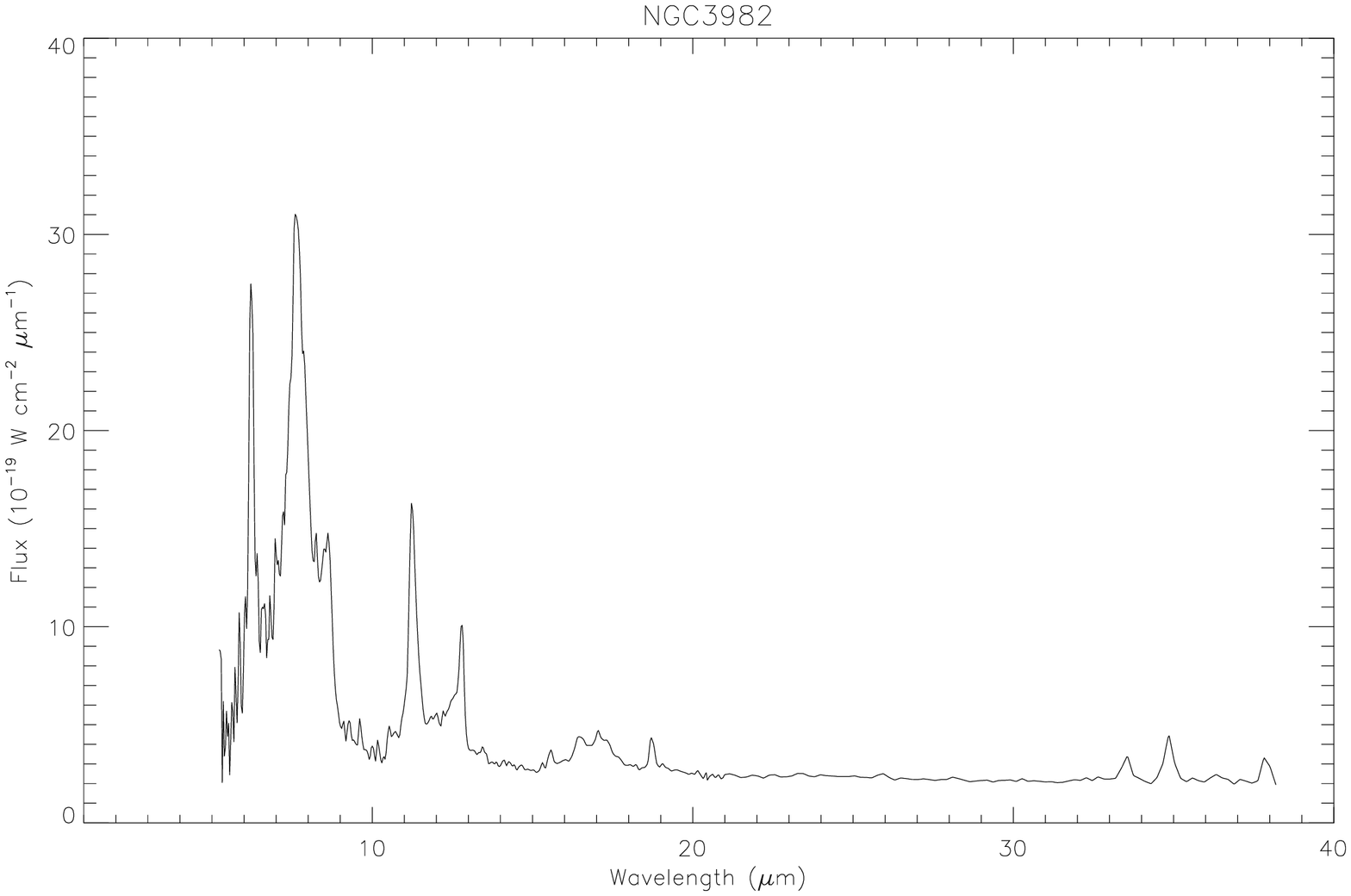} &
    \includegraphics[angle=90, trim=5 5 5
    5,clip,width=0.50\textwidth,height=0.4\textheight]
    {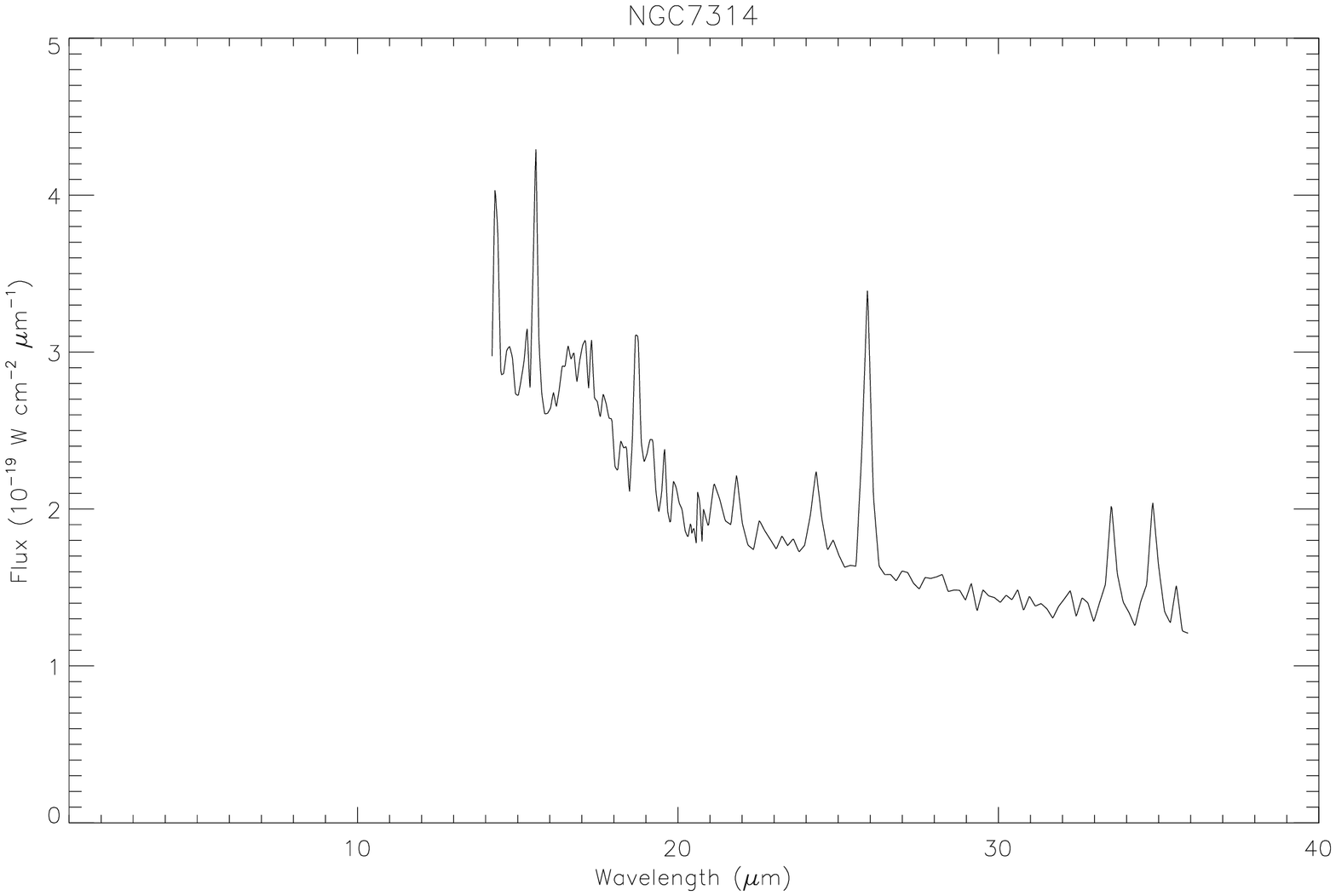}
  \end{array}$
  \caption{{\it Spitzer} IRS spectra of Seyferts (clockwise from top left): NGC 2992, NGC 3786, NGC 3982, and NGC 7314}
  \label{spitzerfig3}
\end{figure}
\clearpage
\rotate
\begin{figure}[!h]
\vspace{-1.0in}
  \centering
  $\begin{array}{l@{\hspace{0.5cm}}r}
    \multicolumn{1}{l}{\mbox{}} &
    \multicolumn{1}{l}{\mbox{}} \\[-10pt]
    \hspace{-0.6cm}
    \includegraphics[angle=90, trim=5 5 5
    5,clip,width=0.5\textwidth,height=0.4\textheight]
    {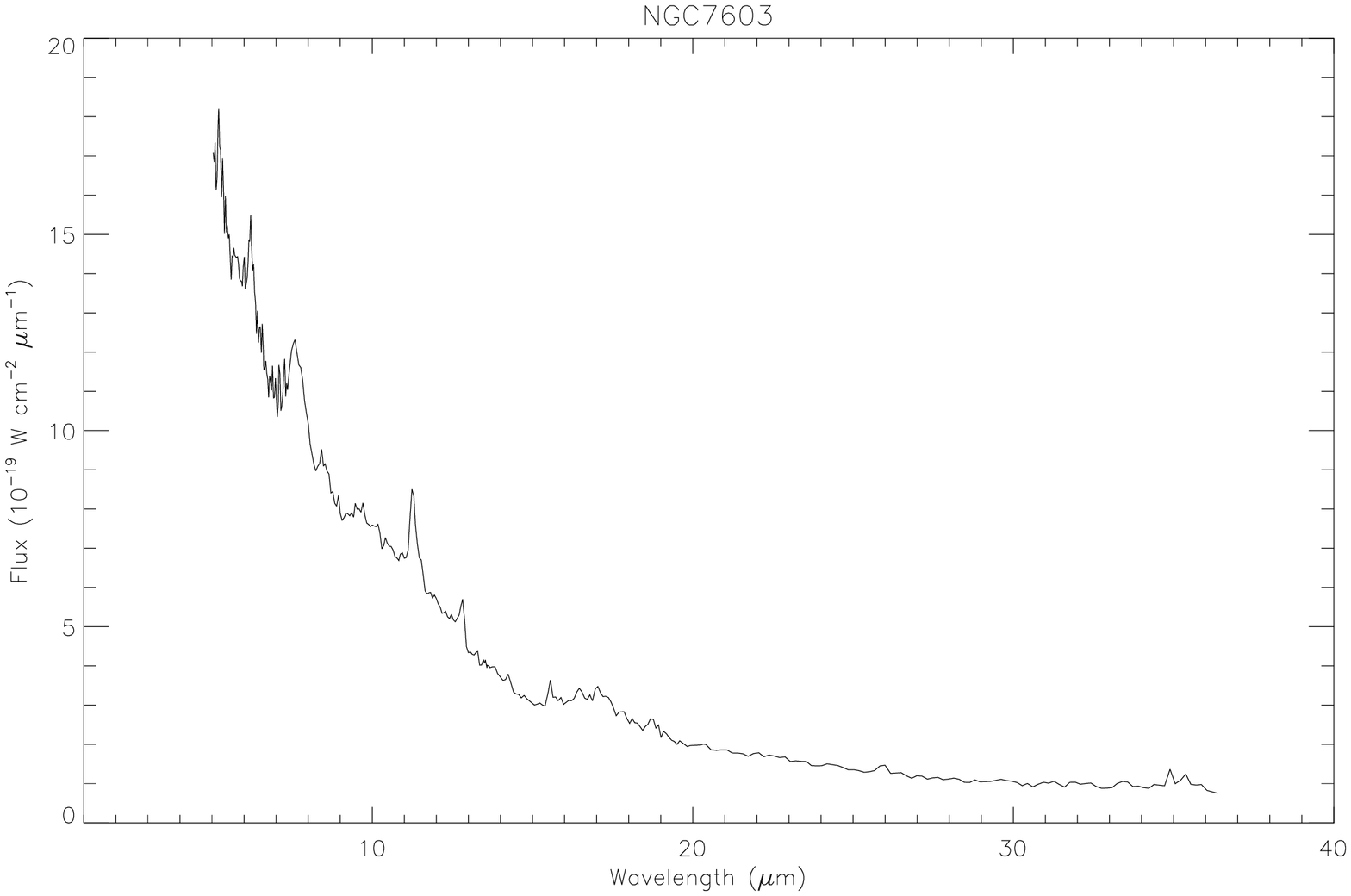} &
    \includegraphics[angle=90, trim=5 5 5
    5,clip,width=0.5\textwidth,height=0.4\textheight]
    {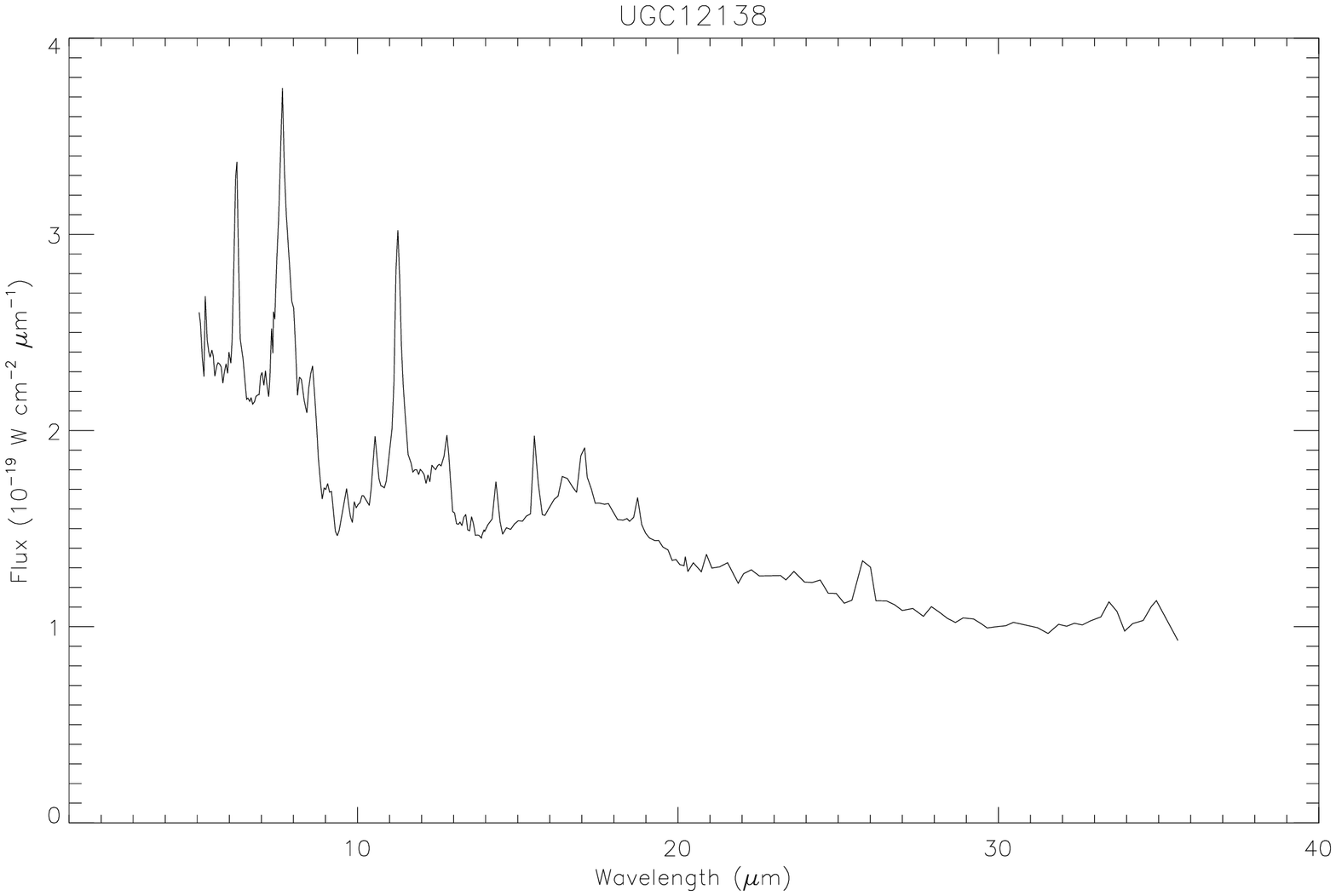}\\[-10pt]
    \multicolumn{1}{l}{\mbox{}} &
    \multicolumn{1}{l}{\mbox{}} \\[-10pt]
    \hspace{-0.6cm}
    \includegraphics[angle=90, trim=5 5 5
    5,clip,width=0.5\textwidth,height=0.4\textheight]
    {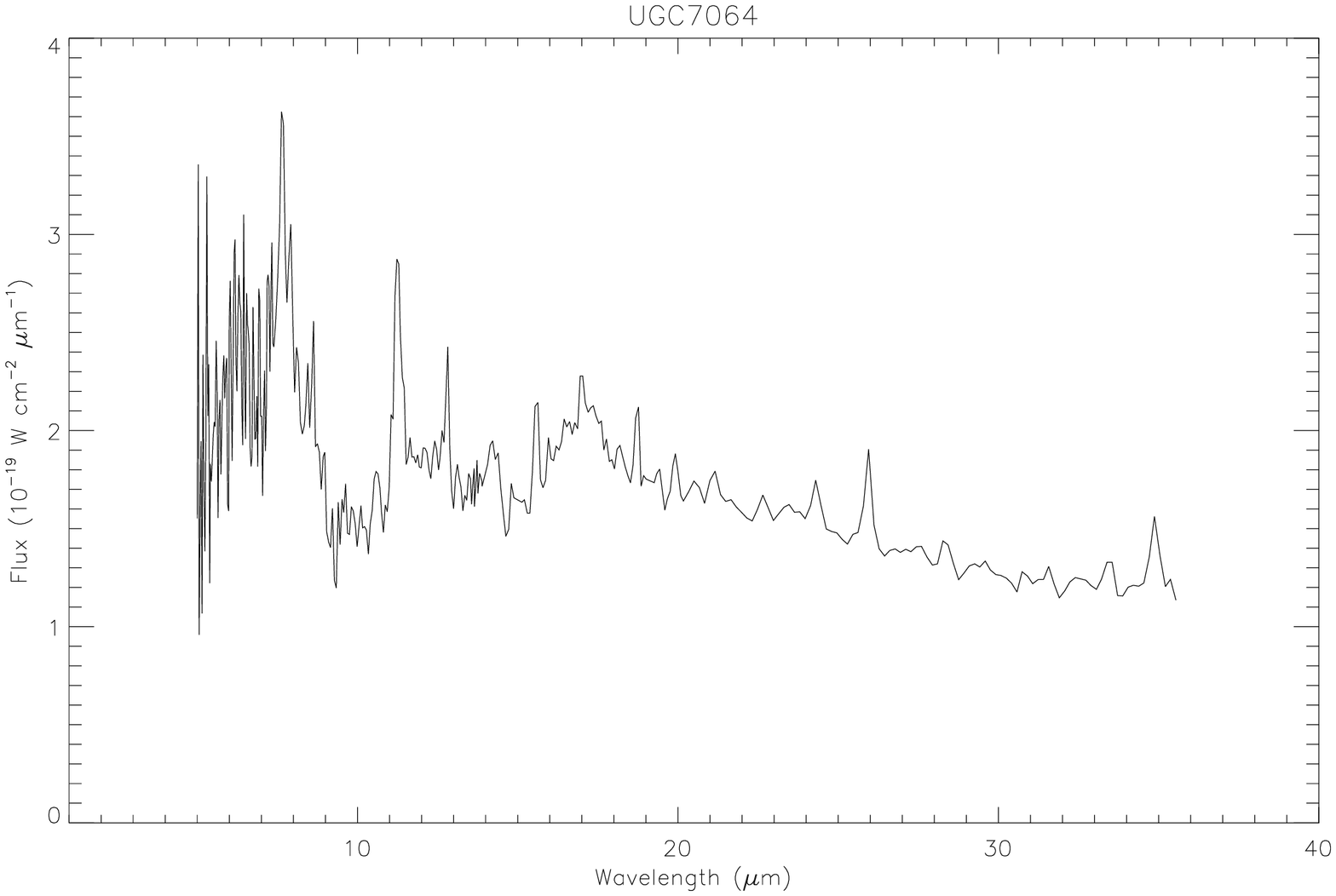} &
    \includegraphics[angle=90, trim=5 5 5
    5,clip,width=0.5\textwidth,height=0.4\textheight]
    {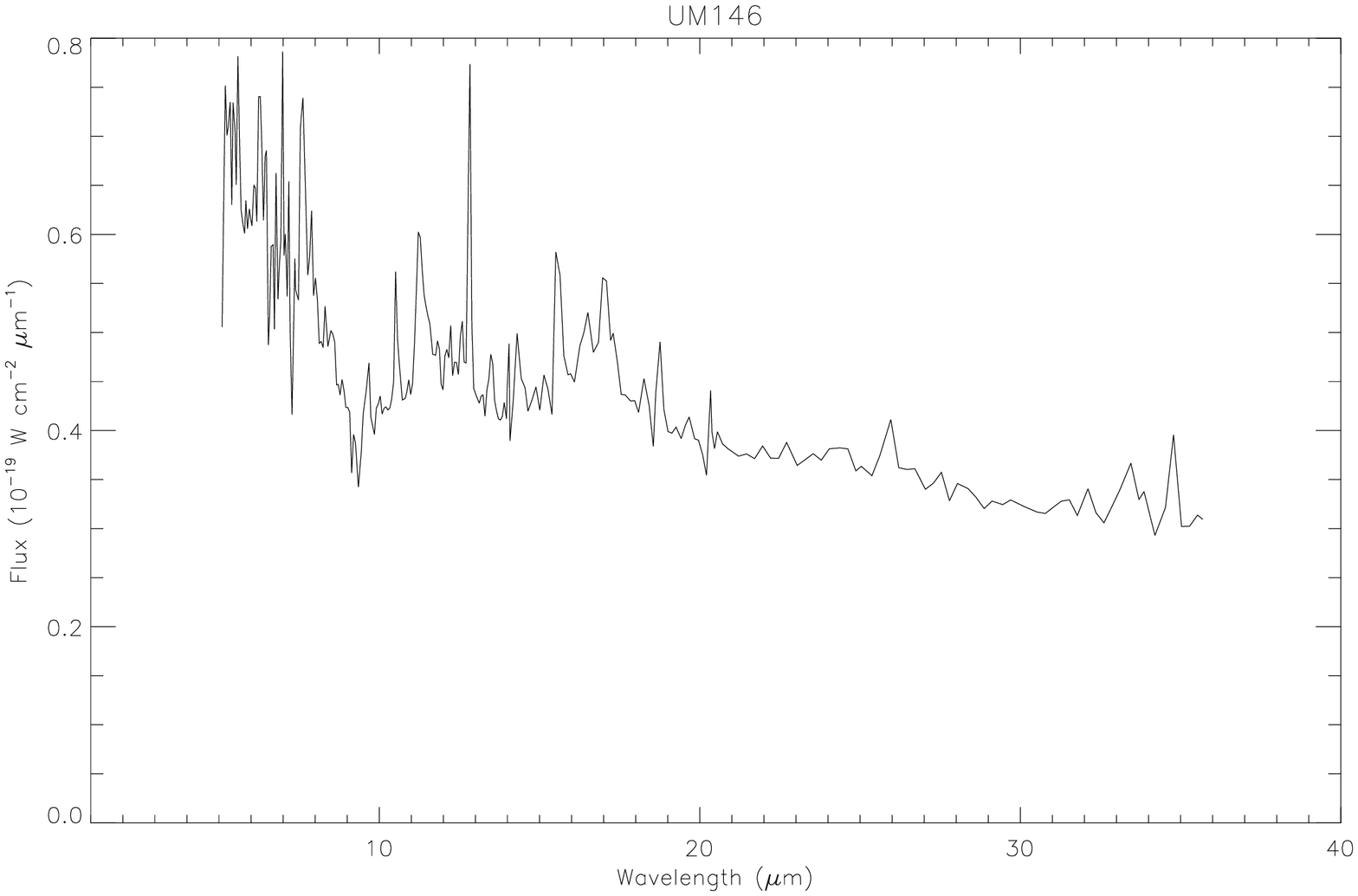}
  \end{array}$
  \caption{{\it Spitzer} IRS spectra of Seyferts (clockwise from top left): NGC 7603, UGC 12138, UGC 7064, and UM 146}
  \label{spitzerfig4}
\end{figure}

\end{document}